\documentclass{cernrep} 
\usepackage{texnames}
\usepackage[T1]{fontenc}
\usepackage[bookmarks, colorlinks=true, linktoc=page, pdftex, linkcolor=red, citecolor=red, urlcolor=red]{hyperref}
\usepackage{slashed,amssymb}
\usepackage{mathtools,graphicx}  
\usepackage[usenames,dvipsnames]{xcolor}
\definecolor{Blu}{rgb}{0.,0.,1.}

\newcommand{\beq}{\begin{equation} }
\newcommand{\eeq}{\end{equation}} 

\DeclareMathOperator{\diag}{diag} 
\DeclareMathOperator{\Imag}{Im}

\begin{document}
\title{Introduction to flavour physics}
 
\author {Jure Zupan}

\institute{Department of Physics, University of Cincinnati, Cincinnati, Ohio 45221,USA}

\begin{abstract}
We give a brief introduction to flavour physics. The first part covers the flavour structure of the Standard Model, how the Kobayashi-Maskawa mechanisem is tested and provides examples of searches for new physics using flavour observables, such as meson mixing and rare decays. In the second part we give a brief overview of the recent flavour anomalies and how the Higgs can act as a new flavour probe. 
\end{abstract}


\maketitle % this produces the title block
 
\section{Introduction}
The term ``flavour'' was coined in 1971 by Murray Gell-Mann and his student at the time, Harald Fritzsch, while sitting at a Baskin-Robbins ice-cream store in Pasadena, CA \cite{Browder:2008em}. Just as ice-cream has both colour and flavour so do quarks.  ``Flavour'' is now used slightly more generally to denote the species of any Standard Model (SM) fermion, both quarks and leptons.  ``Flavour physics'' thus has little to do with one's adventures in kitchen, but rather is a research area that deals with properties of quarks and leptons. 

Grouped according to their QCD and QED quantum numbers, $SU(3)\times U(1)_{\rm em}$, the SM fermions are,
\begin{equation}
\label{eq:fermion:content}
\begin{matrix*}[l]
3_{2/3}: & \text{~~up type quarks;~~} & u,c,t,
\\
3_{-1/3}: & \text{~~down type quarks;~~}& d,s,b,
\\
1_{-1}: & \text{~~charged leptons;~~} & e,\mu, \tau,
\\
1_0: & \text{~~neutrinos;~~} & \nu_e,\nu_\mu, \nu_\tau.
\end{matrix*}
\end{equation}
Each fermion type comes in three copies, i.e., the SM fermions group into three generations.

\begin{figure}
\centering\includegraphics[width=.9\linewidth]{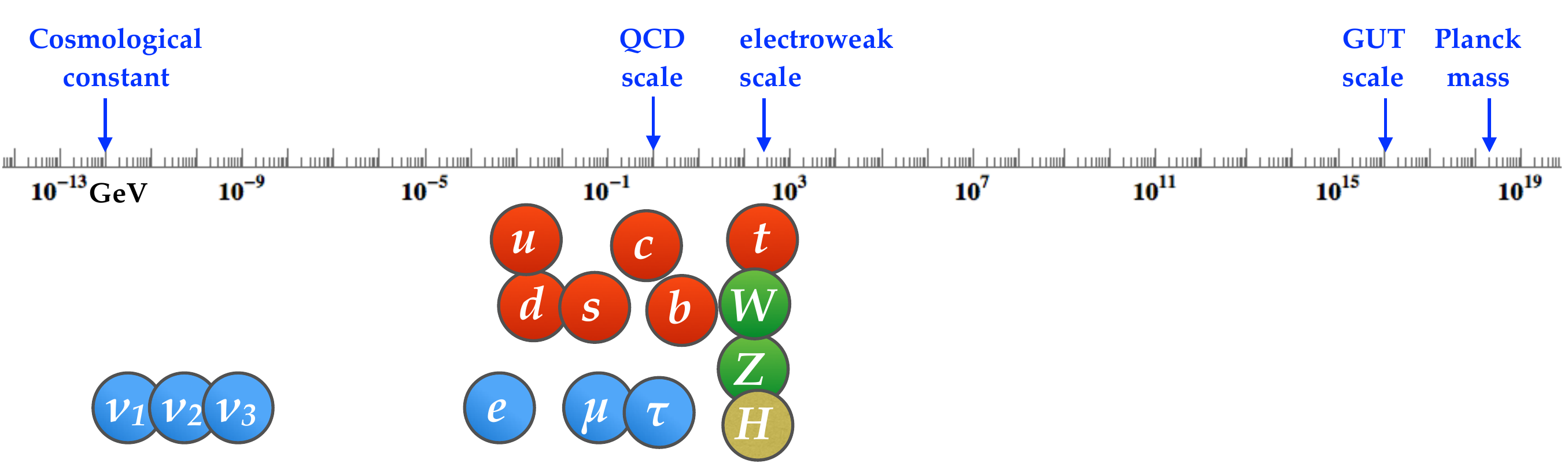}
\caption{The distribution of masses of the elementary particles, along with some of the relevant energy scales. The absolute values of neutrino masses are not known - their placement on the graph is indicative of the upper bound.}
\label{fig:sm_flavour}
\end{figure}

In this brief introduction to flavour physics we will cover some of the classic topics on the subject: the flavour structure of the Standard Model (SM), how the Kobayashi-Maskawa mechanism is tested, as well as the constraints on the New Physics (NP) due to flavour observables such as the meson mixing and decays. We will also touch on the more recent developments: the $B$ physics anomalies and the Higgs as a new probe of flavour. Along the way we will address two major questions currently facing particle physics. The first question is why do the SM fermions exhibit such a hierarchical structure, shown in Fig. \ref{fig:sm_flavour}? This is commonly referred to as the SM flavour puzzle. The other question is what lies above the electroweak scale? Here flavour physics offers a way to probe well above the electroweak scale. 

Other excellent introductions to flavour physics the reader may want to consult include Refs. \cite{Grossman:2017thq,Blanke:2017ohr,Gedalia:2010rj,Nir:2007xn,Kamenik:2017znu,Nierste:2009wg}. Ref. \cite{Grossman:2017thq} in particular is chock full of physics insights without too much burdensome formalism. Section \ref{sec:flavourSM} borrows liberaly from \cite{Nir:2007xn}, which, while slightly outdated, is still a masterful introduction to the basic topics in flavour physics. For a reader that is seeking much more depth a good starting point can be Refs. \cite{Branco:1999fs,Buras:2011we,Kou:2018nap,Cerri:2018ypt}.

%\jz{cite books}  

%Useful references  \jz{explain that I borrowed heavily from Nir, since it is hard to improve on the master, at least in the introductory parts}
%
%In the SM 

\section{The flavour of the Standard Model}
\label{sec:flavourSM}
\subsection{The SM symmetry structure}
%{\it Here we give the basic Lagrangian of the SM, explain the diagonalization, that we think the origin of flavour are the SM Yukawa's, counting of the physical parameters, approximate symmetries}
A renormalizable particle physics model is defined by specifying (i) the gauge group and (ii) the particle field content. The next step is to write down the most general renormalizable Lagrangian. The SM gauge group is 
\beq
\label{eq:SM:gauge:group}
{\cal G}_{\rm SM}= SU(3)_c\times SU(2)_L\times U(1)_Y.
\eeq
Here $SU(3)_c$ is the gauge group of strong interactions, Quantum Chromodynamics (QCD), the $SU(2)_L$ is the gauge group of weak isospin, and $U(1)_Y$ the gauge group of hypercharge. The field content of the SM consists of a single scalar, EW doublet
\beq
\label{eq:H:rep}
H \sim (1,2)_{1/2},
\eeq
and a set of fermion fields, %(right-handed neutrinos may be absent)
\beq
\begin{split}
\label{eq:fermion:rep}
Q_{Li}&\sim (3,2)_{+1/6}, \quad u_{Ri} \sim (3,1)_{+2/3}, \quad d_{Ri}\sim (3,1)_{-1/3},
\\
L_{Li}&\sim (1,2)_{-1/2}, \quad \ell_{Ri}\sim (1,1)_{-1}. 
%\quad ( N_{Ri}\sim (1,1)_0 )
\end{split}
\eeq
Each of the fields comes in three copies (three generations), $i=1,2,3$. To simplify the discussion we will set neutrino masses to zero. The modifications due to nonzero neutrino masses are given in appendix \ref{app:neutrino:masses}.
The ${\cal G}_{\rm SM}$  is spontaneously broken  by the Higgs vacuum expectation value, $\langle H\rangle =(0,v/\sqrt2), v=246$ GeV, down to  
\beq
{\cal G}_{\rm SM}\to SU(3)\times U(1)_{em}.
\eeq
After the electroweak symmetry breaking the field content in \eqref{eq:fermion:rep} splits into up and down quarks, charged leptons and neutrinos as listed in Eq.~\eqref{eq:fermion:content}.

\subsection{The SM Lagrangian}
  The SM Lagrangian is the most general renormalizable Lagrangian that is consistent with the gauge group  ${\cal G}_{\rm SM}$ and the field content \eqref{eq:H:rep}, \eqref{eq:fermion:rep}
  \beq
  {\cal L}_{\rm SM}={\cal L}_{\rm kin}+{\cal L}_{\rm Yukawa}+{\cal L}_{\rm Higgs}.
  \eeq
  The kinetic terms in the Lagrangian are determined by the gauge structure through the covariant derivative
  \beq
  D_\mu\psi=(\partial_\mu+i g_s G_\mu^a t^a+i g W_\mu^i \tau^i+i g' B_\mu Y)\psi.
  \eeq
The strong interaction term is a product of the strong coupling, $g_s$, the eight gluon fields, $G_\mu^a$, and the generators $t^a$ of $SU(3)_c$. For color triplet $\psi$ these are $t^a=\lambda^a/2$, with $\lambda^a$ the eight $3\times3$  Gell-Mann matrices, while for color singlet $\psi$,  $t^a=0$. The $SU(2)_L$ term is a product of the weak coupling, $g$, the three weak gauge bosons, $W_\mu^i$, and the generators of $SU(2)_L$, $\tau^i$ (equal to $\tau^i=\sigma^i/2$ for $\psi$ that is a doublet, with $\sigma^i$ the Pauli matrices, while for singlets $\tau^i=0$). The last term is due to the hypercharge $U(1)_Y$.

The covariant derivatives are flavour blind, i.e., generation independent. For instance, for $Q_L^i$ the kinetic term is
\beq
{\cal L}_{\rm kin}\big|_{Q_L}=i \bar Q_L^i \big( \partial_\mu+i g_s G_\mu^a \tfrac{1}{2}\lambda^a+i g W_\mu^i \tfrac{1}{2}\sigma^i+i \tfrac{1}{6} g' B_\mu \big) \delta^{ij} Q_L^j, 
\eeq
for up quarks it is
\beq
{\cal L}_{\rm kin}\big|_{u_R}=i \bar u_R^i \big( \partial_\mu+i g_s G_\mu^a \tfrac{1}{2}\lambda^a+i \tfrac{2}{3} g' B_\mu \big) \delta^{ij} u_R^j, 
\eeq
and similarly for the other fields. Each of the kinetic terms is invariant under the global $U(3)=SU(3)\times U(1)$ transformations. Thus ${\cal L}_{\rm kin}$ has a global flavour symmetry
\beq\label{eq:Gflavour}
{\cal G}_{\rm flavour}= U(3)_{\rm q}^3\times U(3)_{\rm lep}^2,
\eeq
where 
%\newpage
\begin{align}
U(3)_{\rm q}^3&=U(3)_Q\times U(3)_u \times U(3)_d,
\\
 U(3)_{\rm lep}^2&= U(3)_L \times U(3)_\ell.
\end{align}
That is, each of the five different types of fermions in Eq.~\eqref{eq:fermion:rep} can be separately rotated in flavour space, $\psi^i\to U^i_j \psi^j$, where $U^i_j$ is a unitary $3\times3$ matrix, without changing ${\cal L}_{\rm kin}$.

However,  ${\cal G}_{\rm flavour}$ cannot be an exact symmetry of the whole Lagrangian. We know from observations that, e.g., the top quark differs from the up quark due to their differing masses. The 
part of the Lagrangian that breaks ${\cal G}_{\rm flavour}$ is 
\beq
\label{eq:Yukawa}
{\cal L}_{\rm Yukawa}=- Y_d^{ij}\bar Q_{L}^i H d_R^j- Y_u^{ij} \bar Q_L^i H^c u_R^j - Y_\ell^{ij} \bar L_L^i H \ell_R^j+{\rm h.c.}.
\eeq
The above Yukawa interactions break 
\beq
\label{eq:Gflavour:breaking}
{\cal G}_{\rm flavour}\to U(1)_B\times U(1)_e\times U(1)_\mu\times U(1)_\tau \times U(1)_Y,
\eeq
 where  $U(1)_B$ is the baryon number, and $U(1)_\ell$ are the separate lepton numbers. That ${\cal L}_{\rm Yukawa}$ breaks the flavour symmetry is not surprising, since it is the origin of fermion masses, once the Higgs obtains the vacuum expectation value (vev), $\langle H\rangle = (0,v/\sqrt{2})$, with $v=246$ GeV. 

\subsection{A Standard Model vs. the Standard Model}
Before we proceed further in understanding the breaking pattern in Eq.~\eqref{eq:Gflavour:breaking}, let us make a small detour and elaborate on the difference between {\it a Standard Model} and {\it the Standard Model}. {\it A Standard Model } denotes any model with the SM gauge group \eqref{eq:SM:gauge:group} and the SM field content \eqref{eq:H:rep}, \eqref{eq:fermion:rep}, but with some arbitrary values for the coupling constants in the most general renormalizable Lagrangian. {\it The Standard Model} is a Standard Model with exactly the values of coupling constants observed in nature. {\it A Standard Model} has the exact accidental symmetry $U(1)_B\times U(1)_e\times U(1)_\mu\times U(1)_\tau$. This accidental symmetry is present for any values of the parameters in the renormalizable SM Lagrangian (but can be broken by non-renormalizable terms). It is accidental, since we did not explicitly ask for it -- it is simply present because we cannot write down renormalizable terms that break it, given the field and gauge content in Eqs. \eqref{eq:SM:gauge:group}-\eqref{eq:fermion:rep}. For {\it the Standard Model}, because of the actual values of the parameters, there can be additional approximate symmetries. 

Isospin is an example of such an approximate symmetry. In QCD interactions one can replace $u$ and $d$ quarks without affecting appreciably the results. For instance, the neutron and proton masses are very close to each other even though, $p\sim uud$, while $n\sim udd$. The reason is not that up and down quark masses would be equal to each other but rather that they are both small, cf. Fig. \ref{fig:sm_flavour},
\beq
\frac{|m_u-m_d|}{\Lambda_{\rm strong}}\ll 1.
\eeq
Here ${\Lambda_{\rm strong}}\sim{\mathcal O}(1 {\rm GeV})$ is the typical scale at which QCD becomes nonperturbative and generates the bulk of the mass for proton and neutron. 

\subsection{Counting physical parameters}
\label{sec:counting:physical:params}
The next question we need to address is how one counts the physical parameters. The SM has 19 physical parameters: 3 gauge couplings, 3 lepton masses, 6 quark masses, 4 parameters in the Cabibbo-Kobayashi-Maskawa (CKM) matrix, 2 parameters in the Higgs sector (the Higgs mass and the strength of the self interaction), and the QCD $\theta$ parameter. Physical parameters are parameters that cannot be rotated away by performing phase transformations or flavour rotations.

 Let us understand this on the case of charged lepton masses. The charge lepton Yukawa 
   \beq
   \label{eq:lepton:Yukawa}
{\cal L}_{\rm Yukawa}\supset  - Y_\ell^{ij} \bar L_L^i H \ell_R^j+{\rm h.c.},
\eeq
can always be made diagonal and real positive through a bi-unitary transformation, $L_L\to V_L L_L$, $\ell_R\to V_\ell \ell_R$, which gives
\beq
Y_\ell\to V_L^\dagger Y_\ell V_\ell =\diag(y_e, y_\mu, y_\tau).
\eeq
How many physical parameters are there? The starting point, $Y_\ell$, is described by 9 real and 9 imaginary numbers. The unitary matrices $V_L, V_\ell$ have in total $2\times(3{\rm~real}+6{\rm~im.})$ numbers. When we rotate $L_L^i$ and $\ell_R^i$ by the same phase there is no change in $y_{\ell_i}$. That means that 3 phases (im. numbers) have no effect. Thus we have $9-2\times 3=3$ real, and $9-(2\times 6-3)=0$ imaginary physical parameters. The three real physical parameters are the charged lepton masses, while there are no physical phases. 

Extrapolating from this exercise we can postulate the general rule on how to count the physical parameters \cite{Grossman:2017thq}
\beq
\label{eq:general:rule}
\# \text{~physical parameters}= \# \text{~parameters}- \# \text{~broken symmetry generators}.
\eeq
Let us check this with a simple example: the spin $1/2$ in a magnetic field. If there is no magnetic field the system has an $SO(3)$ symmetry (3 generators), since the spin can be oriented in an arbitrary direction without changing the energy. The system also has two degenerate eigenstates corresponding to spin up and spin down. In the magnetic field the Zeeman effect splits the two states. The splitting depends on the strength of the magnetic field, $B$. There is thus one physical parameter that controls the splitting. However, the magnetic field in general has three components, and is thus described by 3 parameters, $\vec B= B_x \hat x+B_y \hat y +B_z\hat z$. One can use the rotation around $x$ and $y$ axes to align $\vec B$ along the $z$ axis, i.e., set $B_x=B_y=0$. After this is done, making any further rotations around $x$ and $y$ axes would change the $\vec B$ component: there are 2 broken symmetry generators. Using the general rule \eqref{eq:general:rule} gives that there is $3-2=1$ physical parameter, as expected. 

We can now apply \eqref{eq:general:rule} to count the physical parameters in the quark sector of the SM. Using the unitary transformations
\beq
Q_L\to V_Q Q_L, \qquad u_R\to V_u u_R, \qquad  d_R\to V_d d_R,
\eeq
one can bring the Yukawa couplings to the form
\beq
\label{eq:Yd:VYu}
Y_d=\diag(y_d,y_s,y_b), \qquad Y_u=V_{\rm CKM}^\dagger \diag(y_u,y_c,y_t),
\eeq
with $V_{\rm CKM}$ a unitary $3\times 3$ CKM matrix \cite{Cabibbo:1963yz,Kobayashi:1973fv}. How many entries in $V_{\rm CKM}$ are physical? The starting point, the $Y_u, Y_d$ matrices, have $2\times (9{\rm~real}+9{\rm~im.})$ parameters. The three unitary matrices, $V_{Q}, V_u, V_d$ have in total $3\times (3{\rm~real}+6{\rm~im.})$ parameters. Finally, there is one global phase corresponding to common phase change $Q_L\to \exp(i\phi) Q_L, u_R\to \exp(i\phi) u_R,  d_R\to \exp(i\phi) d_R$, which has no effect. That is, there is one unbroken symmetry generator -- the baryon number, while all the other symmetry generators are broken. Using  \eqref{eq:general:rule} we see that there are $2\times 9-3\times 3=9$ real parameters and $2\times 9-(3\times 6-1)=1$ imaginary physical parameter. These are the 6 quark masses, as well as the 3 mixing angles and one phase describing the CKM matrix. 

A conventional parametrization of the CKM matrix is \cite{Chau:1984fp}
\beq
\label{eq:CKM:param}
\begin{split}
V_{\rm CKM}&=
\begin{pmatrix}
1 & 0 & 0 \\
0 & c_{23} & s_{23} \\
0 & -s_{23} & c_{23}
\end{pmatrix}
\begin{pmatrix}
c_{13} & 0 & s_{13} e^{-i\delta}\\
0 & 1 & 0 \\
-s_{13}e^{i\delta} & 0 & c_{13}
\end{pmatrix}
\begin{pmatrix}
c_{12} & s_{12} & 0\\
-s_{12} & c_{12} & 0 \\
0 & 0 & 1
\end{pmatrix}
\\
&=
\begin{pmatrix}
c_{12} c_{13} & s_{12} c_{13} & s_{13} e^{-i\delta}\\
-s_{12}c_{23}-c_{12}s_{23}s_{13}e^{i\delta} & c_{12}c_{23}-s_{12}s_{23}s_{13}e^{i\delta} & s_{23}c_{13} \\
s_{12}s_{23}-c_{12}c_{23}s_{13}e^{i\delta} & -c_{12}s_{23}-s_{12}c_{23}s_{13}e^{i\delta} & c_{23} c_{13}
\end{pmatrix},
\end{split}
\eeq
where $c_{ij}\equiv \cos\theta_{ij}$, $s_{ij}\equiv \sin\theta_{ij}$, so that the CKM matrix is a product of three rotations with one phase inserted in the matrix describing the $\theta_{13}$ rotation. Experimentally, we observe that $\theta_{12}\gg \theta_{23}\gg \theta_{13}$, while $\delta \sim {\mathcal O}(1)$.

As the side benefit of the counting of physical parameters we just performed, we also understand that the flavour breaking due to the Yukawa matrices is as given in Eq. \eqref{eq:Gflavour:breaking}. In more detail, if we were to take nonzero just a single Yukawa coupling matrix at the time, the breaking pattern is
\begin{itemize}
\item since $Y_\ell \not \propto {1}$:  $U(3)_L \times U(3)_\ell \to  U(1)_e\times U(1)_\mu\times U(1)_\tau$, i.e., the charged lepton family numbers,
\item since $Y_u \not \propto {1}$:  $U(3)_Q \times U(3)_u \to  U(1)_u\times U(1)_c \times U(1)_t$, i.e., the up-quark family  numbers,
\item since $Y_d \not \propto {1}$:  $U(3)_Q \times U(3)_d \to  U(1)_d\times U(1)_s \times U(1)_b$, i.e., the down-quark family number,
\item since $[Y_d, Y_u] \ne 0$:   $U(1)_q^6\to U(1)_B$, i.e., the above quark $U(1)$'s further break to a global baryon number.
\end{itemize}
Note that the final $U(1)$'s are composed both from the $U(1)$ factors in the original $[U(3)=SU(3)\times U(1)]$'s, as well as from the $t^3$ and $t^8$ generators of the $SU(3)$'s. In particular, not all of the $U(1)$ factors in  ${\cal G}_{\rm flavour}$ get broken by the Yukawas. The ${\cal G}_{\rm flavour}$ contains five $U(1)$ factors, which can be chosen to be 
 $U(1)^5 =U(1)_Y\times U(1)_B\times U(1)_L \times U(1)_{\rm PQ}\times U(1)_{\ell_R}$.
The $U(1)_Y$ is the hypercharge group, which is gauged, while $B$ and $L$ are the global baryon and lepton numbers. These are not broken by ${\cal L}_{\rm Yukawa}$. The remaining two global  $U(1)$'s can be taken to be  the Peccei-Quinn symmetry $U(1)_{\rm PQ}$  ($H$ and $d_R^i, \ell_R^i$ have opposite charges, all others zero), while under $U(1)_{\ell_R}$ only $\ell_R^i$ is charged. The $U(1)_{\rm PQ}$ is broken by $Y_u\ne 0$, and $U(1)_{\ell_R}$ by $Y_{\ell}\ne 0$. Had we included neutrino masses in the discussion, these would furthermore break the separate lepton numbers to a common lepton number, $U(1)_L$, if the neutrino masses are Dirac, while Majorana masses also break $U(1)_L$, see appendix \ref{app:neutrino:masses}.

%The above symmetry contains 

\subsection{The flavour violation as seen in the mass basis}

\begin{figure}
\centering
\includegraphics[width=.2\linewidth]{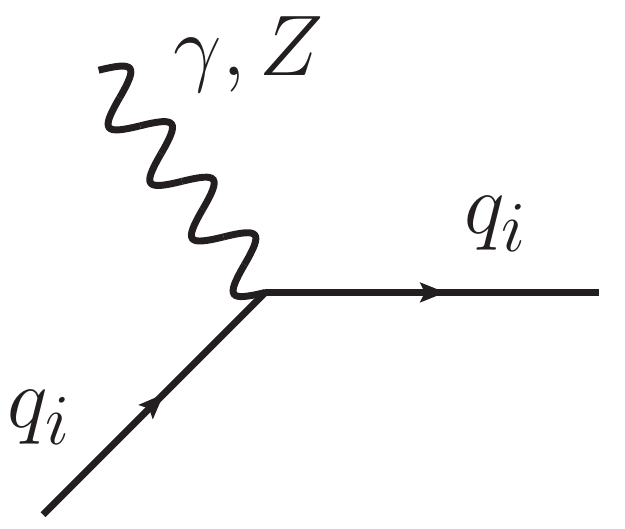}~~~
\includegraphics[width=.2\linewidth]{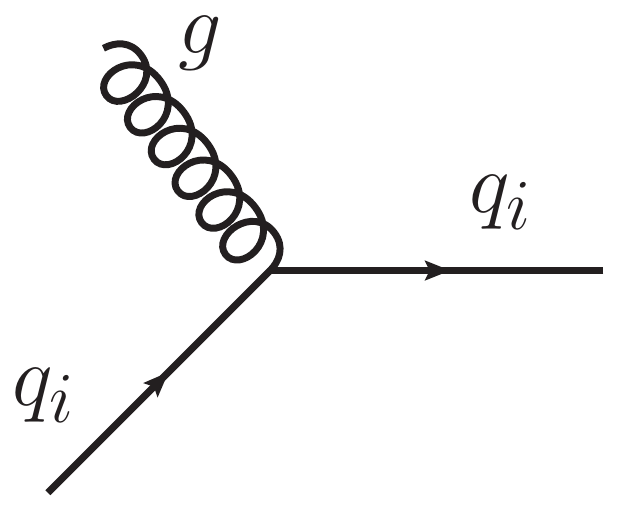}~~~
\includegraphics[width=.2\linewidth]{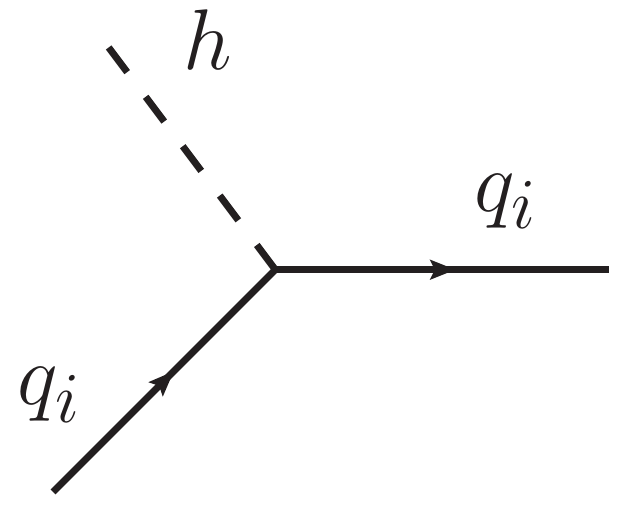}~~~
\includegraphics[width=.2\linewidth]{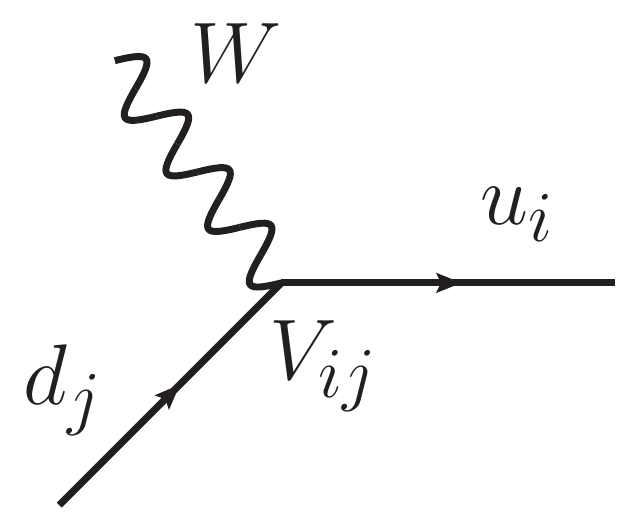}
\\[-6mm]
\includegraphics[width=.1\linewidth]{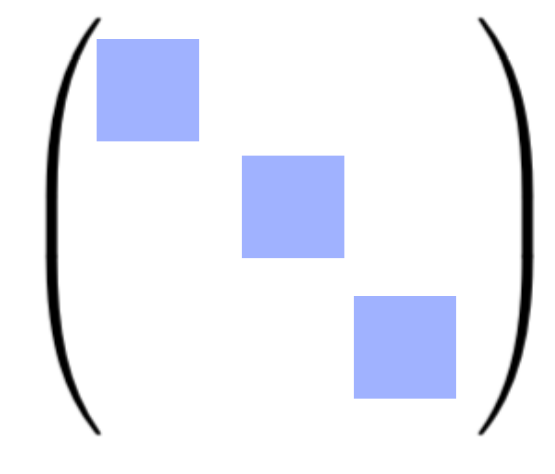}~~~~~~~~~~~~~~~~~~~~
\includegraphics[width=.1\linewidth]{figs/flavour_diag}~~~~~~~~~~~~~~~~~~~~
\includegraphics[width=.1\linewidth]{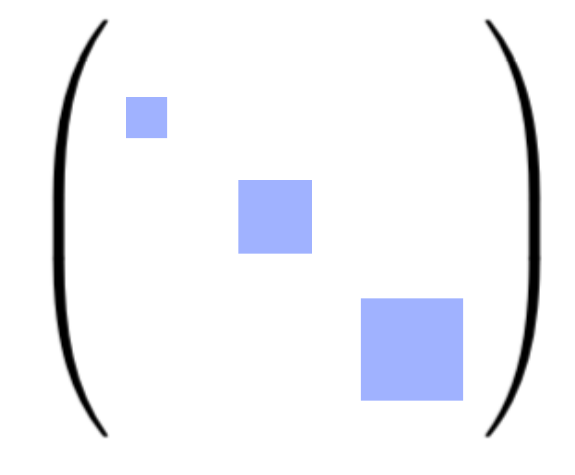}~~~~~~~~~~~~~~~~~~~~
\includegraphics[width=.1\linewidth]{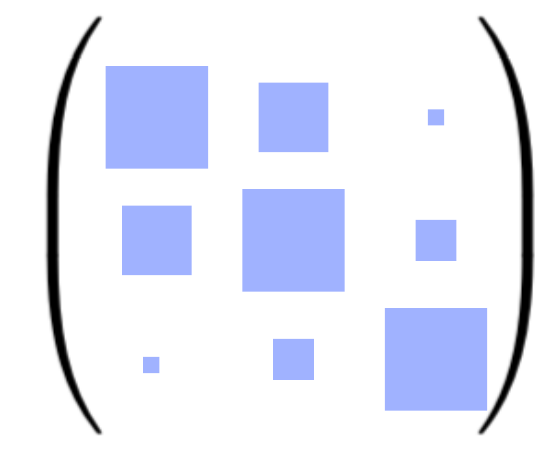}
\caption{The Feynman diagrams for flavour conserving couplings of quarks to photon, $Z$ boson, gluon and the Higgs (the first three diagrams),  and  the flavour changing coupling to the $W$ (the last diagram). The $3\times 3$ matrices are visual representations of couplings in the generation space, with couplings to $\gamma, Z, g$ flavour universal, the couplings to the Higgs flavour diagonal but not universal, and the couplings to $W$ flavour changing and hierarhical. }
\label{fig:Feynman:diagrams}
\end{figure}

The main message of the discussion so far is: in the SM the flavour structure (flavour breaking) resides in the Yukawa sector of the SM Lagrangian, Eq. \eqref{eq:Yukawa}.   If the Yukawa couplings were vanishingly small, the SM would have had a very large  flavour symmetry $
{\cal G}_{\rm flavour}$, Eq.~\eqref{eq:Gflavour}. In general, the flavor breaking can be parametrized as in Eq. \eqref{eq:Yd:VYu}, by 6 diagonal Yukawa couplings, and the elements of the CKM matrix, $V_{\rm CKM}$. 

After Higgs obtains the vev, the Yukawa terms give the quark and charge lepton masses,
\beq
{\cal M}_f=Y_f \frac{(v+h)}{\sqrt2}.
\eeq
With a field redefinition for the left-handed up quark fields
\beq
Q_L\to 
\begin{pmatrix}
V^\dagger u_L
\\
d_L
\end{pmatrix},
\eeq
we can move  the flavour changing interactions to the kinetic term. This
gives the SM Lagrangian for the quarks in the mass basis
\beq
{\cal L}_{\rm SM}\supset (\bar q_i \slashed D_{\rm NC} q_i) +\frac{g}{\sqrt 2} \bar u_L^i \slashed W^+ V_{\rm CKM}^{ij} d_L^j+ m_{u_i} \bar u_L^i u_R^i \Big(1+\frac{h}{v}\Big)+m_{d_i} \bar d_L^i d_R^i \Big(1+\frac{h}{v}\Big) +{\rm h.c.}.
\eeq
The covariant derivative $D_{\rm NC}$ contains flavour (generation) universal couplings of photon, gluon and the $Z$. The Higgs has flavour diagonal, yet non-universal, couplings that are proportional to quark masses, while the flavour changing transitions reside in charged currents, with the strength encoded in the CKM matrix, see Fig. \ref{fig:Feynman:diagrams}.

\subsection{Charged currents vs. neutral currents}
\label{sec:charged:neutral}
In the SM there is a very important distinction  between flavour changing neutral and charged currents. Flavour Changing Neutral Currents (FCNCs)  are processes in which the quark flavour changes, while the quark charge stays the same. The charged currents change both the flavour and the charge of the quark. A glimpse at the PDG booklet \cite{Tanabashi:2018oca} reveals that the probabilities for the two types of processes are strikingly different. The charged currents lead to the dominant weak decays, while the FCNC induced decays are extremely suppressed. Rounding the  experimental results, and not showing the errors, a few representative decays are
\begin{align*}
&&&\text{charged currents:} &&& &\text{neutral currents:} &
\\
&s\to u\mu^- \bar \nu_\mu&: &\quad Br(K^+ \to \mu^+\nu)=64\%,\quad && s\to d\mu^+\mu^-&: & \quad Br(K_L \to \mu^+\mu^-)=7 \times 10^{-9},
\\
&b\to c\ell^- \bar \nu_\ell&: &\quad Br(B^- \to D^0 \ell \bar \nu)=2.3\%,\quad && b\to d\mu^+\mu^-&: & \quad Br(B^-\to K^{*-}\ell^+\ell^-)=5 \times 10^{-7},
\\
&c\to s\mu^+\nu_\mu&: &\quad Br(D^\pm \to K^0 \mu^\pm \nu)=9\%,\quad && c\to u\ell^+\ell^-&: & \quad Br(D^0\to \pi^0\ell^+\ell^-)<1.8\times 10^{-4},
\end{align*}

\begin{figure}
\centering
\includegraphics[width=.22\linewidth]{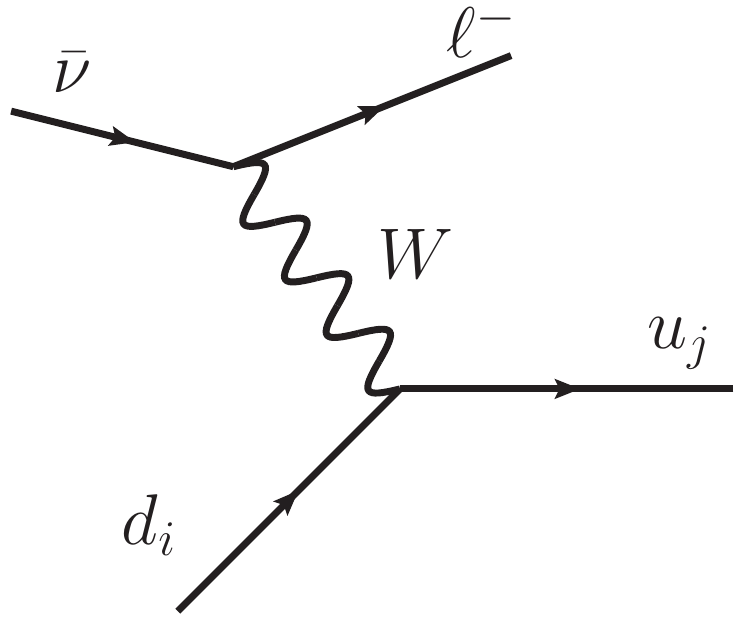}~~~~~~~~
\includegraphics[width=.28\linewidth]{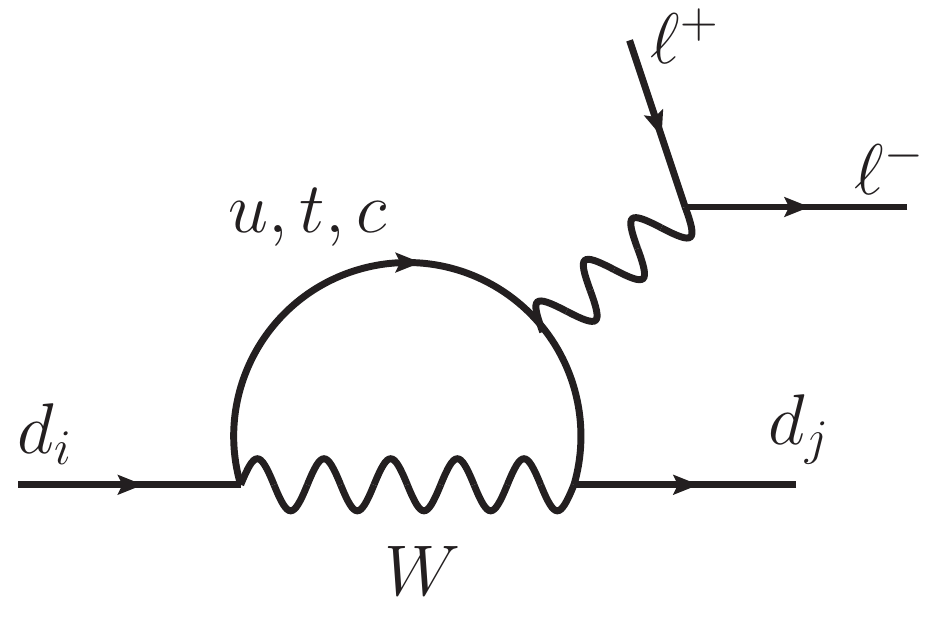}
\caption{Representative tree level charged current diagram (left) and a loop induced  FCNC diagram (right). }
\label{fig:CC:FCNC}
\end{figure}

The reason for such a striking difference is that in the SM the charged currents occur at tree level, while FCNCs are forbidden at tree level and only arise at one loop, see Fig.~\ref{fig:CC:FCNC}. Furthermore, the FCNCs come suppressed by the difference of the masses of the quarks running in the loop, $m_j^2-m_i^2$.  This so called Glashow-Iliopoulos-Maiani (GIM) mechanism \cite{Glashow:1970gm} is a result of the fact  that there is no flavour violation, if all the quark masses are the same.  

\subsection{The CKM matrix}
The Cabibbo-Kobayashi-Maskawa (CKM) matrix is very hierarchical in the SM, 
\beq
V_{\rm CKM}=
\begin{pmatrix}
V_{ud} &V_{us} & V_{ub}
\\
V_{cd} & V_{cs} & V_{cb}
\\
V_{td} & V_{ts} & V_{tb}
\end{pmatrix}
\sim
\begin{pmatrix}
1 & 0.2 & 0.004
\\
0.2  & 1 & 0.04
\\
0.008 & 0.04 & 1
\end{pmatrix}
.
\eeq
In fact, for processes at colliders in many cases the CKM matrix can even be approximated as
\beq
 V_{\rm CKM}
\sim
\begin{pmatrix}
1 & 0 & 0
\\
0  & 1 & 0
\\
0 & 0 & 1
\end{pmatrix}
,\qquad [\text{collider physicist]}
\eeq
i.e., for many processes at high $p_T$ to a good enough precision the generation number is conserved.

We, on the other hand, are interested precisely in the off-diagonal entries in $V_{\rm CKM}$. These entries roughly obey a power scaling in $\lambda\equiv |V_{us}|\simeq 0.22$, giving the Wolfenstein parametrization of the CKM matrix \cite{Wolfenstein:1983yz},
\beq\label{eq:VCKM:Wolfenstein}
V_{\rm CKM}=
\begin{pmatrix}
1-\lambda^2/2 & \lambda  & A \lambda^3 (\rho - i \eta)
\\
-\lambda & 1-\lambda^2/2 & A \lambda^2
\\
A \lambda^3 (1-\rho-i \eta) & -A \lambda^2 & 1
\end{pmatrix}
+{\mathcal O}\big(\lambda^4\big).
\eeq
This parametrization also encodes that the CKM matrix is unitary,  $V_{\rm CKM}^\dagger V_{\rm CKM}=V_{\rm CKM} V_{\rm CKM}^\dagger=1$. The CKM matrix depends on 3 real parameters and 1 phase. In parametrization of Eq. \eqref{eq:CKM:param} these were the three mixing angles and the phase $\delta$. In the Wolfenstein parametrization, Eq. \eqref{eq:VCKM:Wolfenstein}, these are the three real parameters $\lambda$, $A$, $\rho$, and one imaginary parameter, $\eta$, all counted as being ${\mathcal O}(1)$.  A global fit to the flavour observables gives \cite{Charles:2004jd}
\beq
\label{eq:CKM:values}
A=0.825(9), \qquad \lambda=0.2251(3), \qquad \bar \rho=0.160(7), \qquad \bar \eta=0.350(6),
\eeq
where the modified $\rho,\eta$ parameters were introduced as $\bar \rho+ i \bar \eta=-{V_{ud}V_{ub}^*}/({V_{cd}V_{cb}^*})$, valid to all orders in $\lambda$. To ${\mathcal O}(\lambda^4)$ we have $\bar \rho=\rho(1-\lambda^2/2)$ and $\bar \eta=\eta(1-\lambda^2/2)$. Note that numerically $\bar \rho, \bar \eta$ are maybe closer to $\bar \rho, \bar \eta\sim {\mathcal O}(\lambda)$ than $\bar \rho, \bar \eta\sim {\mathcal O}(1)$, while at the time when  Wolfenstein parametrization was written down this was not known. This can be incorporated in modified expansions \cite{Ahn:2011fg}, though the change in counting only matters at higher orders, not for the leading order expressions in Eq.~\eqref{eq:VCKM:Wolfenstein}.

\subsection{Origin of CP violation in the SM}
\label{sec:CPV:SM}
The SM Lagrangian is invariant under the discrete CP symmetry, apart from the Yukawa terms.\footnote{There is another CP violating parameter, the strong CP phase multiplying the QCD anomaly term, $ g^2/(32 \pi^2) \theta G^{a\mu\nu} \tilde G^a_{\mu\nu}$. It is bounded experimentally to be small, $\theta\lesssim 10^{-10}$ and, even if eventually found to be nonzero, is negligible for all the processes discussed in these lectures.} These transform as  (writing explicitly also the hermitian conjugate terms)
\beq
Y_{ij} \bar \psi_L^i H \psi_R^j+Y_{ij}^* \bar \psi_R^j H^\dagger \psi_L^i \xrightarrow{\text{CP}} Y_{ij} \bar \psi_R^j H^\dagger \psi_L^i+Y_{ij}^* \bar \psi_L^i H \psi_R^j.
\eeq
The CP is conserved, if Yukawa couplings are real,
\beq
Y_{ij}^*=Y_{ij}.
\eeq
Since there is only one physical phase in the CKM, in the SM the CP violation (CPV) is controlled by one parameter, the ``CKM phase'', which in the Wolfenstein parametrization is the parameter $\eta$. CP is thus violated only, if $\eta\ne0$. This origin of the observed CPV is called the Kobayashi-Maskawa (KM) mechanism \cite{Kobayashi:1973fv}.
Furthermore, CPT is conserved in any Lorentz invariant Quantum Field Theory, and therefore also in the SM. This means that CPV is equivalent to having T violation -- the time reversal is also violated in the SM.

For the existence of CPV in the SM it is crucial that there are at least 3 generations of quarks. Repeating the counting of physical parameters from Sec. \ref{sec:counting:physical:params} we can easily convince ourselves that it is possible in the case of 2 generations  to make CKM real through field redefinitions. Furthermore, if $Y_u$ and $Y_d$ are ``aligned'', meaning that they are diagonalized with the same left-handed rotation, then $V_{\rm CKM}=1$. This means that in the SM, if there is no flavour violation, there is also no CP violation (ignoring the flavour universal, but numerically negligible $\theta$ term). 

The above insights can be encoded in a measure of CP violation, the Jarlskog invariant \cite{Jarlskog:1985ht}
\beq
J_Y\equiv \Imag\big(\det\big[Y_dY_d^\dagger, Y_u Y_u^\dagger\big]\big).
\eeq
The $J_Y$ is invariant under flavour transformations, ${\cal G}_F$, Eq. \eqref{eq:Gflavour}, and is thus basis independent. The 
CP is conserved, if $J_Y=0$. We can also write $J_Y$ as 
\beq
J_Y=J_{\rm CP} \prod_{i>j}\frac{m_i^2-m_j^2}{v^2/2}\simeq {\mathcal O}(10^{-22}),
\eeq
where the invariant measure of CP violation is
\beq
J_{\rm CP}=\Imag\big[V_{us}V_{cb} V_{ub}^* V_{cs}^*\big]=c_{12}c_{23}c_{13}^2 s_{12} s_{23}s_{13} \sin\delta_{\rm KM} \simeq \lambda^6 A^2 \eta\simeq {\mathcal O}(10^{-5}).
\eeq
The product of masses is
\beq
\prod_{i>j}\frac{m_i^2-m_j^2}{v^2/2}=\frac{(m_t^2-m_c^2)}{v^2/2}\frac{(m_t^2-m_u^2)}{v^2/2}\frac{(m_c^2-m_u^2)}{v^2/2} \frac{(m_b^2-m_s^2)}{v^2/2}\frac{(m_b^2-m_d^2)}{v^2/2}\frac{(m_s^2-m_d^2)}{v^2/2}.
\eeq
It would vanish, if any of the two pairs of masses were equal, in which case CP would have been conserved.

\section{Tests of the CKM structure}
\subsection{The standard CKM unitarity triangle}

\begin{figure}
\centering
\includegraphics[width=.4\linewidth]{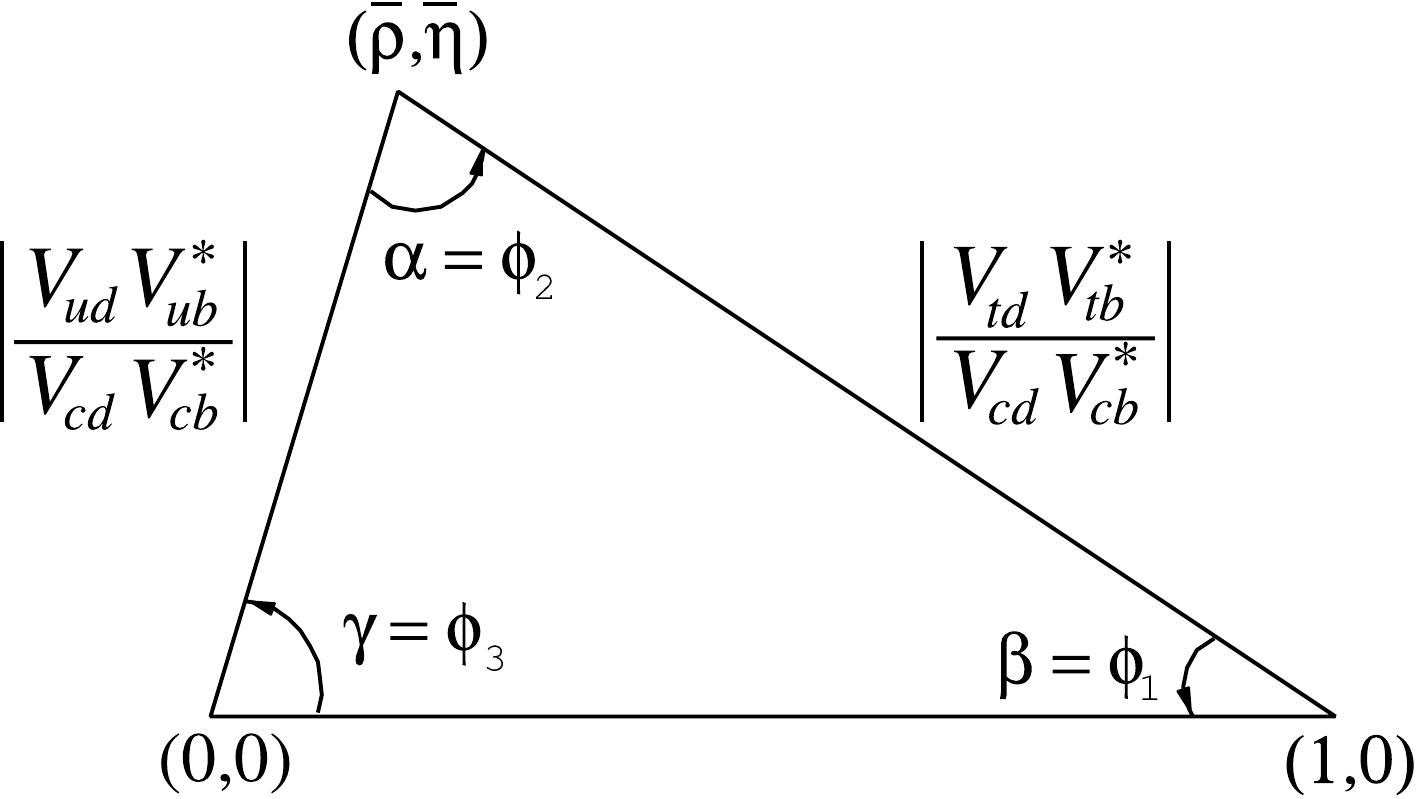}~~~~~~~~
\caption{The standard CKM unitarity triangle (from  \cite{Tanabashi:2018oca}).}
\label{fig:CKMtriangle}
\end{figure}

\begin{figure}
\centering
\includegraphics[width=.32\linewidth]{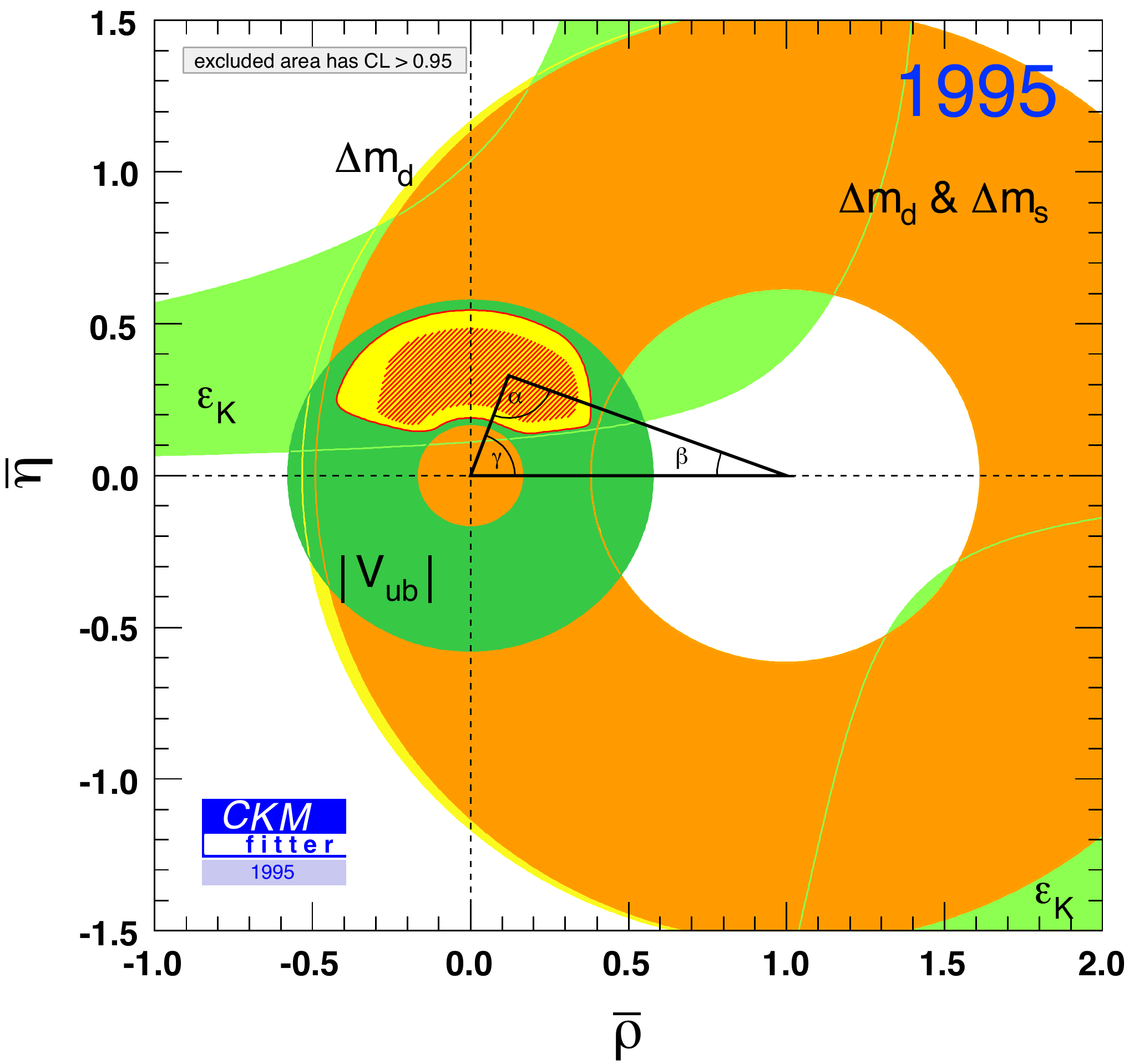}~~~~
\includegraphics[width=.32\linewidth]{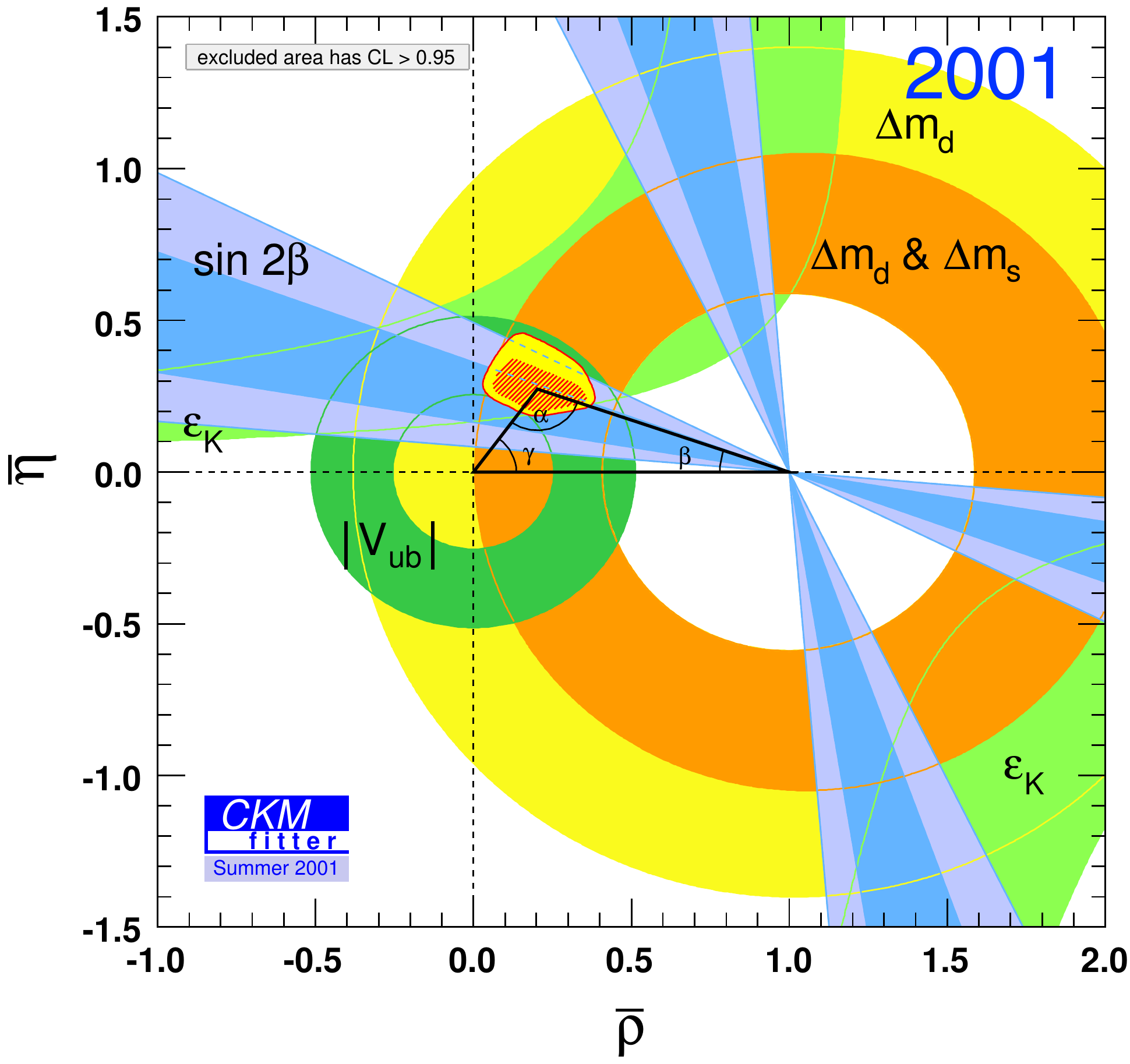}~~~~
\includegraphics[width=.32\linewidth]{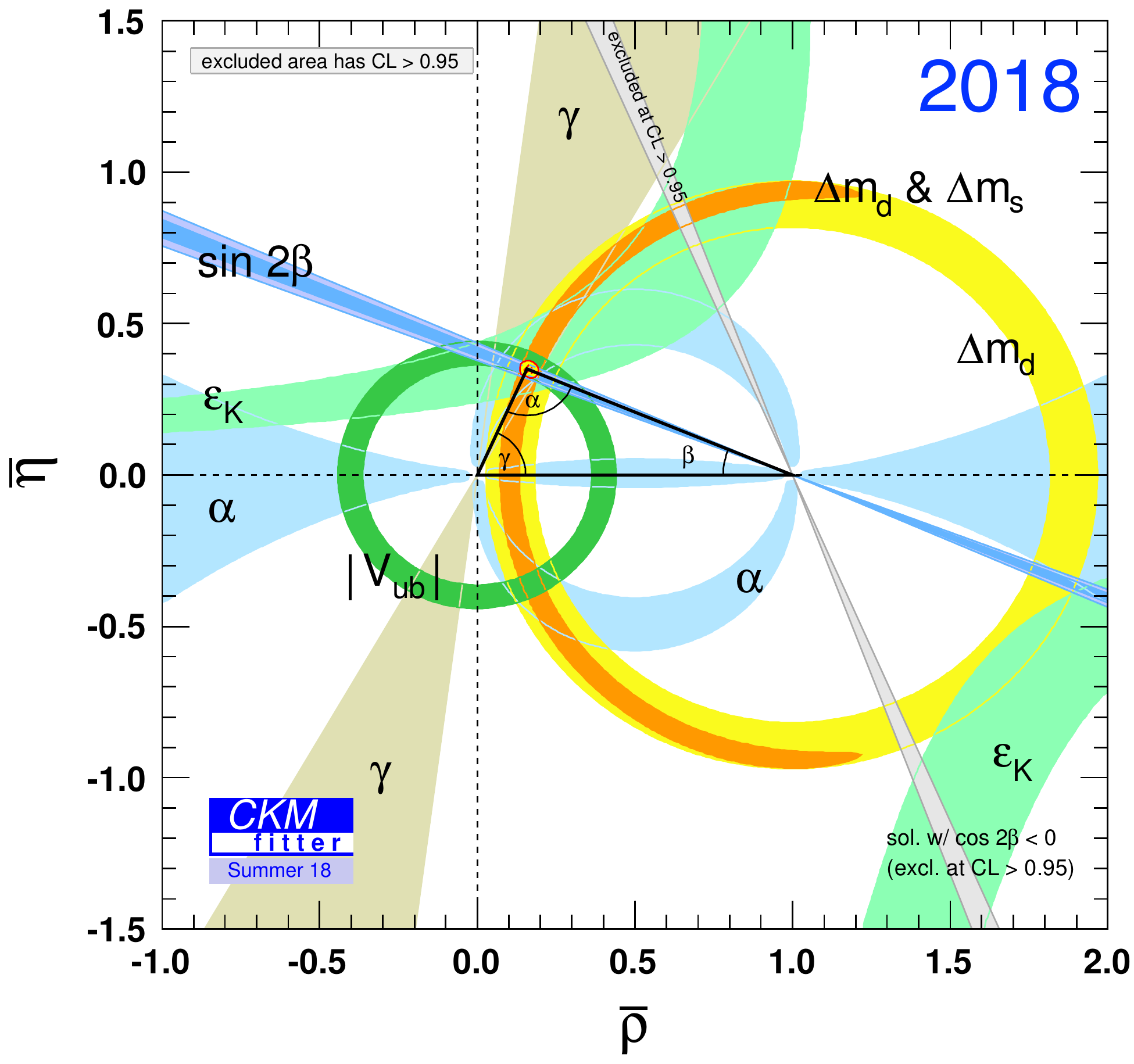}
\caption{The evolution of the constraints in the standard CKM unitarity triangle plane from 1995 (left), to just after the start of $B$ factories (middle), to the present (right). Taken from the ckmfitter website \cite{Charles:2004jd}.}
\label{fig:CKMtriangle:evolution}
\end{figure}

All flavour transitions in the SM depend on only  4 fundamental parameters, $\lambda$, $A$, $\rho$, and $\eta$. We can test the Kobayashi-Maskawa mechanism by making many measurements, over-constraining the system. One way to visualize a subset of experimental constraints is through the standard CKM unitarity triangle, which tests one out of nine  unitarity equations, $V_{\rm CKM} V_{\rm CKM}^\dagger=1$. The standard CKM unitarity triangle is obtained from a product of the first and the third column of the CKM matrix
\beq
V_{ud}V_{ub}^*+V_{cd}V_{cb}^*+V_{td} V_{tb}^*=0,
\eeq
which we can rewrite as
\beq
\label{eq:UT}
\frac{V_{ud}V_{ub}^*}{V_{cd}V_{cb}^*}+1+\frac{V_{td} V_{tb}^*}{V_{cd}V_{cb}^*}=0.
\eeq
In terms of the Wolfenstein parameters this sum rule is
\beq
\label{eq:rhoeta:UT}
-\big(\bar \rho+ i \bar \eta\big)+1+\big(-1+\bar \rho +i \bar \eta\big)=0.
\eeq
The relation \eqref{eq:UT} can be interpreted as a sum of three complex numbers that are the sides of a triangle, shown in Fig. \ref{fig:CKMtriangle}. There are two common notations for the angles of the standard CKM unitarity triangle: either $\alpha$, $\beta$, $\gamma$ or $\phi_1, \phi_2, \phi_3$, used by the two $B$-factories, BaBar and Belle, respectively. The Belle experiment (1999-2010) at KEK, Japan  produced about $\sim 1.5\times 10^9$ $B$ mesons, while BaBar experiment 1999-2008) at SLAC, USA collected about $\sim 0.9 \times 10^9$ $B$ mesons. The two experiments established that the KM mechanism is the main source of CP violation in the SM. The progression of constraints in the CKM unitarity triangle plane is shown in Fig. \ref{fig:CKMtriangle:evolution}. We see that there was a big qualitative jump after the start of the $B$ factories, and a very impressive set of improvements in the constraints since then.

\begin{figure}
\centering
\includegraphics[width=.9\linewidth]{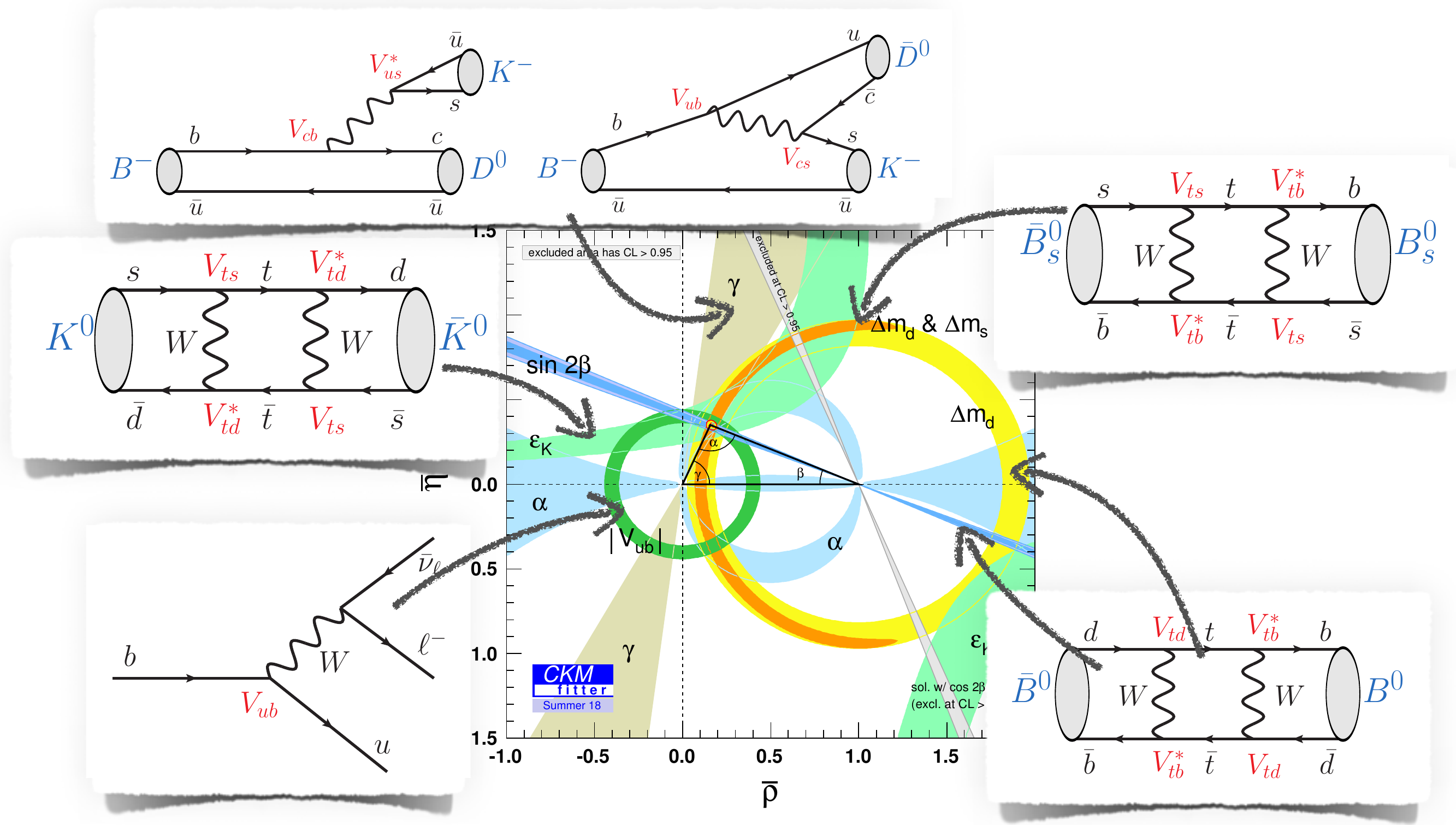}
\caption{Some of the main CKM constrains and the respective SM diagrams.}
\label{fig:CKMtriangle:diagrams}
\end{figure}

The constraints on the standard CKM unitarity triangle are coming from several different meson systems, the $B_d^0, B^+$ mesons from measurements at Belle, BaBar and LHCb, the $B_s$ meson and $\Lambda_b$ baryon from measurements at LHCb, and the kaon physics experiments.  Different constraints in the standard CKM unitarity triangle plane are shown in Fig.~\ref{fig:CKMtriangle:diagrams}, together with the relevant SM diagrams. 
The upshot of these results is that the KM mechanism is the dominant origin of CPV. The measurements point to a consistent picture of flavour violation, described by four parameters, $A$, $\lambda$, $\bar \rho$, $\bar \eta$, with the values given in Eq.~\eqref{eq:CKM:values}. Since $\bar \rho\lesssim \bar \eta$ the CKM phase is large, ${\mathcal O}(1)$. It is given by 
\beq
\gamma=\arctan\big(\bar \eta/{\bar \rho}\big) = \arg(V_{ub}^*),
\eeq
where in the last equality we used the common parametrization of the CKM matrix, where the weak phase is moved to the $V_{ub}$ and $V_{ts}$ CKM elements, \eqref{eq:VCKM:Wolfenstein}. Experimentally~\cite{Charles:2004jd}, 
\beq
\gamma=(65.4\pm1.1)^\circ,
\eeq
so that the weak phase is indeed ${\mathcal O}(1)$ when measured in radians. 

\begin{figure}[t]
\begin{center}
\includegraphics[width=0.49\textwidth]{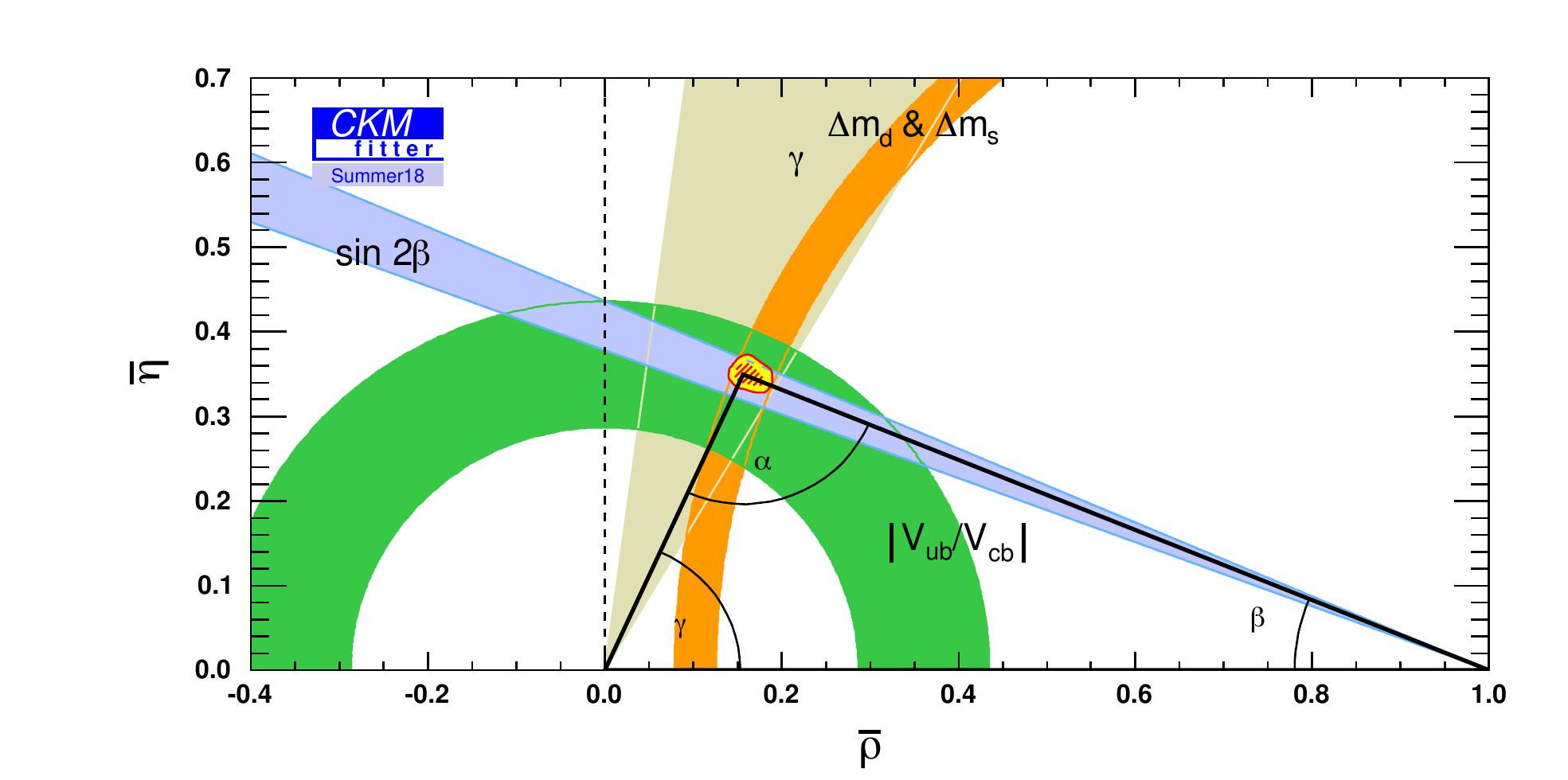}
~
\begin{minipage}{0.49\textwidth}
\includegraphics[width=\textwidth]{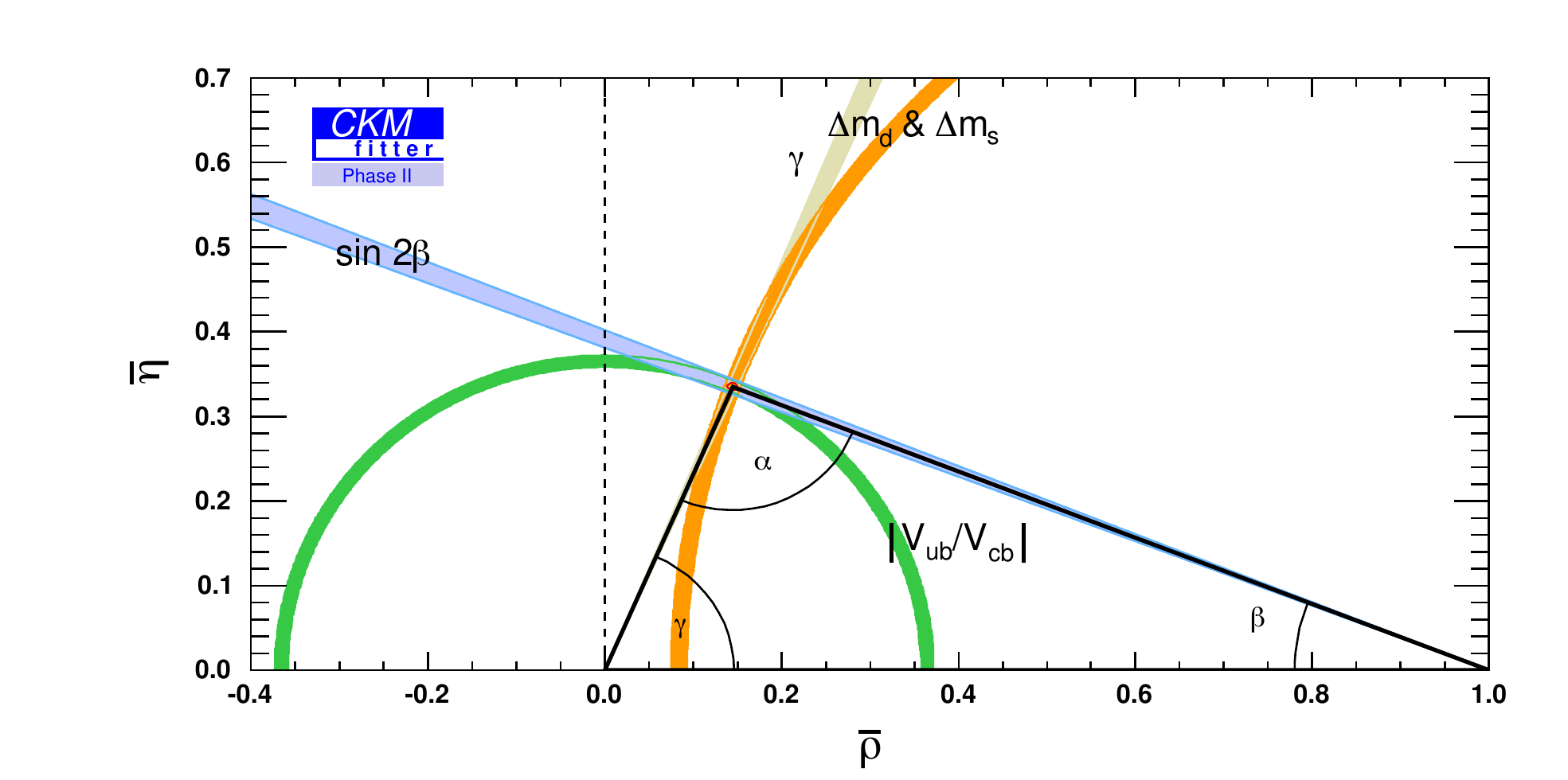}\vspace{-3.49cm}
\mbox{\hspace{5.2cm}\includegraphics[width=0.26\textwidth]{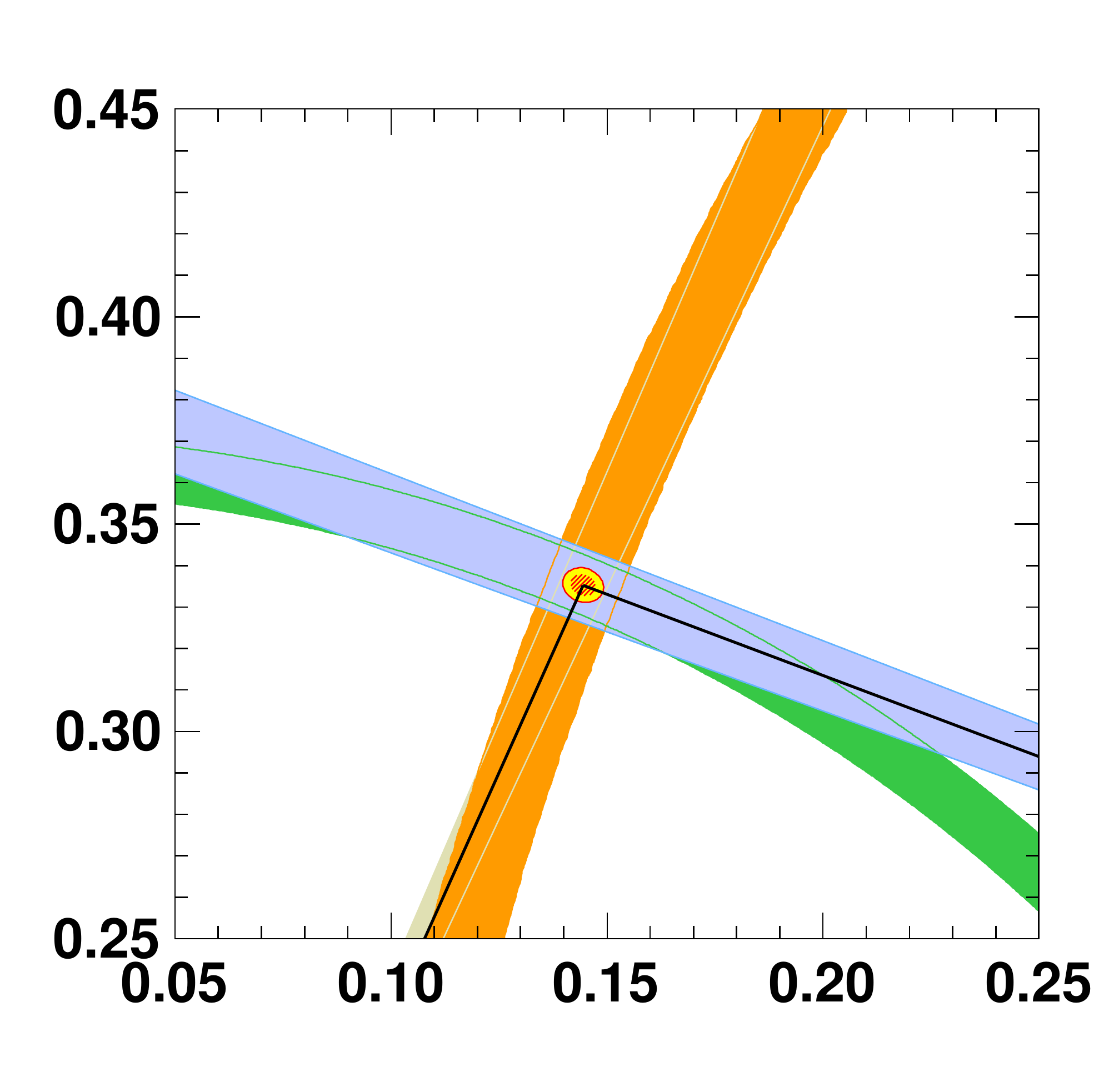}}
\vspace{5.10cm}
\end{minipage}
\vspace{-3.80cm}
\caption{\small  Evolving constraints in the $\bar{\rho}-\bar{\eta}$ plane from LHCb measurements and improvements in lattice QCD calculations, alone, with current inputs (2018), and the anticipated improvements from the data accumulated by 2035 (300\,fb$^{-1}$ of integrated luminosity). More information on the fits may be found in \cite{Cerri:2018ypt,Bediaga:2018lhg}.
 }
\label{fig:UTprojection}
\end{center}
\end{figure}

The field is undergoing a big upgrade in available statistics. The successor to Belle experiment, called Belle II, is ramping up right now, with the first physics run expected in early 2019 \cite{Kou:2018nap}. Belle II aims to collect about $\sim 8 \times 10^{10}$ $B$ mesons by about 2025,  roughly $50\times$ more than Belle did. The LHCb experiment also has ambitious upgrade plans \cite{Bediaga:2018lhg}. After the end of Upgrade II in 2035 it may have the statistics that corresponds to roughly $\sim 10^{11}$ or more useful $B$'s (because of hadronic environment this number fluctuates from channel to channel), as well as $B_s$ mesons and heavy baryons, which are also produced in the $pp$ collisions. The constraints on the elements of the CKM matrix are thus set to become much more precise in the future. Fig.~\ref{fig:UTprojection} (right) shows the improvements that can be achieved by using just the LHCb measurements alone at the end of the high luminosity LHC programme. A similar projection for the improvements using Belle II measurements can be found in Ref. \cite{Charles:2013aka}.

The constraints in the standard CKM unitarity triangle plot are of two types: the tree level transitions, which are less likely to be affected by new physics, and the loop level transitions, which are more likely to be affected by new physics. In the rest of this section we will choose an example transition for each of the two types of transitions and look at it in detail. This will then lead us to the discussion of new physics searches in Section \ref{sec:NP:searches}. However, before we do that, we need to introduce several new concepts.

\subsection{The meson mixing}
The term {\it mixing} denotes that the flavour eigenstates do not equal mass eigenstates, i.e., that the eigenstates of the SM Hamiltonian are composed of states with different flavour compositions. For instance, $B^0\sim \bar b d$ and $\bar B^0\sim b\bar d$ are flavour eigenstates but are not mass eigenstates. The mass eigenstates are admixtures of $B^0$ and $\bar B^0$. 

The term {\it oscillations} denotes that the initial  flavour eigenstate time evolves to a different flavour eigenstate. The reason for this is that the flavour eigenstates are composed from two mass eigenstates, each of which evolves slightly differently. The oscillation frequency is the energy splitting, $\omega=\Delta E$. In the rest frame this equals the mass splitting, $\Delta E=\Delta m$, which means that the oscillations are an excellent way to measure small mass splittings. 

What kind of mixings between states are possible? A general rule that applies here is: what is not explicitly forbidden is allowed \cite{Grossman:2017thq}. Using this important rule let us look at two examples:
\begin{itemize}
\item
Can $B^+\sim \bar b u$ and $B^-\sim b\bar u$ mix? The answer is no, since the electric charge is conserved. That is, the $U(1)_{\rm em}$ gauge symmetry forbids such mixings to all orders in perturbation theory.
\item
Can $B^0\sim \bar b d$ and $\bar B^0\sim b\bar d$ mix? In this case the answer is yes, since nothing forbids it. That is, there is no exact symmetry that forbids this mixing to all orders, so at some order in perturbation theory the mixing will occur. Such FCNCs are forbidden at three level in the SM, but are allowed at 1 loop. 
\end{itemize}

\begin{figure}
\centering
\includegraphics[width=.42\linewidth]{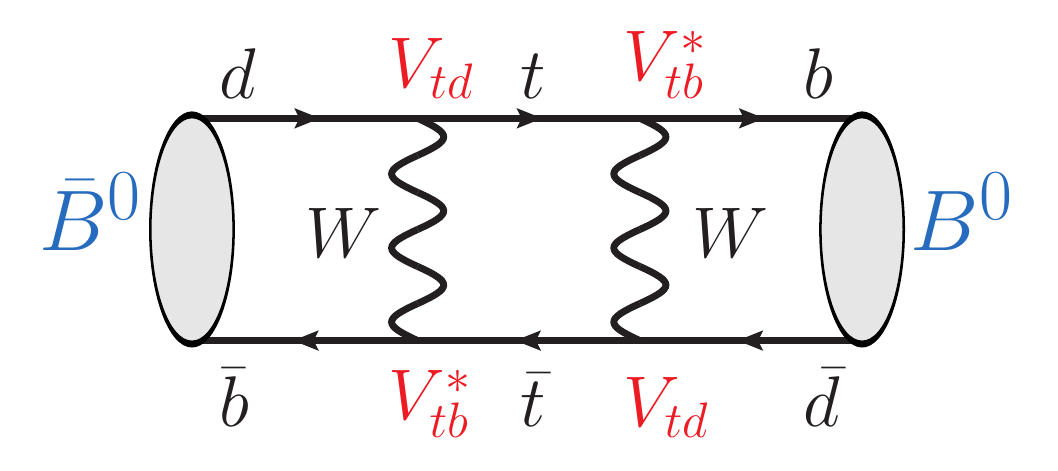}
\includegraphics[width=.40\linewidth]{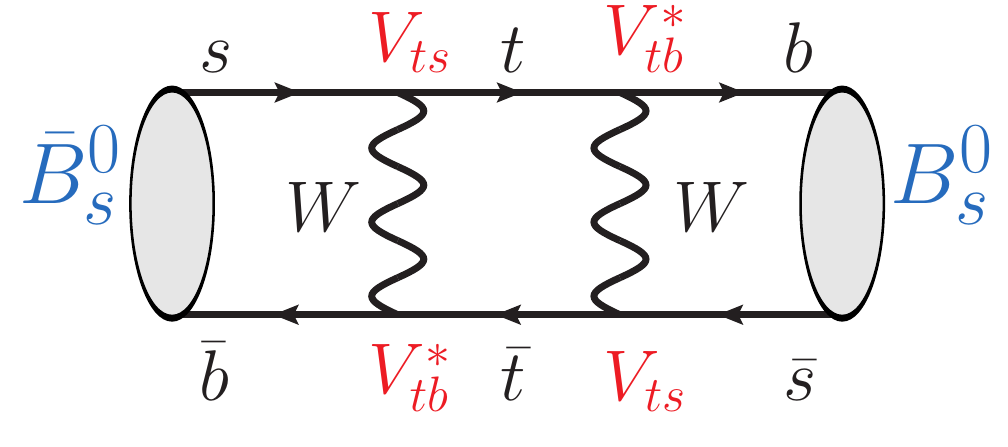}
\caption{The SM diagrams leading to $B_d$ (left) and $B_s$ mixing (right).}
\label{fig:Bd:Bs:mixing}
\end{figure}

A representative 1 loop weak interactions diagram that mixes $\bar B^0 \sim \bar bd$ and $B^0 \sim b \bar d$ in the SM , is shown in Fig. \ref{fig:Bd:Bs:mixing} (left). These diagrams contribute to the flavour off diagonal elements in the Hermitian $\bar B^0$, $ B^0$ mass matrix, see e.g.,  \cite{Nierste:2009wg},\footnote{Note that our phase conventions differ from  \cite{Nierste:2009wg} by a sign. The results in  \cite{Nierste:2009wg} are obtained by replacing $q\to -q$, $|\bar B^0\rangle\to -|\bar B^0\rangle$.}
\beq
{\cal M}=
\begin{pmatrix}
M_{11}  & M_{12} 
\\
M_{21} & M_{22}
\end{pmatrix},
\eeq
written in the flavour basis
\beq
\begin{pmatrix}
|B^0\rangle
\\
|\bar B^0 \rangle
\end{pmatrix}.
\eeq
The off-diagonal elements are much smaller than the diagonal ones, so that the mass matrix has the form,
\beq \label{eq:M:epsilon}
{\cal M}\propto
\begin{pmatrix}
~1~  & ~\epsilon~ 
\\
~\epsilon~ & ~1~
\end{pmatrix}.
\eeq
CPT guarantees $M_{11}=M_{22}$. If CP is conserved, then also $M_{12}=M_{21}$. Numerically, $M_{11}=M_{22}\simeq m_B$, while $M_{12,21}\ll M_{11}$.  If CP is conserved, the mass eigenstates are 
\beq
|B_{L,H}\rangle =\frac{1}{\sqrt 2}\big( |B^0\rangle \pm  |\bar B^0\rangle\big), 
\eeq
where we used the phase convention
\beq
CP |B^0\rangle =|\bar B^0\rangle, \qquad CP |\bar B^0\rangle =|B^0\rangle,
\eeq
such that
\beq
CP|B_{L,H}\rangle=\pm |B_{L,H}\rangle.
\eeq
That is, the mass eigen-states are maximally mixed, exactly what we are used to for eigenstates of the matrices of the form in Eq. \eqref{eq:M:epsilon}.

For $B^0-\bar B^0$ meson system the discussion deviates from the above results in two important ways. The first complication is that the CP is violated (but CPT still conserved). Then $M_{12}\ne M_{21}$, while $M_{11}=M_{22}$, in which case the two mass eigenstates are
\beq
\label{eq:BLH}
|B_{L,H}\rangle=p|B^0\rangle\pm q |\bar B^0 \rangle.
\eeq
For CP conserving case $p=q=1/\sqrt 2$. 

The other complication is that $B^0$ and $\bar B^0$ decay. We can describe this through a non-unitary evolution of a two-state system, given by a non-hermitian Hamiltonian
\beq
{\cal H}=M+i \Gamma,
\eeq
so that the time evolution of a two-state system is described by
\beq
i \frac{d}{dt}
\begin{pmatrix}
|B^0(t)\rangle
\\
|\bar B^0(t) \rangle
\end{pmatrix}
={\cal H} 
\begin{pmatrix}
|B^0(t)\rangle
\\
|\bar B^0(t) \rangle
\end{pmatrix}
=
\begin{pmatrix}
M_{11} +i \Gamma_{11}, & M_{12} +i \Gamma_{12}
\\
M_{21} +i \Gamma_{21}, & M_{22} +i \Gamma_{22}
\end{pmatrix}
\cdot
\begin{pmatrix}
|B^0(t)\rangle
\\
|\bar B^0(t) \rangle.
\end{pmatrix}
\eeq
The $\Gamma$ matrix encodes the effects of $B^0$ and $\bar B^0$ decays on the time evolution. The non-unitary evolution describes the ``disappearance'' of $B^0$ and $\bar B^0$ states due to decays into final particles, i.e., outside of the two-state system, $|B^0\rangle$, $|\bar B^0\rangle$. The eigenstates of ${\cal H}$ are still given by Eq. \eqref{eq:BLH}, though now in general $|B_{L}\rangle$ and $|B_{H}\rangle$ are no longer orthogonal. 

\subsection{Different ways of measuring the CP violation}
\begin{figure}
\centering
\includegraphics[width=.4\linewidth]{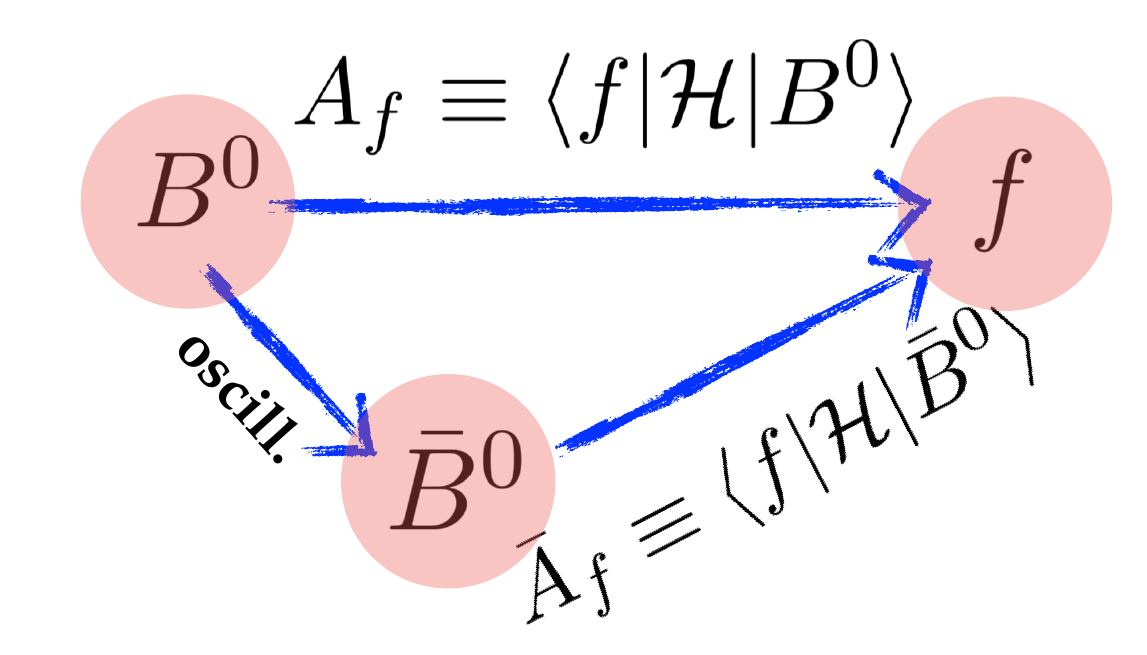}~~~~~~~~~~
\includegraphics[width=.32\linewidth]{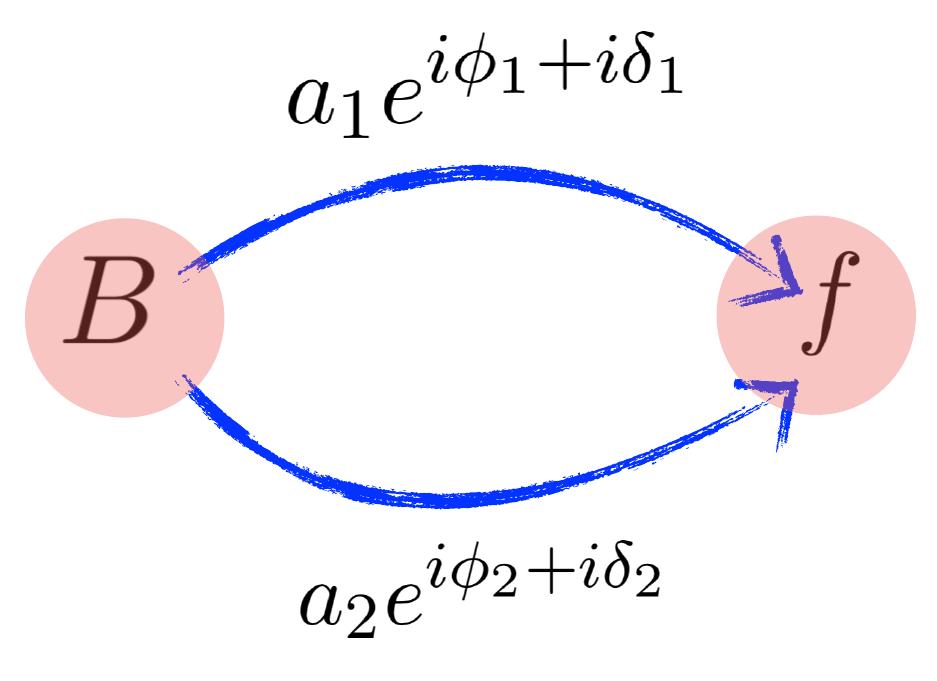}
\caption{Left: The two different paths for a $B^0$ meson to decay to a final sate $f$. Right: two interfering amplitudes are required to have direct CPV.}
\label{fig:CPV:interf}
\end{figure}

CP violation is an inherently quantum mechanical effect. As we saw in Section \ref{sec:CPV:SM} it is intimately tied to the existence of a physical phase in the Lagrangian.  In order to be sensitive to a phase an interference is needed. Thus, CP violating observables necessarily require some kind of interference.  Depending on the type of interference there are three distinct categories of CP violating observables
\begin{enumerate}
\item
{\it CPV in the decay}, also called {\it direct CPV}, occurs when there is interference between different contributions to the decay amplitudes so that
\beq
|A_f|\ne |\bar A_f|.
\eeq
Here we used the short-hand notation
\beq
\label{eq:Af:not}
A_f\equiv \langle f| {\cal H}|B^0\rangle, \qquad \bar A_f\equiv \langle f| {\cal H}|\bar B^0\rangle.
\eeq
\item
{\it CPV in mixing} occurs when there is interference between $M_{12}$ and $\Gamma_{12}$ in the time evolution of the two-state system. This arises when
\beq
|q/p|\ne 1,
\eeq
and corresponds to interference between different ways to oscillate between $B^0$ and $\bar B^0$ states, either through dispersive matrix elements or through absorptive ones. 
\item
{\it CPV in interference between decays with and without mixing}, arises when 
\beq
\label{eq:im:lambdaf}
\Imag \lambda_f\ne0
\eeq
where
\beq
\lambda_f \equiv \frac{q}{p}\frac{\bar A_f}{A_f}.
\eeq
Here the interference is between two different paths of $B^0$ to decay to the final states $f$, see Fig. \ref{fig:CPV:interf}. The two paths are either through direct decay, proportional to $A_f$, or by first oscillating to $\bar B^0$, which then decays to $f$, giving a contribution proportional to $(q/p) \bar A_f$. 
\end{enumerate}

In the next few subsections we will look at two examples: the CPV in the decay, essential to measure the angle $\gamma$ of the standard CKM unitarity triangle, and the determination of angle $\beta$, which relies on the CPV in interference between decays with and without mixing for $B_d$ mesons.

When discussing CPV, it is important to remember that not all of the phases are CP violating. An elementary example is the double slit experiment, in which the interference pattern arises because there is a phase difference between two waves due to different paths, $e^{i\delta}=e^{ik \Delta r}$. This phase difference is not CP violating, since it does not depend on whether the double slit experiment is done with particles or antiparticles.

We thus distinguish two different types of phases. The {\it weak phases} are the (physical) phases that appear in the Lagrangian. The weak phases violate CP, just as the CKM phase in the weak interaction part of the SM Lagrangian violates CP. The {\it strong phase} is the name used for CP conserving phases. An example of such a strong phase is, for instance, the phase shift resulting from rescattering of particles due to QCD/strong interactions. Imagine a thought experiment, in which we collide two pion beams with the total center of mass energy close to the rho meson mass. The $\pi^+\pi^0\to \rho^+\to \pi^+\pi^0$ and $\pi^-\pi^0\to \rho^-\to \pi^-\pi^0$ scatterings both result in the same complex Breit-Wigner scattering amplitude  
\beq
A\propto \frac{1}{p^2-m^2+i m\Gamma}.
\eeq
The imaginary term in the propagator is due to on-shell rescattering through decay products of $\rho$. These are CP conserving processes, and so is the resulting phase, $\arg(A)$. This phase does not change sign, when we exchange $\pi^+\leftrightarrow \pi^-$.

\subsection{CPV in the decay}
We start with the CPV in the decay (or the so called direct CP violation). The CPV observable is the decay asymmetry 
\beq
\label{eq:direct:CPV}
{\cal A}_f\equiv \frac{\Gamma(\bar B\to \bar f)- \Gamma(B\to f)}{\Gamma(\bar B\to \bar f)+ \Gamma(B\to f)}=\frac{1-|A_f/\bar A_f|^2}{1-|A_f/\bar A_f|^2},
\eeq
where $A_f$ are defined in \eqref{eq:Af:not}.
In order to have non-vanishing CP asymmetry, ${\cal A}_f\ne0$, the $B\to f$ decay amplitude needs to receive contributions from (at least) two different  terms with differing weak, $\phi_{1,2}$, and strong phases, $\delta_{1,2}$, see also Fig. \ref{fig:CPV:interf} (right)
\begin{align}
A_f&= a_1 e^{i\phi_1+i\delta_1}+ a_2 e^{i\phi_2+i\delta_2},
\\
\bar A_f &= a_1 e^{-i\phi_1+i\delta_1}+ a_2 e^{-i\phi_2+i\delta_2}.
\end{align}
The weak phases are due to the CKM phase in the SM Lagrangian and change the sign under CP transformation, while the strong phases are due to on-shell rescattering of particles (pions, etc) and are thus CP even, the same as QCD interactions. The CP asymmetry is, in the simplifying limit $a_2/a_1\ll 1$, 
\beq
{\cal A}_f= \frac{a_2}{a_1} \sin(\phi_2-\phi_1)\sin(\delta_2-\delta_1)+{\mathcal O}(a_2^2/a_1^2).
\eeq
The CP asymmetry vanishes in the limit where either (i) there is only one contribution to the amplitude, $a_2\to 0$, and/or (ii) if the weak phase difference vanishes, $\phi_2-\phi_1\to0$, and/or (iii) if the strong phase difference vanishes, $\delta_2-\delta_1\to0$.

\subsection{Measuring the CKM angle $\gamma$}

\begin{figure}
\centering
\includegraphics[width=.4\linewidth]{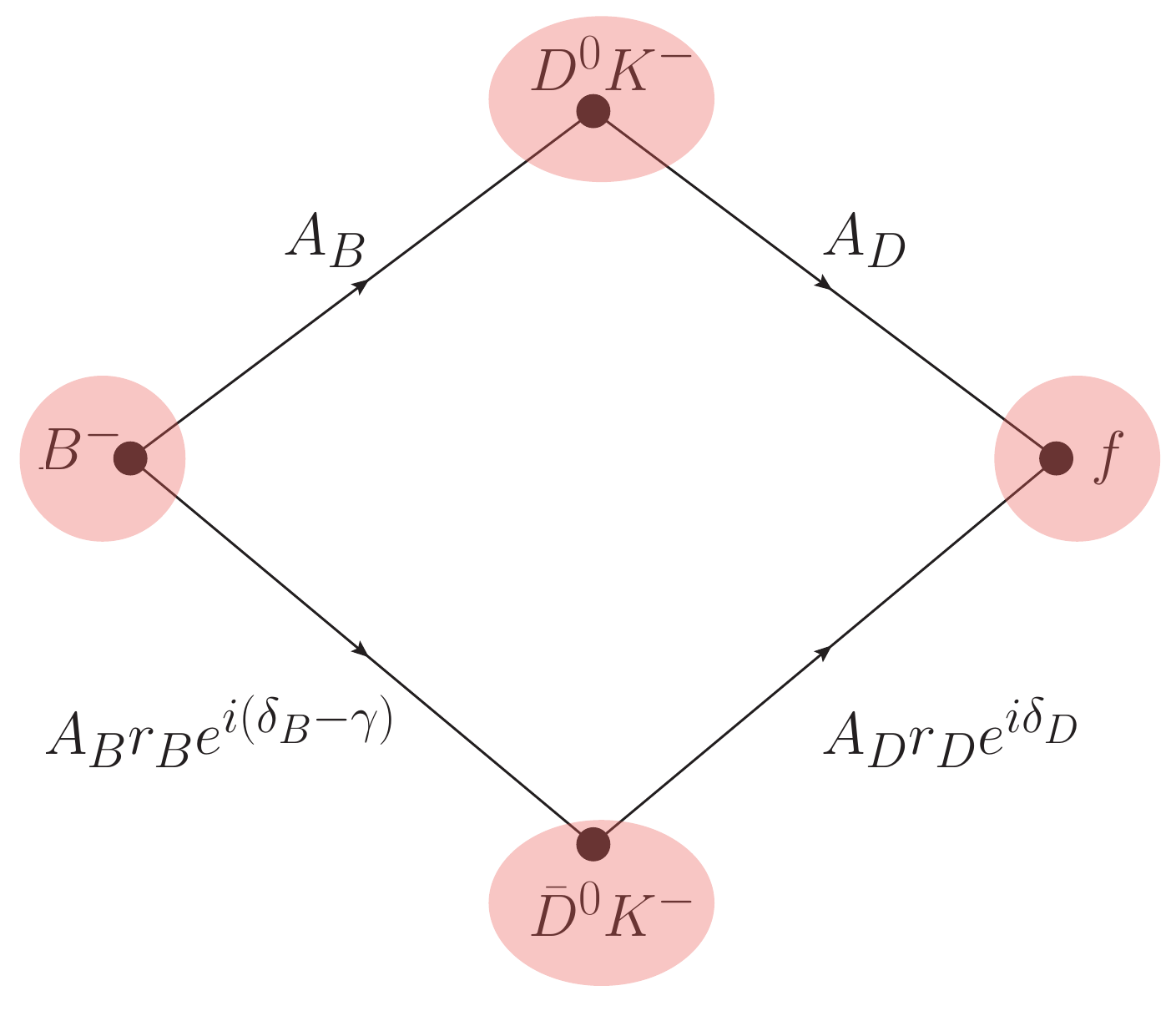}
\caption{The two interfering amplitudes in the $B^-\to D[\to f]K^-$ decays. The decay amplitudes for CP conjugate decay $B^+\to D[\to \bar f]K^+$ are obtained by exchanging $\{B^-,K^-, f\}\to \{B^+,K^+,\bar f\}$, $D^0\leftrightarrow \bar D^0$, and $\gamma\to -\gamma$ in the above.  }
\label{fig:BDK:2paths}
\end{figure}

The measurements of CKM unitarity triangle angle $\gamma$ use the decays in which there is interference between $b\to c\bar u s$ and $b\to u \bar c s$ transitions \cite{Gronau:1991dp,Gronau:1990ra,Giri:2003ty,Atwood:1996ci}.  This happens for instance in the $B^-\to [D\to f]K^-$ decay chain. The $B^- \to D^0 K^-$ decay is due to the $b\to c \bar u s$ transition, while the $B^-\to \bar D^0K^-$ decay is mediated by the $b\to u \bar c s$ transition, which is proportional to $V_{ub}\propto e^{-i\gamma}$, see the top two diagrams in Fig. \ref{fig:CKMtriangle:diagrams}. If the $D^0$ and $\bar D^0$ decay to the same final state, such as $f=\pi^+\pi^-$, $K^+K^-$, $K_S\pi^+\pi^-$, the two decay amplitudes interfere, giving sensitivity to the phase $\delta_B- \gamma$. Our notation is defined in Fig. \ref{fig:BDK:2paths}, with $\delta_{B,D}$ the strong phases, while $\gamma$ is a weak phase and changes sign under CP conjugation. 

In order to extract $\gamma$ both the rates for $B^-\to [D\to f]K^-$ and its CP conjugated mode $B^+\to [D\to \bar f]K^+$ need to be measured. The interference terms in the two rates are proportional to $\delta_B+\delta_D-\gamma$ and $\delta_B+\delta_D+\gamma$, respectively. The difference of the two thus gives the quantity we are after, $\gamma$, if the hadronic parameters, $A_{B,D}, r_{B,D}, \delta_{B,D}$ are known. Note that the direct CPV asymmetries, Eq. \eqref{eq:direct:CPV},
\beq
{\cal A}_f \propto r_B r_D \sin(\delta_B+\delta_D)\sin\gamma,
\eeq
are crucial.
If ${\cal A}_f$ vanish, so thus the sensitivity to $\gamma$. The measurement of $\gamma$ requires both $r_B, r_D$ and the strong phases to be nonzero. 

Amazingly, all the hadronic inputs can be measured experimentally. The $A_D$ and $r_D$ are obtained from $D^{*+}\to [D^0\to f]\pi^+$ decays where the charge of $\pi^+$ tags the flavour of $D^0$, i.e., $\bar D^0$ would be accompanied by a $\pi^-$. Choosing $N_f$ different final states leaves us with $4+N_f$ unknowns: $\gamma, A_B, r_B, \delta_B, \delta_D$. On the other hand, we can measure $2 N_f$ decay branching ratios, $B^-\to [D\to f]K^-$ and $B^+\to [D\to f]K^+$ (taking $f$ not to be CP conjugate final state for simplicity, such as bins in $K_S\pi^+\pi^-$ Dalitz plot). For $N_f\geq4$ there is enough information to extract all the unknowns. The situation is in fact even better, since $\delta_D$ can be measured at CLEO and BESS III from entangled decays $\psi(3770)\to D^0\bar D^0$, improving the precision with which $\gamma$ is extracted.

That all the hadronic uncertainties can be obtained experimentally makes this approach a very powerful tool. It means that the angle $\gamma$ can be extracted with basically no theory uncertainties. The theoretical corrections arise only from one loop electroweak corrections, limiting the ultimate precision with which $\gamma$ can be extracted up to miniscule $\gamma_{\rm th}<10^{-7}$ \cite{Brod:2013sga,Brod:2014qwa}. The experimental error bars will be larger than this for a long time. At present they are at $\delta\gamma\lesssim 6^\circ$ \cite{LHCb-CONF-2017-004}.

\subsection{CPV in interference between decays with and without mixing}

The state that is created at $t=0$ as the $\bar B^0$ [or $B^0$] time evolves according to
\beq
\frac{d}{dt}\Gamma(\bar B^0(t)[B^0(t)]\to f_{\rm CP}) \propto e^{-\Gamma t}\Big[ \frac{1}{2}\Big(1+|\lambda_f|^2\Big) \pm S_f \sin(\Delta m t) \mp  C_f \cos(\Delta m t)\Big],
\eeq
with,
\beq
\label{eq:Sf:Cf}
S_f\equiv \frac{2 \Imag \lambda_f}{1+|\lambda_f|^2}, \qquad  C_f\equiv \frac{1-|\lambda_f|^2}{1+|\lambda_f|^2}.
\eeq
Here we assumed for simplicity that the final state is a CP eigenstate, $f_{CP}$, such as $f_{CP}=J/\psi K_S$. We also used that the mass splitting between the two mass eigenstates is much bigger than the difference between the two decay widths, $\Delta \Gamma\ll \Delta m$, so that it can be neglected, setting $|q/p|=1$. The time evolution is plotted in Fig. \ref{fig:JPsi} (left). The exponential decay is modulated by an oscillatory behaviour as $B^0$ converts to $\bar B^0$ and back (and vice versa), with the frequency of the oscillations given by the mass splitting, $\Delta m$. 

\begin{figure}
\centering
\includegraphics[width=.45\linewidth]{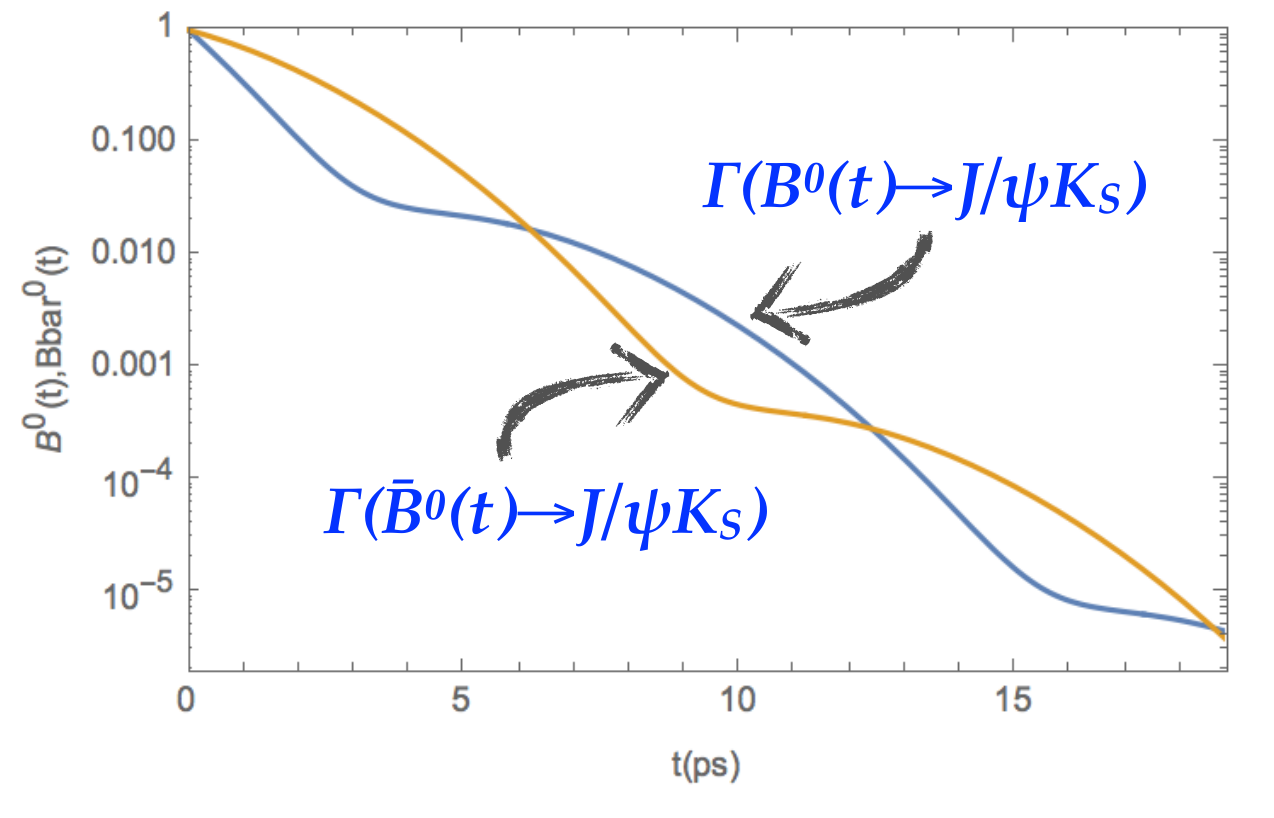}~~~~~~~~~~~
\includegraphics[width=.3\linewidth]{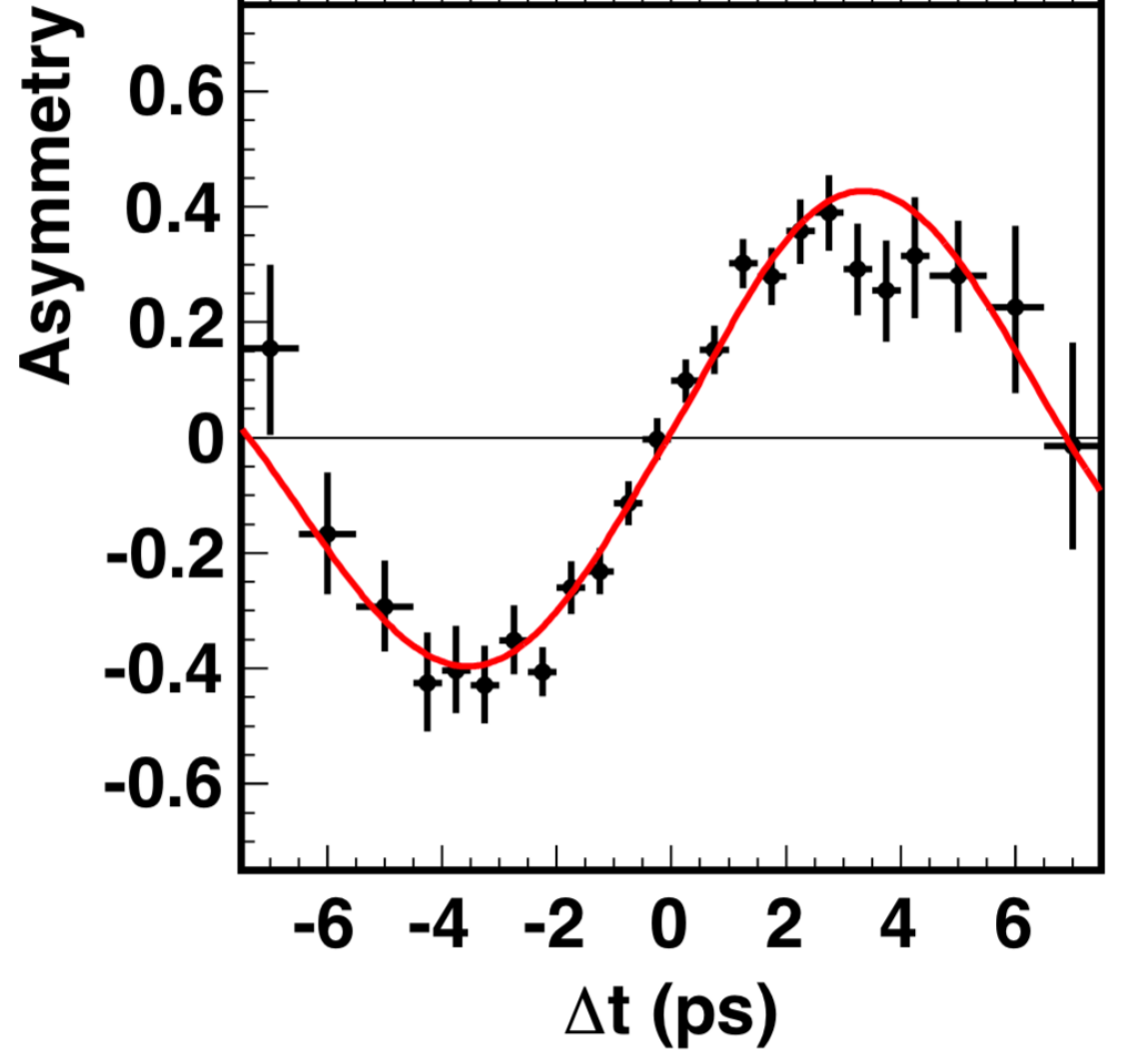}
\caption{Left: the time dependent $\bar B^0(t)\to J/\psi K_S$ (orange) and  $B^0(t)\to J/\psi K_S$ (blue) decay rates. Right: the time dependent CP asymmetry (taken from \cite{Adachi:2012et}). }
\label{fig:JPsi}
\end{figure}

The difference between the two decay rates is the time dependent CP asymmetry
\beq
{\cal A}_{f_{CP}}(t)\equiv \frac{ \frac{d}{dt}\Gamma[\bar B^0(t)\to f_{\rm CP}]-\frac{d}{dt}\Gamma[B^0(t)\to f_{\rm CP}]}
{ \frac{d}{dt}\Gamma[\bar B^0(t)\to f_{\rm CP}]+\frac{d}{dt}\Gamma[B^0(t)\to f_{\rm CP}]},
\eeq
and is described by a purely oscillatory behaviour, see Fig. \ref{fig:JPsi} (right),
\beq
{\cal A}_{f_{CP}}(t)= S_f \sin(\Delta m t) - C_f \cos(\Delta m t).
\eeq
The coefficient of $\cos(\Delta m t)$ is nonzero, if there is direct CPV, since $C_f=-A_f$ for $|q/p|=1$. The coefficient of $\sin(\Delta m t)$ is nonzero if there is CPV in interference between decays with and without mixing, cf. Eqs. \eqref{eq:im:lambdaf} and \eqref{eq:Sf:Cf}. We will see that $S_f$ is an important observable in searches for New Physics (NP). In the SM it is a measure of the CKM unitarity triangle angle $\beta$. 

\subsection{The measurement of angle $\beta$}
The $q/p$ does not depend on the final state $f$, and is the property of the $B^0-\bar B^0$ system. In the SM it is given by the ratio of one loop diagram in Fig. \ref{fig:Bd:Bs:mixing} and its complex conjugated version, so that
\beq
\frac{q}{p}=e^{-i \phi_B}=\frac{V_{tb}^* V_{td}}{V_{tb} V_{td}^*},
\eeq
with hadronic matrix elements cancelling in the ratio.

The decay amplitudes $A_f$, $\bar A_f$ do depend on the final state. However, a simplification occurs for $B^0\to J/\psi K_S$ and other decays that are dominated by a single amplitude, in this case due to the tree level $b\to c\bar c s$ transition. In the ratio $\bar A_f/A_f$ the hadronic matrix elements largely cancel. To a good approximation it is given by the ratio of the CKM elements
\beq
\frac{\bar A_{J/\psi K_S}}{A_{J/\psi K_S}}=\eta_{f} \frac{V_{cb} V_{cs}^*}{V_{cb}^* V_{cs}}+\cdots,
\eeq
with $\eta_f=-1$ the CP of $J/\Psi K_S$, and the ellipses the corrections due to penguin diagrams that depend on a different product of CKM elements. 
Therefore,
\beq
\lambda_{J/\psi K_S}=\eta_f \frac{V_{tb}^* V_{td}V_{cb} V_{cs}^*}{V_{tb} V_{td}^*V_{cb}^* V_{cs}}=\eta_f e^{-i 2\beta},
\eeq
and thus
\beq
\Imag \lambda_{J/\psi K_S}=\sin 2\beta.
\eeq
The measurement of $\sin 2\beta$ was the flagship measurement of the $B$ factories which showed that the CP violating phase in the SM is large, cf. Fig. \ref{fig:CKMtriangle:evolution}. 

\section{New Physics searches}
\label{sec:NP:searches}
So far we looked at the measurements of the SM parameters. We now turn to a different question: how does one search for New Physics (NP)? Before we tackle this question let us first answer a seemingly unrelated question: why is the weak force weak? The weak and strong interactions are similar in many respects. They are both nonabelian gauge interactions, and at high energies, of a few 100 GeV, they even have coupling constants that are not that different in size. However, at low energies they exhibit very different strengths. The strong force gives rise to a strong binding potential, while the weak force results only in a very weak short range potential. The decays that proceed through strong interactions such as $\rho^+\to \pi^+\pi^0$ occur at times scales $\sim 10^{-23}$ s, while the weak decays are much slower, from $\sim 10^{-12}$ s for $B$ decays to hundreds of seconds in the case of  neutron beta decay. 

The reason for this disparity is that the strength of the interaction is governed both by the size of the couplings and the mass of the force carriers. The more massive the carrier the shorter the range of the potential, and the weaker the interaction at low energies. The weak force is weak because the force carriers, $W$ and $Z$ are heavy, with masses equal to $80.4$ GeV and $91.2$ GeV, respectively. The neutron beta decay width is highly suppressed, because the available energy in the decay, $\sim (m_p-m_n)$ up to corrections from electron mass, is much smaller than the  mass of the force carrier, the $W$ boson,
 \beq
\Gamma(n\to p e \bar \nu_e)\propto \frac{(m_p-m_n)^5}{m_W^4}\sim 10^{-20} (m_p-m_n).
\eeq

This detour lead us to an important insight: through rare (or slowly occurring) processes we can probe heavy mediators. Historically, the weak nuclear decays were the first sign of a new force with a heavy mediator, the $W$ boson. Other processes could, in a similar way,  hint at new forces beyond the SM. We thus arrived at the recipe for indirect searches: identify processes that are rare in the SM and then search for deviations from the SM predictions. 

\begin{figure}
\centering
\includegraphics[width=.23\linewidth]{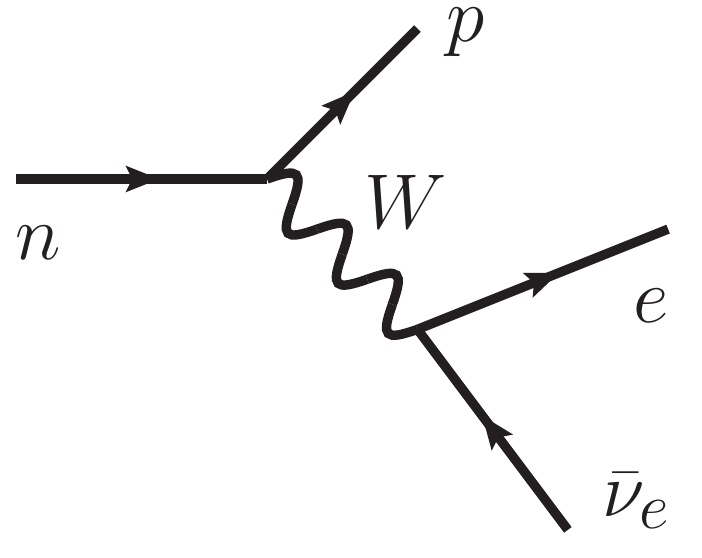}
\caption{The neutron beta decay proceeds through a tree level exchange of the $W$ boson. }
\label{fig:JPsi}
\end{figure}

A good target are the Flavour Changing Neutral Current (FCNC) processes. In the SM there are no FCNCs at tree level -- the gluon, photon, $Z$, and Higgs tree level exchanges  are strictly flavour conserving, cf. Fig. \ref{fig:Feynman:diagrams}. The FCNC processes, such as meson mixings, arise only at loop level and are thus suppressed, see Section \ref{sec:charged:neutral}. The FCNC processes can be easily modified by NP, either through tree level or loop level NP contributions. Taking $B_s$ mixing as an example the tree level NP contributions will have the form $\propto g_{sb}^2/M_{\rm NP}^2$, where $g_{sb}$ is the NP coupling to $b$ and $s$ quarks, and $M_{\rm NP}$ the mass of the new mediator. The NP contributions thus vanish if the NP is very heavy, $M_{\rm NP}\to \infty$, or if the flavour violating coupling constants are small, $g_{sb}\to0$.

In the rest of this section we explore in more detail the two main ways of searching for beyond the SM physics in flavour: by measuring the meson mixing amplitudes, and by measuring rare decays such as $b\to s\ell^+\ell^-$.

%\begin{figure}
%\centering
%\includegraphics[width=.23\linewidth]{figs/beta_decay}
%\caption{The neutron beta decay proceed through a tree level exchange of the $W$ boson. }
%\label{fig:JPsi}
%\end{figure}

\subsection{New physics searches using meson mixings}
There are four neutral meson systems that mix through weak interactions at 1 loop: $K^0-\bar K^0$ ($\bar s d \leftrightarrow s\bar d$), $D^0-\bar D^0$ ($c \bar  u \leftrightarrow \bar c u$), $B^0-\bar B^0$ ($\bar b d \leftrightarrow b \bar d$), and $B^0_s-\bar B^0_s$ ($\bar b s \leftrightarrow b \bar s$). We will mainly focus on $B^0-\bar B^0$ and $B^0_s-\bar B^0_s$ systems, which are dominated by the $W$--top quark loop, Fig \ref{fig:Bd:Bs:mixing}. 

Since $m_{t,W}\gg m_B$ the top and $W$ can be integrated out, leading to the $B_d-\bar B_d$ mixing effective weak Hamiltonian \cite{Nierste:2009wg}
\beq
\label{eq:Heff:v1}
{\cal H}_{\rm eff}^d=\frac{G_F^2}{16\pi^2} m_W^2 \eta_B S_0 \, \big(V_{tb}^*V_{td}\big)^2  \big(\bar b d\big)_{V-A}  \big(\bar b d\big)_{V-A} +{\rm h.c.},
\eeq
where $G_F\simeq 1.166 \cdot 10^{-5} {\rm~GeV}^{-2}$ is the Fermi constant, and $\eta_B S_0\simeq 1.26$ is the product of a properly normalized loop function and the QCD correction factor. The effective weak Hamiltonian is local, i.e., it corresponds to the potential that acts only at a point. This is a result of taking the weak mediators to be infinitely heavy.  
Another way of writing the effective Hamiltonian is
\beq
{\cal H}_{\rm eff}^d=\frac{1}{\Lambda_{\rm MFV}^2} \big(V_{tb}^*V_{td}\big)^2  \big(\bar b_L \gamma^\mu d_L\big)  \big(\bar b_L \gamma_\mu d_L\big) +{\rm h.c.},
\eeq
where the dimensionful prefactor, 
\beq
\Lambda_{\rm MFV}=\frac{2\pi}{G_F m_W \sqrt{\eta_B S_0}}\simeq 6.0{\rm ~TeV},
\eeq
is significantly larger than the weak scale, $m_W\simeq 80.2$ GeV, because we absorbed in it  the loop factor, $1/16 \pi^2$ (up to a factor of $4$ that went into a redefinition of the operator).
For $B_s$ mixing the CKM factors in the weak vertices change, cf. Fig. \ref{fig:Bd:Bs:mixing}, so that one has  instead 
\beq
{\cal H}_{\rm eff}^s=\frac{1}{\Lambda_{\rm MFV}^2} \big(V_{tb}^*V_{ts}\big)^2  \big(\bar b_L \gamma^\mu s_L\big)  \big(\bar b_L \gamma_\mu s_L\big) +{\rm h.c.}.
\eeq

The top and $W$ in the loop are much heavier than the available energy in the mixing -- the $B$ meson mass. The top and $W$ lines in the diagram are thus always off-shell, and so the dominant diagram only contributes to the dispersive part of the mixing amplitude,
\beq
M_{12}^d=\frac{1}{2m_B}\langle \bar B_d^0| {\cal H}_{\rm eff}^d|B_d^0\rangle^*.
\eeq
The absorptive part of the mixing amplitude, $\Gamma_{12}$, receives contributions from the subleading amplitudes with $c$ and $u$ quarks running in the loop. 

\begin{figure}[t]
\begin{center}
\includegraphics[width=0.41\linewidth]{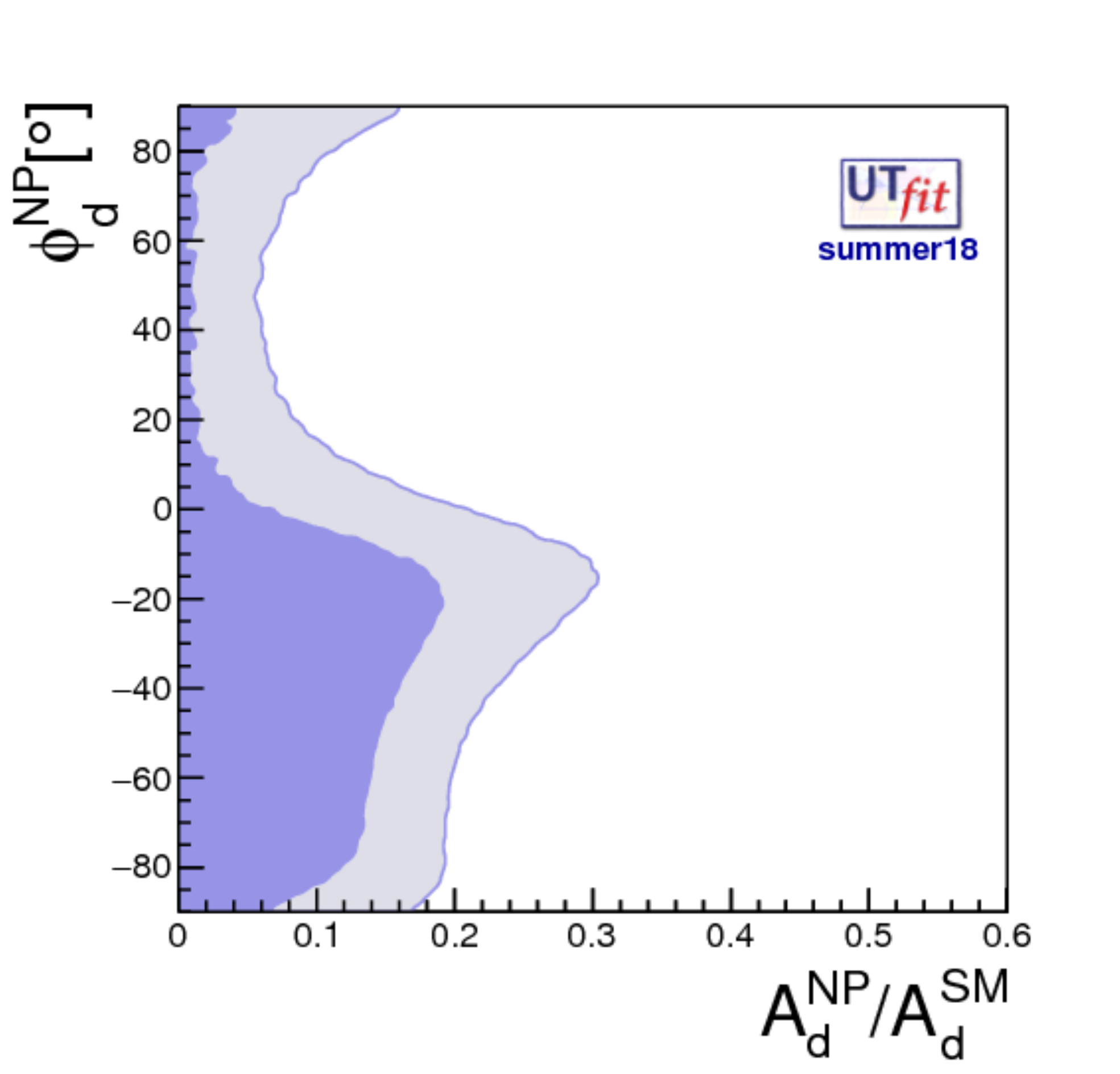}~~~~~~~~
\includegraphics[width=0.41\linewidth]{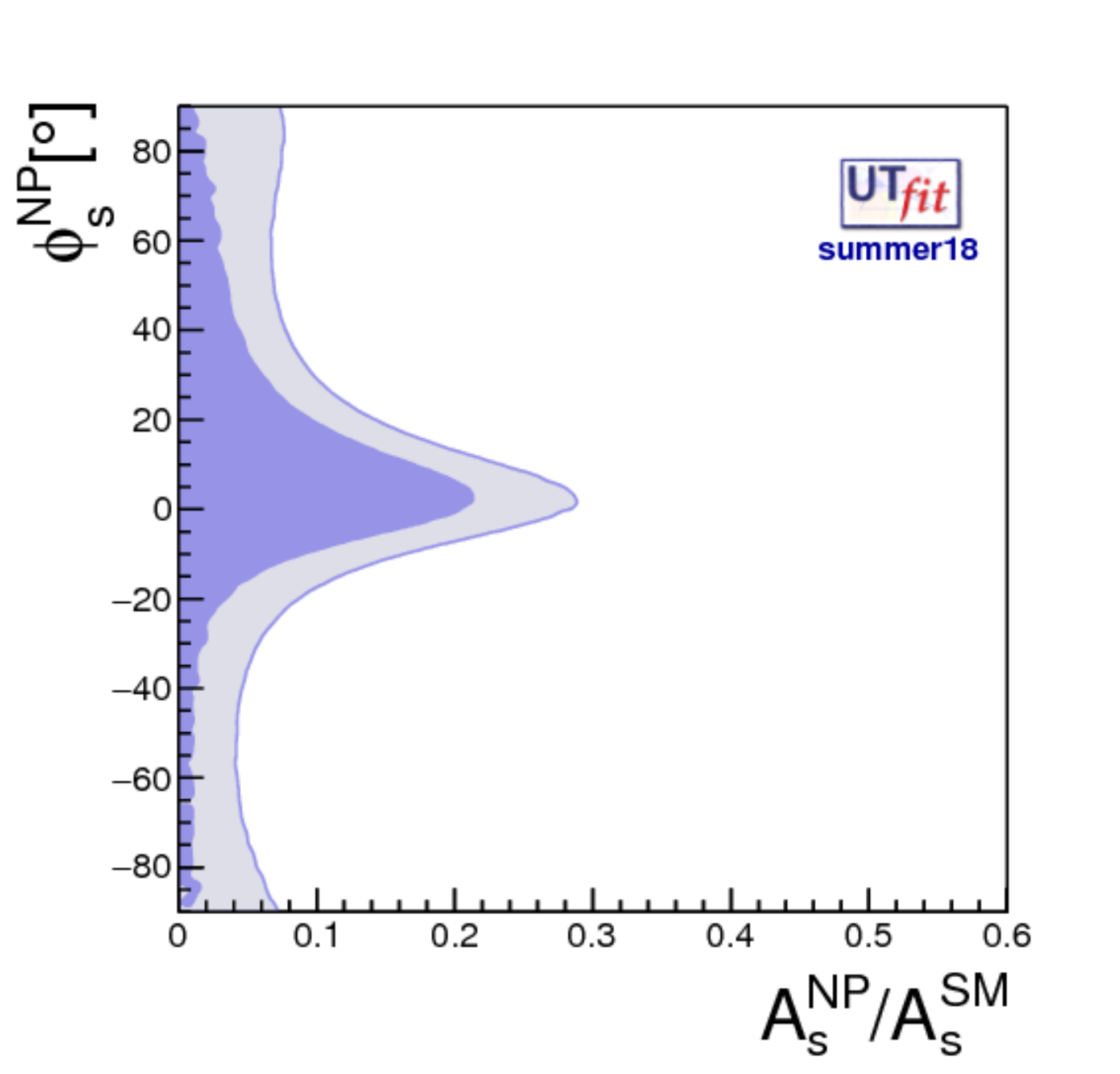}
\\
\includegraphics[width=0.41\linewidth]{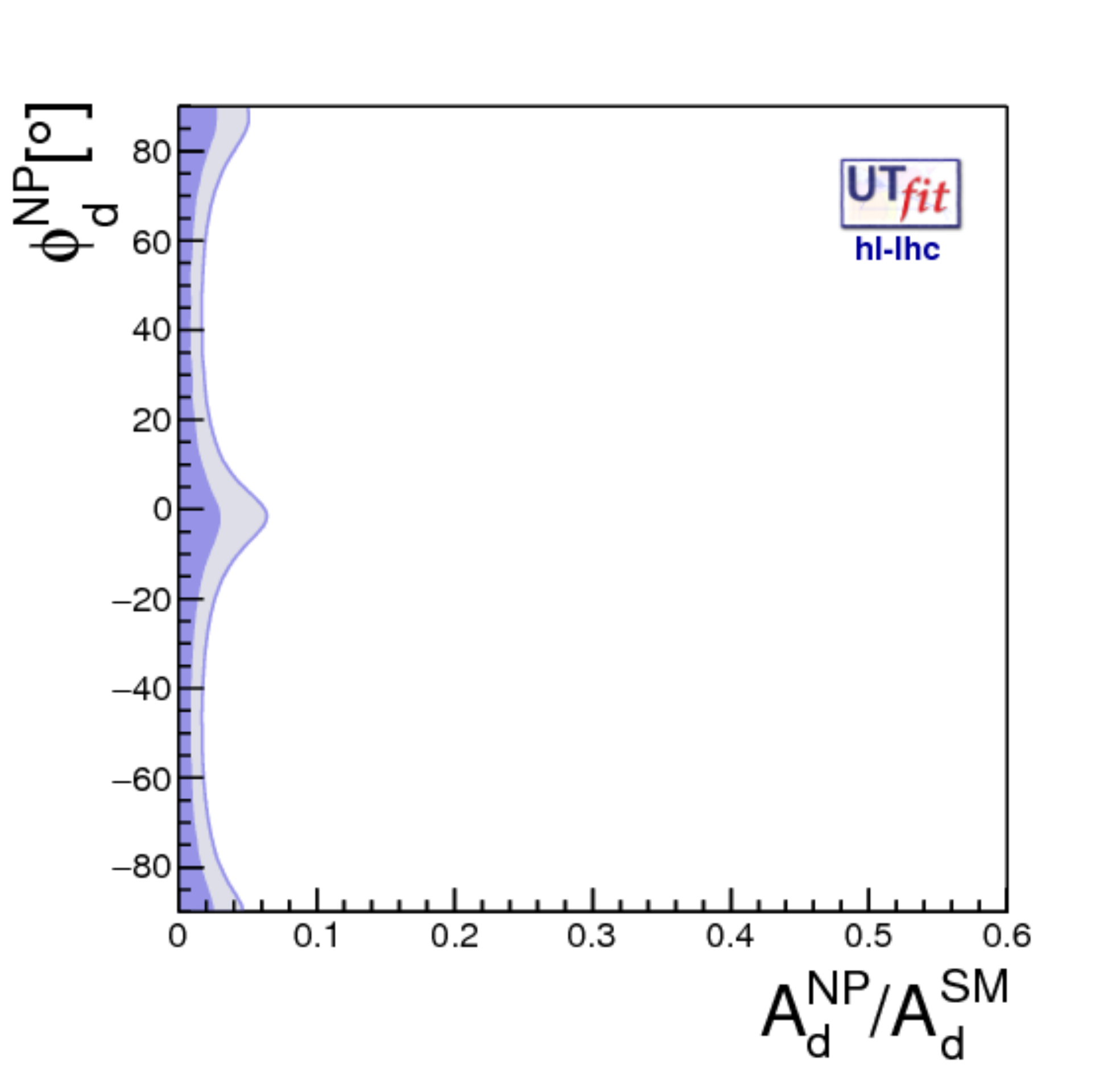}~~~~~~~~
\includegraphics[width=0.41\linewidth]{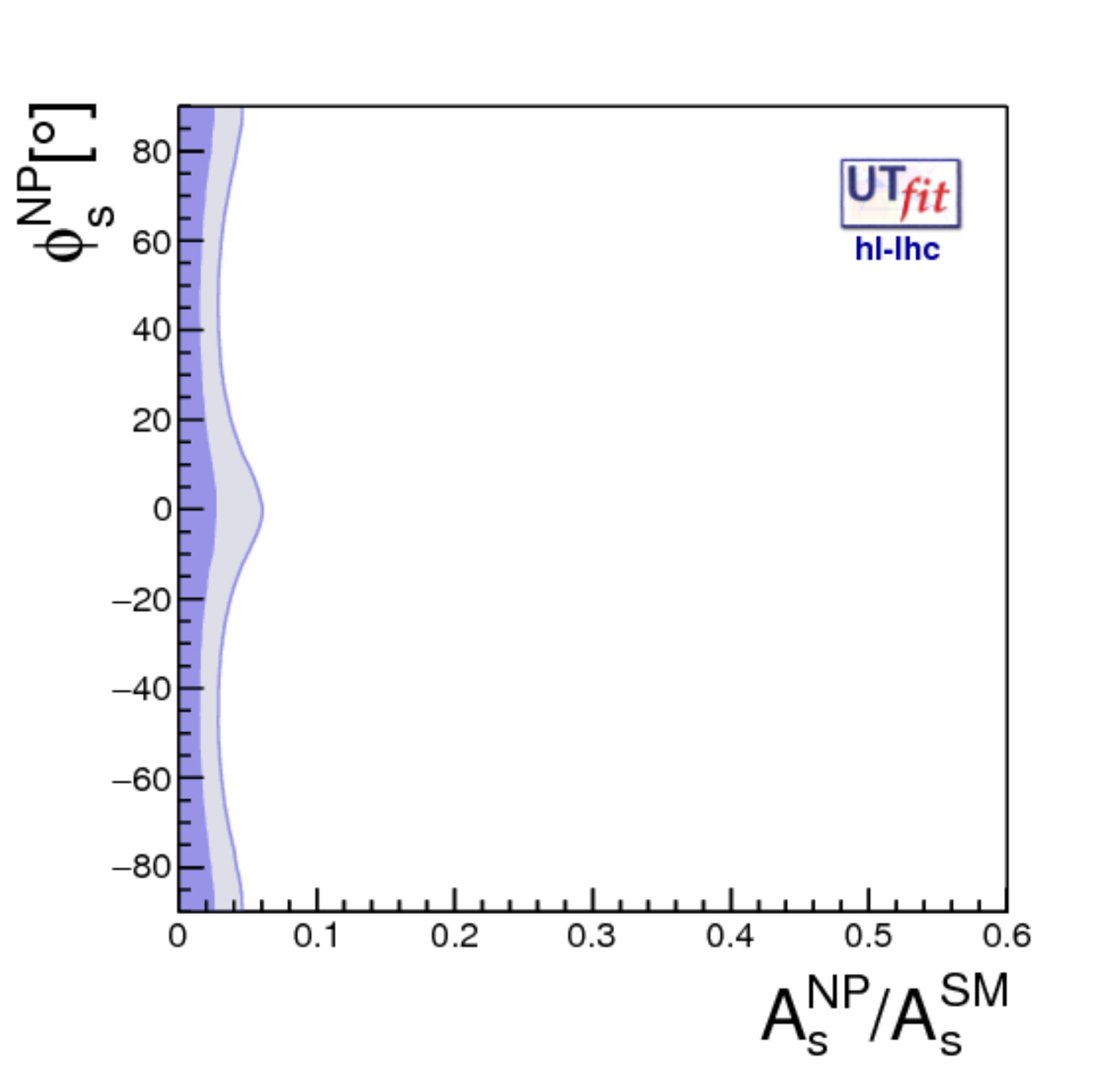}
\caption{Present (upper panels) and future (lower panels) constraints, at the end of Belle II, with $50~{\rm ab}^{-1}$ of integrated luminosity,  and LHCb with $300~{\rm fb}^{-1}$, on the NP contributions to the $B_d$ (left) or $B_s$ (right) mixing amplitudes \cite{Ciuchini:2018}. }
\label{fig:mixing projections}
\end{center}
\end{figure}

When NP is present, then  ${\cal H}_{\rm eff}^q={\cal H}_{{\rm eff},q}^{{\rm SM}}+{\cal H}_{{\rm eff},q}^{{\rm NP}}$, and we can write
\beq
M_{12}^q=M_{12,q}^{{\rm SM}}+M_{12,q}^{{\rm NP}}=M_{12,q}^{{\rm SM}}\Big[1+ \big(A_q^{\rm NP}/A_q^{\rm SM}\big) e^{i\phi_{q}^{\rm NP}}\Big], \qquad q=d,s,
\eeq
for $B_d$ and $B_s$ systems, respectively. 
%An equivalent often used parametrization is (see, e.g., \cite{Charles:2013aka})
%\beq
%h_q=A_q^{\rm NP}/A_q^{\rm SM}, \qquad \sigma_{q}=\phi_{q}^{\rm NP}.
%\eeq
The above parametrization is completely general, as long as NP is heavy, so that it does not appear in the decays of $B_{d,s}$ mesons. There is no NP contribution, when $A_{d,s}^{\rm NP}=0$. If $\phi_{d,s}^{\rm NP}\ne 0$ this means that there are new CP violating phases in the NP contribution, beyond the CKM one. At present $A_{d,s}^{\rm NP}/A_{d,s}^{\rm SM}$ of about 0.2 are still allowed, depending on the NP phase. With the future measurements at Belle II and LHCb this will be drastically improved to less than about 0.05, see Fig. \ref{fig:mixing projections}.

What does this mean in terms of bounds on NP masses? Let us assume that NP has the same $(V-A)\times (V-A)$ structure as the SM, so that the effective Hamiltonian is ($q=d,s$) 
\beq
{\cal H}_{\rm eff}=\biggr(\frac{ \big(V_{tb}^*V_{tq}\big)^2}{\Lambda_{\rm MFV}^2} +\frac{C_{\rm NP}}{\Lambda_{\rm NP}^2}\biggr) \big(\bar b_L \gamma^\mu q_L\big)  \big(\bar b_L \gamma_\mu q_L\big) +{\rm h.c.}.
\eeq
For instance, the new physics contribution, $C_{\rm N}/\Lambda_{\rm NP}^2$ could be due to the $Z'$ exchange. This would give for the effective Hamiltonian
\beq
{\cal H}_{\rm eff}= i (i g_{Z'})^2 \big(\bar b_L \gamma_\mu q_L\big) \frac{-i g^{\mu\nu}}{q^2-m_{Z'}^2}  \big(\bar b_L \gamma_\nu q_L\big) \to \frac{g_{Z'}^2}{m_{Z'}^2} \big(\bar b_L \gamma^\mu q_L\big)  \big(\bar b_L \gamma_\mu q_L\big),
\eeq
where $g_{Z'}$ is the flavour violating $Z'$ coupling to quarks, $m_{Z'}$ the $Z'$ mass, and in obtaining the last expression we used that $q^2\ll m_{Z'}^2$. For the NP Wilson coefficient we thus have
\beq
\frac{C_{\rm NP}}{\Lambda_{\rm NP}^2}=\frac{g_{Z'}^2}{m_{Z'}^2}.
\eeq
If $g_{Z'}=1$, then $\Lambda_{\rm NP}$ can be identified with $m_{Z'}$ for $C_{\rm NP}=1$.   

\begin{figure}[t]
\begin{center}
\includegraphics[width=0.49\linewidth]{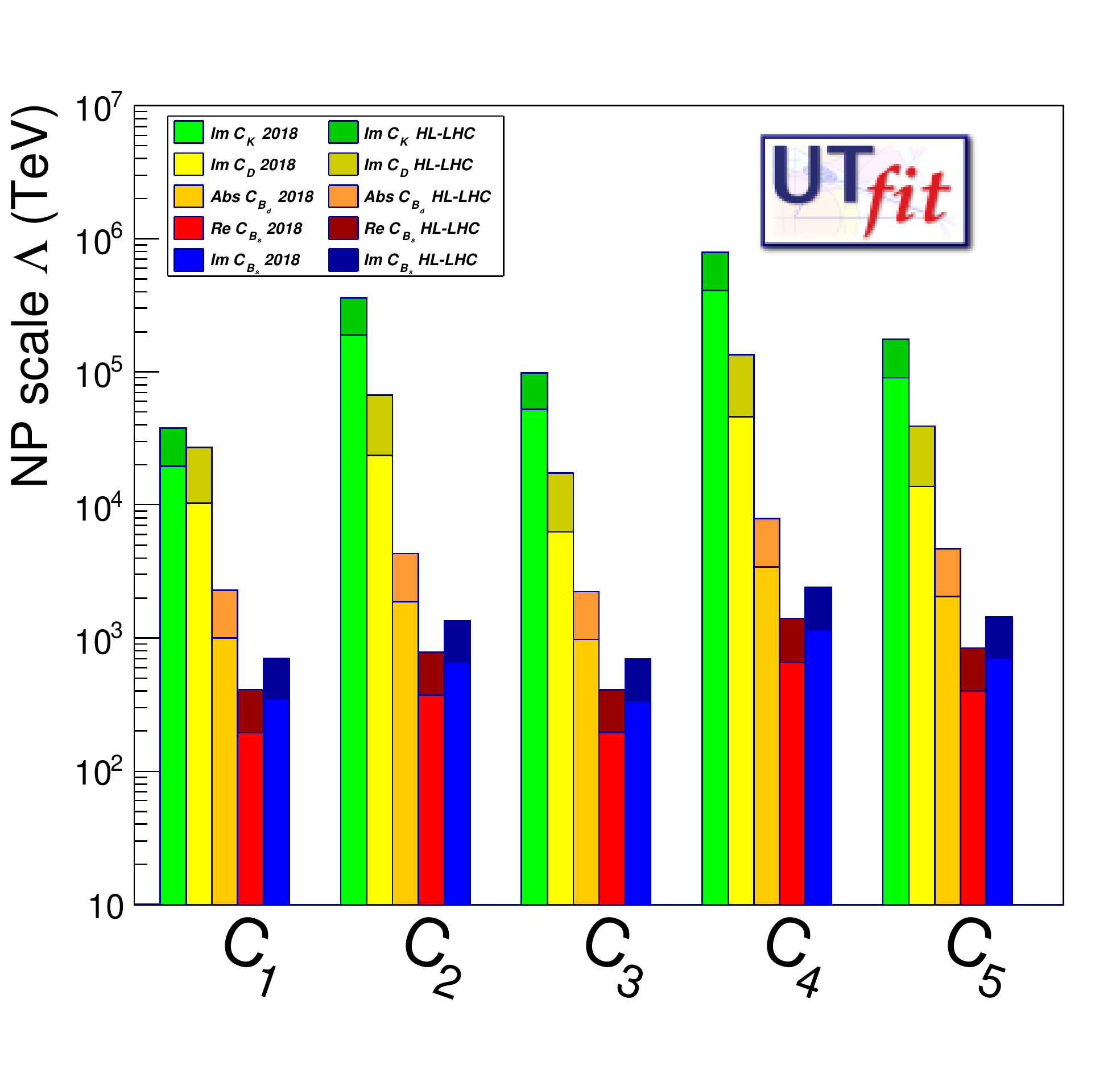}
\includegraphics[width=0.49\linewidth]{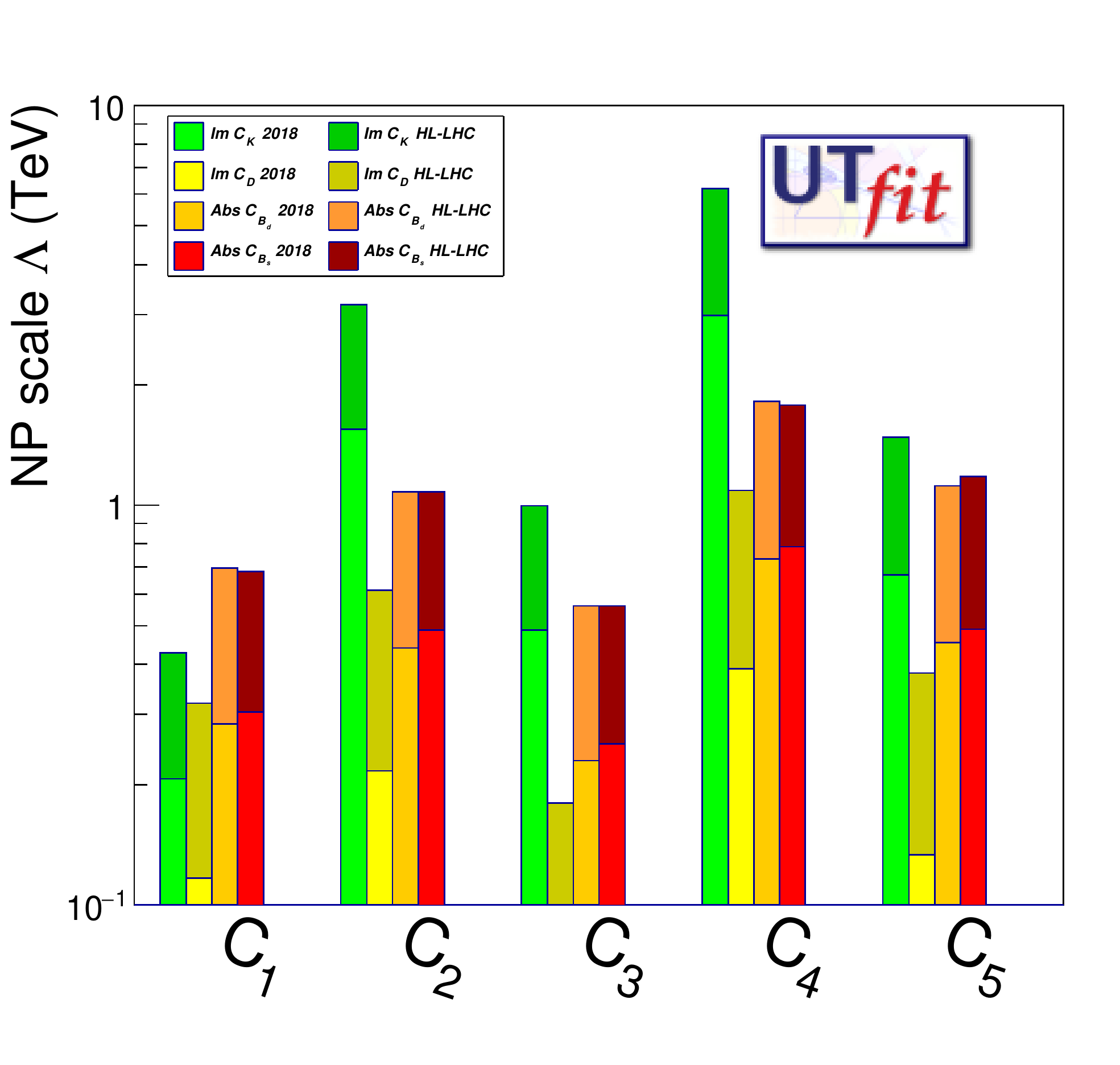}
\caption{Present constraints (lighter) and expected constraints at the end of LCHb upgrade II (darker) on the NP scale, $\Lambda_{\rm NP}$, from the UTfit NP analysis for different meson systems. The right panel shows constraints assuming NP is weakly coupled, has MFV structure of couplings, and enters observables only at one loop, see text for details (from \cite{Cerri:2018ypt}). }
\label{fig:DF2Lambda}
\end{center}
\end{figure}

In general NP will not have the $V-A$ structure. However, the choice of possible operator structure is still quite limited. 
The general dimension 6 operator basis for meson mixing contributions is \cite{Bona:2007vi}
\beq
{\cal H}_{\rm eff}^{\rm NP}=\sum_i \frac{C_i}{\Lambda_{{\rm NP},B_q}^2} Q_{i,q},
\eeq
where
\beq
\begin{split}
Q_{1,q}&=(\bar b_L\gamma^\mu q_L)(\bar b_L\gamma^\mu q_L),
\\
Q_{2,q}&=(\bar b_R q_L)(\bar b_R q_L),
\\
Q_{3,q}&=(\bar b_R^\alpha q_L^\beta)(\bar b_R^\beta q_L^\alpha)
\\
Q_{4,q}&=(\bar b_R q_L)(\bar b_L q_R),
\\
Q_{5,q}&=(\bar b_R^\alpha q_L^\beta)(\bar b_L^\beta q_R^\alpha),
\end{split}
\eeq
along with three other operators obtained from $Q_{i,q}$, $i=1,2,3$ by replacing $L\leftrightarrow R$ (the bounds on these parity related operators are the same as for $Q_{i,q}$, $i=1,2,3$, though).  The operators for other meson systems are obtained through trivial replacements of quark flavours. Taking $|C_i|=1$, the present bounds on the NP scale, $\Lambda_{\rm NP}$ are shown with ligher colors in Fig. \ref{fig:DF2Lambda} (left). The future projections to the end of LHCb Upgrade II are shown with darker colors. A jump in the mass reach is clearly visible even on the logarithmic scale.

%The experimentally allowed deviations of $h_{d,s}\lesssim 0.2$ when give, for $C_{\rm NP}=1$,
%\beq
%\Lambda_{{\rm NP}, B_d} \gtrsim 1500 {\rm~TeV}, \qquad \Lambda_{{\rm NP}, B_s} \gtrsim 300 {\rm~TeV}.
%\eeq
%The difference between the two is entirely from $V_{ts}\simeq 5 V_{td}$.
%The $K- \bar K$ and $D-\bar D$ mixing more complicated, as there are long distance contributions. 

Different colours in Fig.~\ref{fig:DF2Lambda} denote different meson systems: green bars denote the constraints from $K^0-\bar K^0$ mixing, yellow from $D^0-\bar D^0$ mixing, in both cases assuming maximal  new weak phase relative to the SM; orange bar denotes constraints from $B^0-\bar B^0$, marginalized over the weak phase; red (blue) from $B_s^0-\bar B_s^0$ system assuming no (maximal) NP phase. Extraction of  constraints from $K- \bar K$ and $D-\bar D$ mixing is more complicated than for $B_q-\bar B_q$, since in these two cases the long distance contributions from light quarks running in the loop are important. 

The bounds on $\Lambda_{\rm NP}$ are strikingly different for the various meson systems. This is easy to understand by considering the CKM suppression of the SM contributions, since the precision of experimental measurements and theoretical predicitons is typically at a fraction of the SM amplitude. 
Demanding for illustration that the contribution from NP is at most 20\% of the short-distance SM this would give, for the operator $Q_{1,q}$
\beq
\begin{split}
\text{for NP=20\% SM,~~~~~} K-\bar K: & \qquad (\underbrace{V_{ts}^*}_{\lambda^2} \underbrace{V_{td}}_{\lambda^3})^2 \quad \Rightarrow\quad  \Lambda_{\rm NP} \gtrsim 4 \cdot 10^4{\rm ~TeV},
\\
\text{for NP=20\% SM,~~~}B_d-\bar B_d: & \qquad (\underbrace{V_{tb}^*}_{1} \underbrace{V_{td}}_{\lambda^3})^2 \quad \Rightarrow\quad  \Lambda_{\rm NP} \gtrsim 1.5 \cdot 10^3{\rm ~TeV},
\\
\text{for NP=20\% SM,~~~\,}B_s-\bar B_s: & \qquad (\underbrace{V_{tb}^*}_{1} \underbrace{V_{ts}}_{\lambda^2})^2 \quad \Rightarrow\quad  \Lambda_{\rm NP} \gtrsim 3 \cdot 10^2{\rm ~TeV}.
\end{split}
\eeq
roughly in agreement with the constraints shown in Fig.~\ref{fig:DF2Lambda}. 

Note that the interpretation of the bounds in term of NP scale crucially depends  on the assumed flavour structure in the dimensionless Wilson coefficient, $C_i$. If the NP contribution also follows the SM CKM suppression, this is referred to as Minimal Flavour Violation (MFV). Fig. \ref{fig:DF2Lambda} (right) shows the bounds for the case of MFV type NP running in the loop, i.e., the Wilson coefficients were set to $C_a= (V_{ti}^* V_{tj})^2 g^4/16 \pi^2$, with $V_{ti}, V_{tj}$ the appropriate SM CKM coefficients, and $g$ the weak coupling constant. We see that even for a weakly coupled NP that has the MFV flavour structure and only contributes at 1 loop, the bounds are in the few 100 GeV to few TeV range. 

\subsection{New physics searches using rare decays}
We turn next to the other main pathway to searching for new physics - searching for deviations in rare decays. Here the benefit is that there are many observables in flavour physics: the branching ratios, asymmetries, distributions, ... There is also a choice of different parent particles as well as many possible final states. The abundance of observables is clearly illustrated by opening the ``bible'' of particle physics, the Particle Data Group (PDG) book \cite{Tanabashi:2018oca}. Even the condensed version, the PDG booklet, clocks out at more than 170 pages. 

To shorten the discussion we will focus on the processes that are at present showing deviations from the SM expectations. The present  experimental situation can then succinctly be described in the following way. There are many different transitions that were measured, all of which agree with the SM expectations within experimental and theoretical errors. There are only two sets of quark level transitions that are showing $\sim 4 \sigma$ deviations from the SM: the $b\to c\tau \nu$ and $b\to s\mu^+\mu^-$ transitions.\footnote{There are other interesting deviations, e.g., the $\sim 3\sigma$ deviation in $\epsilon'/\epsilon$,  see, e.g., \cite{Buras:2015yba,Blum:2015ywa,Buras:2015jaq}.}  The apparent NP scale that explains the deviations is quite different in the two cases. For instance, if the NP is due to the following $V-A$ operator
\beq
{\cal L}_{\rm NP}\supset \frac{1}{\Lambda_{\rm NP}^2}\big(\bar Q_i \gamma^\mu \sigma^A Q_j\big)\big(\bar L_k \gamma_\mu \sigma^A L_l\big), 
\eeq
then $\Lambda_{\rm NP}\sim 3$ TeV in order to explain the deviations in $b\to c\tau \nu$ transitions, and $\Lambda_{\rm NP}\sim 30$ TeV in order to explain the $b\to s\mu^+ \mu^-$ anomalies. We discuss next the possible NP explanations for each of the two.

%The SM diagrams for these are shown in Fig.~\ref{fig:RD:diagrams} (left) and Fig.~\ref{fig:bsmumu:diagrams} (left), respectively. 

\subsection{New physics searches in $b\to s\mu^+ \mu^-$ transitions}
\label{Sec:bsmumu}
The upshot of the observed $b\to s \mu^+\mu^-$ anomaly is: choosing only the theoretically clean observables the excess is at the $\sim 4 \sigma$ level. From the NP perspective the scale required to explain the anomaly makes sense, since it is high enough to avoid many of the experimental constraints. The models that explain the anomaly do, however, face I.I. Rabi's question:`` Who ordered that?'', when the muon was first discovered \cite{NYT:1987}. 

\begin{figure}[t]
\begin{center}
\includegraphics[width=0.95\linewidth]{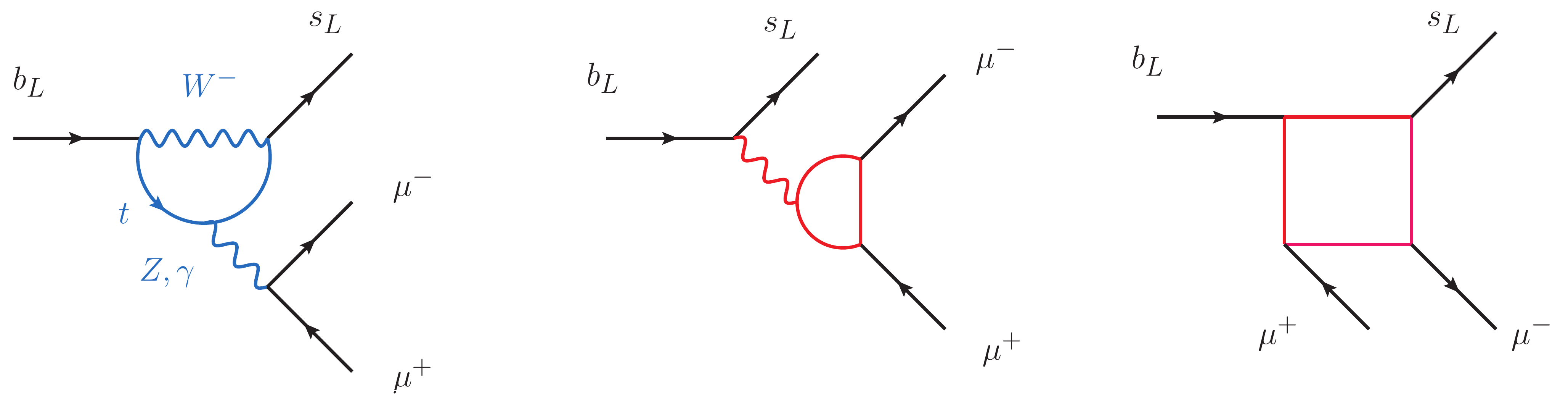}
\caption{A representative SM diagram for $b\to s \mu^+ \mu^- $ transition (left), and the representative loop level NP contributions (middle and right).}
\label{fig:bsmumu:diagrams}
\end{center}
\end{figure}

The FCNC $b\to s\ell^+\ell^-$ transitions are generated at 1-loop in the SM. A representative diagram in the SM is shown in Fig.~\ref{fig:bsmumu:diagrams} (left). Integrating out the heavy degrees of freedom, $W, Z, t$, gives the following effective Hamiltonian \cite{Grinstein:1988me,Buchalla:1995vs,Chetyrkin:1996vx}
\beq
\label{eq:bsmumu:eff:SM}
{\cal H}_{\rm eff}=G_F V_{tb} V_{ts}^* \frac{\alpha}{4\pi}\Big[ C_9 \big(\bar s_L \gamma^\mu b_L\big)\big(\bar \ell \gamma_\mu \ell\big)+ C_{10} \big(\bar s_L \gamma^\mu b_L\big)\big(\bar \ell \gamma_\mu \gamma_5 \ell\big)\Big],
\eeq
where in the SM $C_9^{\rm SM}\simeq -C_{10}^{\rm SM}$, i.e., the SM diagrams give to a good approximation a $V-A$ structure of the leptonic current. 

Another prediction of the SM is that the rates for the $b\to s e^+ e^-$ and $b\to s \mu^+\mu^-$ transitions should be equal to each other as soon as we are reasonably far above the muon production threshold so that the effect of muon mass on the available phase space can be neglected. The SM prediction of Lepton Flavour Universality (LFU) is deeply engrained in the structure of the theory, since it is a consequence of the fact that the electroweak gauge group is the same for all three generations. 

The prediction of LFU can be tested experimentally by forming theoretically clean observables such as the ratios of $b\to s\mu\mu$ to $b\to s ee $ rates,  
\beq
R_{K^{(*)}}=\frac{Br(B\to K^{(*)}\mu\mu)}{Br(B\to K^{(*)} ee)}.
\eeq
In the ratios the uncertainties from hadronic inputs (the form facors) cancel to a very good approximation. Above the muon threshold they are equal to 1 within a percent, and are also presicely predicted close to the muon threshold \cite{Bordone:2016gaq,Descotes-Genon:2015uva,Capdevila:2016ivx,Capdevila:2017ert,Serra:2016ivr,Straub:2015ica,Altmannshofer:2017fio,Jager:2014rwa}.  Experimentally, on the other hand $R_{K^{(*)}}\sim 0.7$ \cite{Aaij:2014ora,Aaij:2017vbb,Wei:2009zv,Lees:2012tva}, violating LFU with a significance of $2.2-2.6\sigma$ in each of the three most precise measurements (the measurements are at different dilepton invariant masses). A combined significance for the discrepancy with the SM is $\sim4 \sigma$ \cite{DAmico:2017mtc,Geng:2017svp,Altmannshofer:2017yso,Capdevila:2017bsm,Hurth:2017hxg,Hiller:2017bzc}. The most precise measurements are due to LHCb, which dominates the world averages for $R_{K^{(*)}}$.

\begin{figure}[t]
\begin{center}
\includegraphics[width=0.45\linewidth]{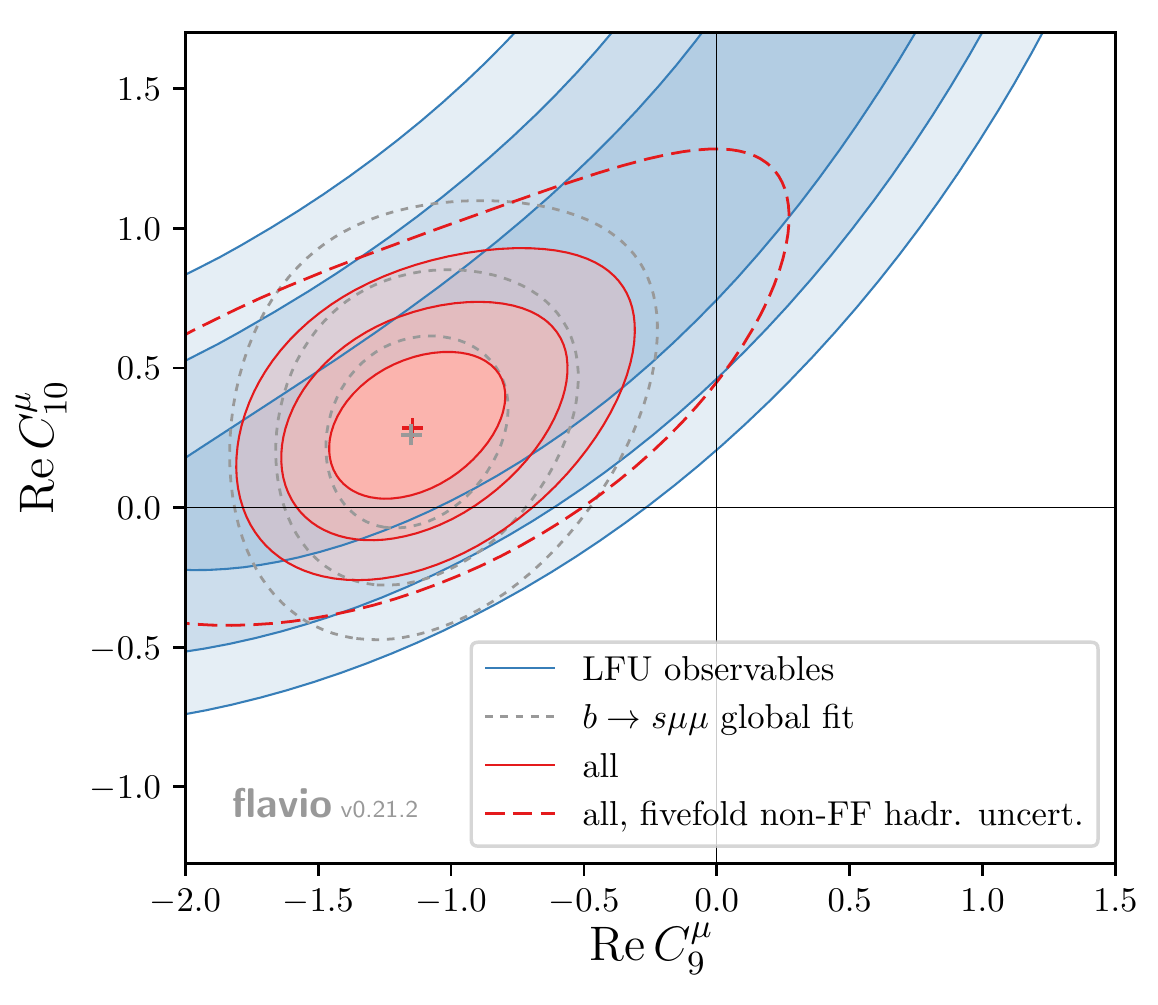}
\caption{The fit for NP contributions $C_{9,10}^\mu$ to Wilson coefficients in \eqref{eq:bsmumu:eff:SM} from LFU only observables (red) or including all the observables (blue). The SM point is $C_{9,10}^\mu=0$. (from \cite{Altmannshofer:2017yso}.) }
\label{fig:bsmumu:global}
\end{center}
\end{figure}

LFU ratios is not the only experimental information about the $b\to s \ell\ell$ transitions. In principle there is much more information available, branching ratios for different choices of initial and final state mesons, $Br(B\to K^{(*)}\mu\mu)$, $Br(B_s\to \phi \mu\mu)$, $Br(B\to X_s\mu\mu)$, angular observables in $B^0\to K^{*0}\mu\mu, B_s\to \phi \mu\mu$, etc. However, the interpretation of these is much more sensitive to hadronic inputs. It requires form factor predictions (now coming from QCD sum rules), the estimate of charm loops, nonfactorizable contributions, etc. Using the best available estimates for these inputs the favored interpretation is that the NP is mostly in muons~\cite{DAmico:2017mtc,Geng:2017svp,Altmannshofer:2017yso,Capdevila:2017bsm,Hurth:2017hxg,Hiller:2017bzc}. Furthermore, the picture obtained from such global fits to data seems to be in agreement with the LFU only determination, see Fig.~\ref{fig:bsmumu:global} for a simultaneous fit to NP contributions in $C_{9,10}$.

If the anomaly is due to NP we thus already have a significant amount of information about it. First of all, there are only four dimension 6 operators that can explain $R_K$ \cite{Alonso:2014csa}
\beq
{\cal O}_9^{(')\ell}=\frac{\alpha}{4\pi}\big(\bar s \gamma^\mu P_{L(R)}b\big)\big(\bar \ell \gamma_\mu \ell\big),  \qquad 
{\cal O}_{10}^{(')\ell}=\frac{\alpha}{4\pi}\big(\bar s \gamma^\mu P_{L(R)}b\big)\big(\bar \ell \gamma_\mu \gamma_5 \ell\big).
\eeq
The other operators are either constrained by $B_s\to \ell\ell$ as is the case for scalar currents, or come from further suppressed dimension 8 operators before electroweak symmetry is broken, as is the case for tensor operators. 

\begin{figure}[t]
\begin{center}
\includegraphics[width=0.43\linewidth]{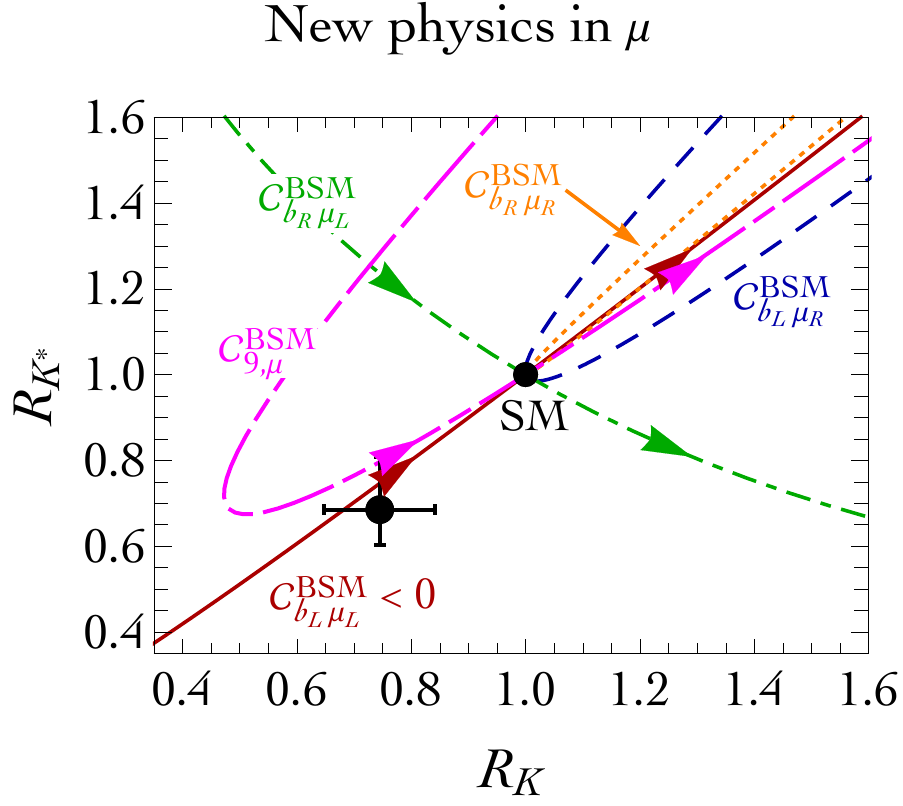}~~~~~
\includegraphics[width=0.43\linewidth]{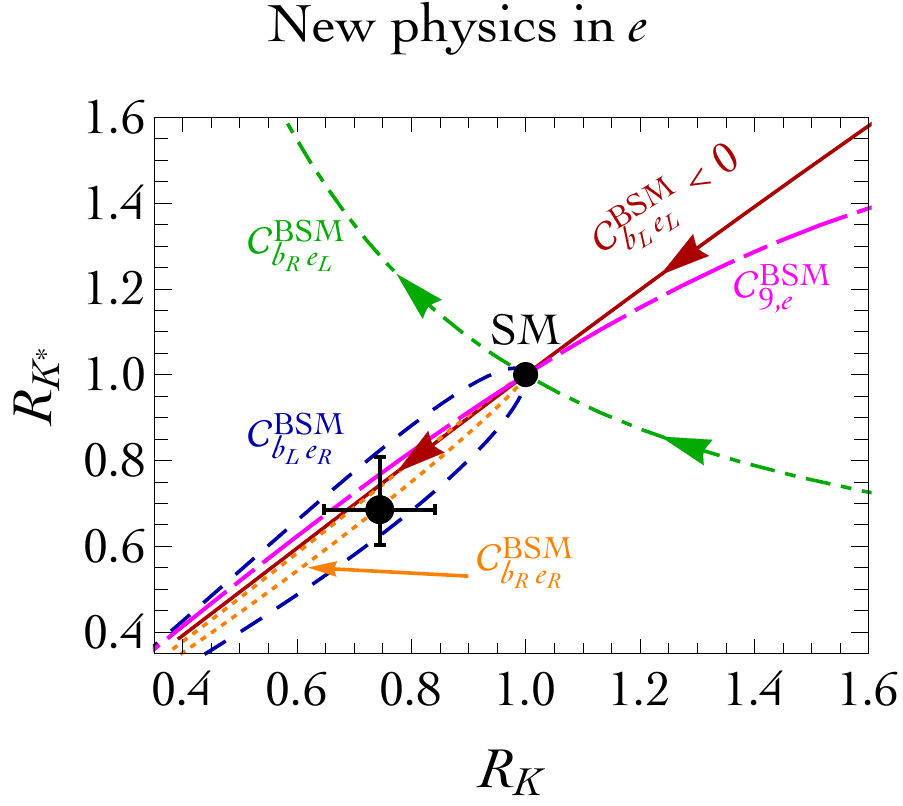}
\caption{The predictions for the central $q^2$ bins in $R_K$ and $R_{K^*}$ for several NP scenarios, assuming NP is only in muons (left) or only in electrons (right). The experimental values are given by the black error bars, the SM value by a point (from \cite{DAmico:2017mtc}). }
\label{fig:RKRKs:interp}
\end{center}
\end{figure}

Since the $K$ and $K^*$ in the final states differ in their spin-parity quantum numbers, one is pseudoscalar, the other vector meson, the ratios $R_K$ and $R_{K^*}$ give complementary information. For instance, for the central $q^2$ bins we have \cite{DAmico:2017mtc}
\beq
R_K\simeq 1+2 \frac{{\rm Re} C_{b_{L+R}(\mu-e)_L}^{\rm BSM}}{C_{b_L\mu_L}^{\rm SM}}, \qquad 
R_{K^*}\simeq R_K -4 p \frac{{\rm Re} C_{b_{R}(\mu-e)_L}^{\rm BSM}}{C_{b_L\mu_L}^{\rm SM}},
\eeq
 when expanded to linear order in the BSM contributions to the Wilson coefficients (here $p\simeq0.86$ is the polarization  fraction of $K^*$). The resulting predictions for several choices of chirality in the NP contributions to the Wilson coefficients are shown in Fig. \ref{fig:RKRKs:interp}. Using just the ``clean'' observables, $R_K$ and $R_{K^*}$, NP can be either due to a deficit in muons or an increase in the electron channel. In both cases the operators with $(\bar s \gamma^\mu b)_L$ current can explain the anomaly, with significant freedom for the chirality of the leptonic current. For electrons also the NP due to $(\bar s \gamma^\mu b)_R(\bar e \gamma_\mu e)_R$ is possible. In this case the NP contribution enters only quadratically to a good approximation, since there is almost no interference  with the SM predominantly $V-A$ leptonic current.  The NP thus increases the rate for the electron channel, reducing the $R_{K^{(*)}}$ ratios below 1. It is only once additional observables, such as the absolute branching ratios, are taking into account that the possibility of NP right-handed currents is disfavored. These additional observables do require theoretical inputs and are subject to hadronic uncertainties. 

For the remainder of this section let us assume that there is NP in $b\to s \mu^+\mu^-$. What kind of NP can explain it? There is significant freedom in the NP interpretations, since the associated scale is quite high. The Wilson coefficients shown in Fig. \ref{fig:bsmumu:global} are normalized to 
\beq
V_{tb} V_{ts}^* \frac{\alpha}{4\pi v^2}C_I=\frac{C_I}{(36 {\rm~TeV})^2}.
\eeq
The NP scale of $\sim 30$ TeV is high enough that the NP can enter either at tree level, or even only at one loop level. The tree level NP models are of two distinct types. The mediator can be {\em (i)} a $Z'$, either an $SU(2)_L$ singlet or part of a triplet \cite{Altmannshofer:2013foa,Altmannshofer:2014cfa,Greljo:2015mma}, or {\em (ii)} a leptoquark with either spin 0 or spin 1 \cite{Hiller:2017bzc,Hiller:2014ula}. The diagrams for the two types of mediators are shown in Fig.~\ref{fig:RD:diagrams} middle and right, respectively. 

There are 4 different possible charge assignments under the SM gauge group $SU(3)_c\times SU(2)_L\times U(1)_Y$ for a scalar leptoquark, and 3 for a vector leptoquark \cite{Hiller:2017bzc}. However, only one scalar leptoquark, $S_3\sim (\bar 3, 3, 1/3)$, and only two vector leptoquarks, $V_1\sim (3,1,2/3)$ and $V_3\sim (3,3,-2/3)$, lead to $R_K\simeq R_{K^*}<1$ in agreement with the data. All three predict $C_9^\mu=-C_{10}^\mu$. At 1-loop the leptoquarks contribute to $B_s-\bar B_s$ mixing, correcting the mass splitting by $\Delta m_{B_s} \propto (YY^*)^2/M^2$, where $M$ is the leptoquark mass, and $Y$ the relevant couplings to the SM fermions. The corrections to $R_{K^{(*)}}$, on the other hand, scale as $R_{K^{(*)}}-1\propto YY^*/M^2$. This means that the value of $Y$ required to explain $R_{K^{(*)}}$ grows faster with leptoquark mass than does the value of $Y$ still allowed by the $B_s-\bar B_s$ mixing constraints. In other words, the bound on allowed NP in $B_s-\bar B_s$ mixing  implies un upper bound on the leptoquark mass, $M\lesssim 40 {\rm~TeV}, 45 {\rm~TeV}, 20 {\rm~TeV}$, for leptoquarks $S_3, V_1, V_3$, respectively \cite{Hiller:2017bzc}. 

The bounds on allowed NP contributions to $B_s$ mixing also imply a nontrivial constraint on the models with $Z'$, since this contributes at tree level, giving \cite{Altmannshofer:2013foa,Altmannshofer:2014rta}
\beq
\frac{g_{bsZ'}}{m_{Z'}}\lesssim \frac{0.01}{2.5{\rm~TeV}}.
\eeq
 Thus a $2.5$ TeV $Z'$ has to have a relatively small, but not extremely small, flavour violating coupling, $g_{bsZ'}\lesssim 0.01$ (comparable, for instance, with  $|V_{ts}|\simeq 0.04$). This also means that the $Z'$ has to have sizeable couplings to muons. If the coupling is to left-handed muons, this implies nontrivial constraints from neutrino trident production in neutrino scattering on nuclei, i.e., from bounds on the process $\nu N \to \nu N \mu^+\mu^-$ mediated by a $Z'$ \cite{Altmannshofer:2014pba,Altmannshofer:2014cfa}. For couplings to left-handed muons the $b\to s \mu^+\mu^-$ is also accompanied by a $b\to s \bar \nu\nu$ signal, giving stringent constraints on the parameter space. Another important constraint are the $Z'$ searches at the LHC. 
 
 To recap, the NP explanations of the $b\to s \mu^+\mu^-$ anomaly should lead to new signals in a number of observables. The present constraints give meaningful bounds on the models already, but they are not too constraining. For instance, simply raising the mass of $Z'$ avoids the high $p_T$ constraints at the LHC. The bounds are more stringent for loop induced models \cite{Kamenik:2017tnu,Belanger:2016ywb,Gripaios:2015gra,Bauer:2015knc,Becirevic:2017jtw}, Fig. \ref{fig:bsmumu:diagrams} (middle and right), since there the NP particles need to be lighter, below about a TeV.

\subsection{New physics searches in $b\to c\tau \nu$ transitions}
\label{Sec:bctaunu}
\begin{figure}[t]
\begin{center}
\includegraphics[width=0.28\linewidth]{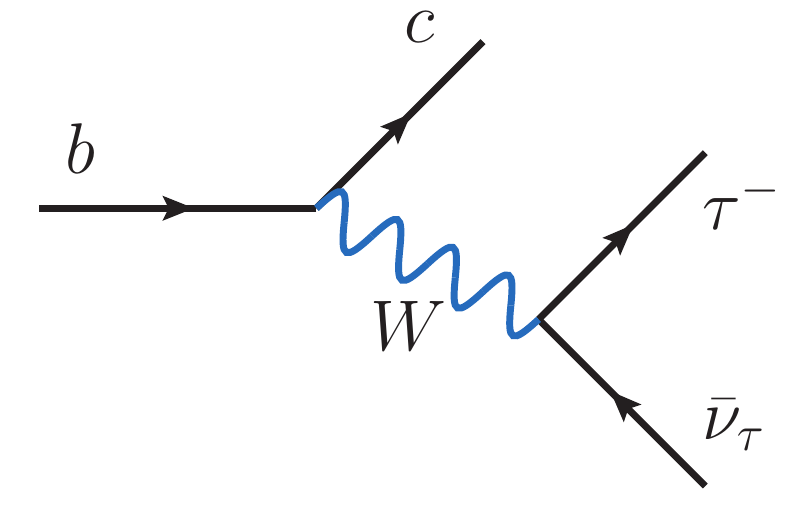}~~~~~
\includegraphics[width=0.29\linewidth]{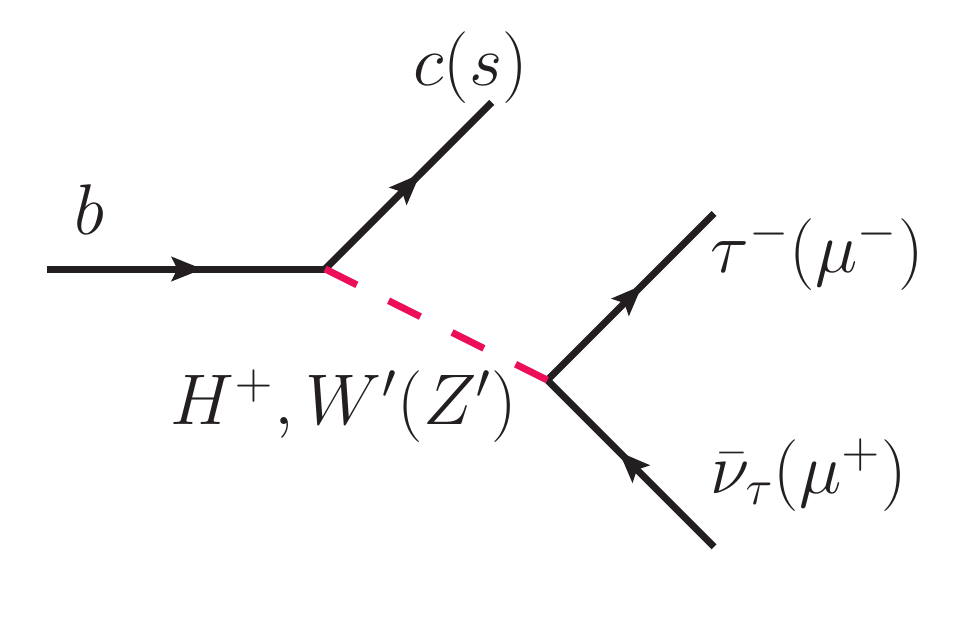}~~~~~
\includegraphics[width=0.28\linewidth]{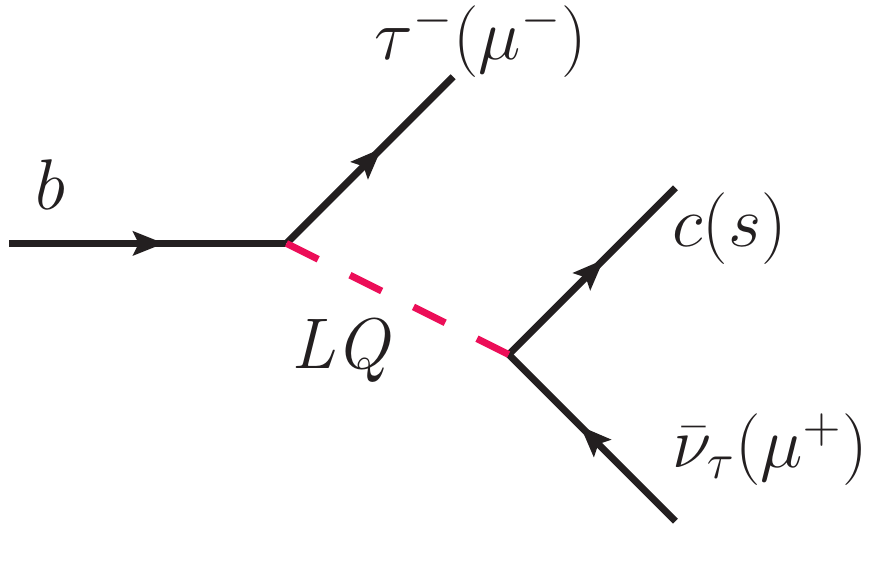}
\caption{The SM diagrams for $b\to c \tau \nu $ transition (left), and the possible tree level NP contributions to $b\to c \tau\nu$ or $b\to s \mu\mu$ transitions (middle and right). }
\label{fig:RD:diagrams}
\end{center}
\end{figure}

The $b\to c\tau\nu$ flavour anomaly is similarly  very clean theoretically \cite{Fajfer:2012vx}, and the disagreement with the SM predictions is also about $\sim 4 \sigma$. However, the NP effect is large, ${\mathcal O}(20\%)$ of the SM tree level contribution given in Fig.~\ref{fig:RD:diagrams} (left). This means that the scale of NP needs to be low, and consequently the NP interpretations are often in conflict with the other constraints. 

\begin{figure}[t]
\begin{center}
\includegraphics[width=0.85\linewidth]{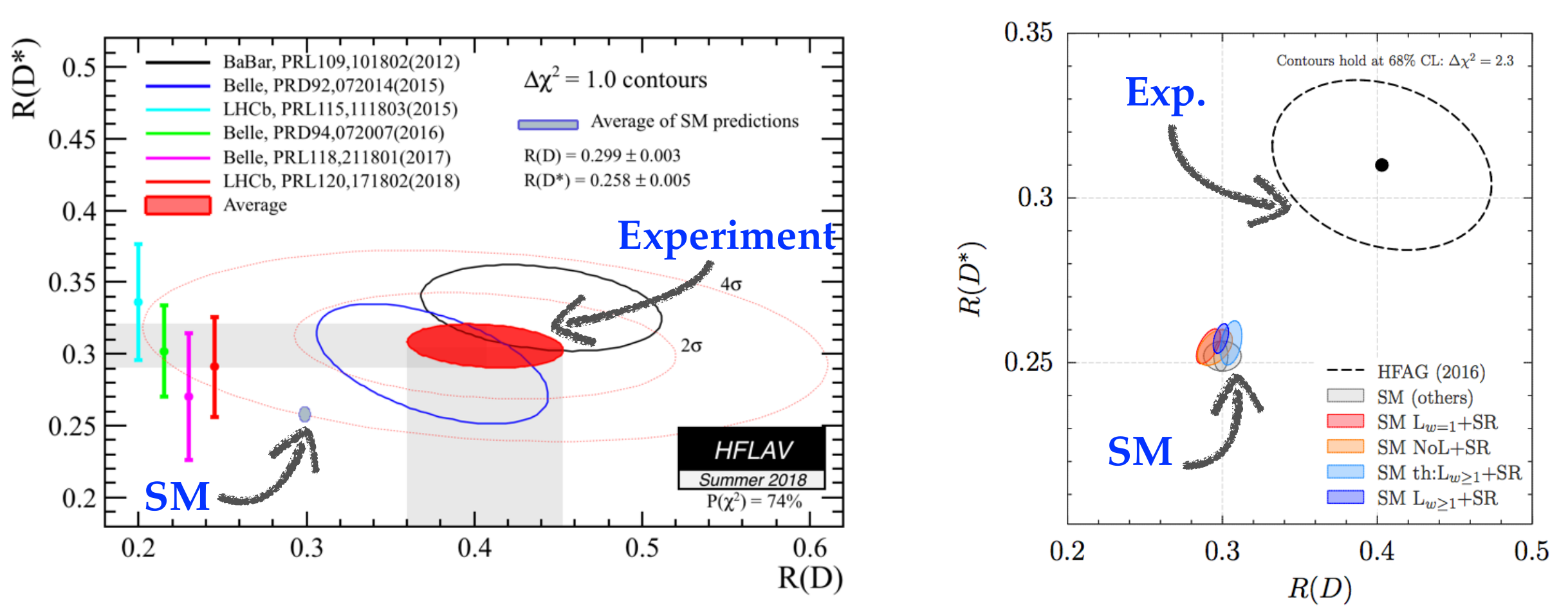}
\caption{Left panel: the measurements of $R(D)$ and $R(D^*)$ by different experiments \cite{Lees:2012xj,Lees:2013uzd,Huschle:2015rga,Sato:2016svk,Aaij:2015yra,Hirose:2016wfn,Hirose:2017dxl,Aaij:2017deq},  with the world average shown in red and the SM prediction in blue \cite{Bigi:2016mdz,Bernlochner:2017jka,Bigi:2017jbd,Jaiswal:2017rve}. Right panel shows that the variation in the SM predictions is small even when some of the theoretical constraints are relaxed (from \cite{Bernlochner:2017jka}).}
\label{fig:RD:exp}
\end{center}
\end{figure}

The two main observables are 
\beq
R(D^{(*)})=\frac{\Gamma(\bar B\to D^{(*)}\tau \bar \nu)}{\Gamma(\bar B\to D^{(*)}\ell \bar \nu)}, \qquad \ell=\mu, e,
\eeq
where $\bar B^+\sim b\bar u$, $D\sim c\bar u$, etc.
The SM predictions are shown in \ref{fig:RD:exp}. Even though these are flavour universality ratios, the SM predictions are well below 1, because the $b\to c\tau \nu$ decays have much less final state phase space available due to the large $\tau$ mass.  The thing to note is that the trend $R(D^{(*)})_{\rm exp}> R(D^{(*)})_{\rm SM}$ is seen in several experiments. Furthermore, the theoretical predicitons are well under control. Another comment is that, since the neutrino is not seen in the expriments, it does not need to be the SM neutrino. It could be a new state, possibly even of right-handed chirality \cite{Asadi:2018wea,Greljo:2018ogz,Becirevic:2016yqi,Cvetic:2017gkt}.

 What kind of NP could explain this anomaly? The most obvious candidates are ruled out. Theoretical bias would have been that the new charged currents are either due to a charged Higgs, $H^+$, or a new vector boson, $W'$, see Fig. \ref{fig:RD:diagrams} (middle). The charged Higgs option is in conflict with total $B_c$ lifetime \cite{Alonso:2016oyd}, the $b\to c \tau \nu$ leptonic mass distributions, and searches in $pp\to \tau^+\tau^-$ \cite{Faroughy:2016osc}. The  $W'$ is excluded by $pp\to\tau$+MET searches at the LHC \cite{Greljo:2018tzh}, while in addition the $pp\to \tau^+\tau^-$ in conjunction with $B$ mixing constraints exclude the related $Z'$.
 
 There are several viable leptoquark solutions, both with SM neutrinos \cite{Freytsis:2015qca} and for right-handed neutrinos \cite{Robinson:2018gza}. The vector leptoquark $V_1$ also allows to simultaneously explain the $b\to c\tau\nu$ and the $b\to s \mu^+\mu^-$ anomalies \cite{Buttazzo:2017ixm}. A simultaneous explanation is also possible, if there are more than one scalar leptoquarks contributing \cite{Crivellin:2017zlb}. 
 
 \subsection{Other modes}
Besides the two quark level transitions that are showing experimental discrepancies there are a number of other rare decays that are important probes of NP.  The useful rare decays are such that we can predict them precisely and that NP contributions are possible or even likely. The modes with only one final state hadron, $K\to \pi \nu\nu, B\to K\ell\ell$,..., fall into this category. The hadronic matrix elements for these decays are easier to predict than for the fully hadronic decays. Another example are inclusive decays, where one sums over all hadronic final states, which are also easier to predict theoretically. We look at one important example for each of these two categories. 

\begin{figure}[t]
\begin{center}
\includegraphics[width=0.24\linewidth]{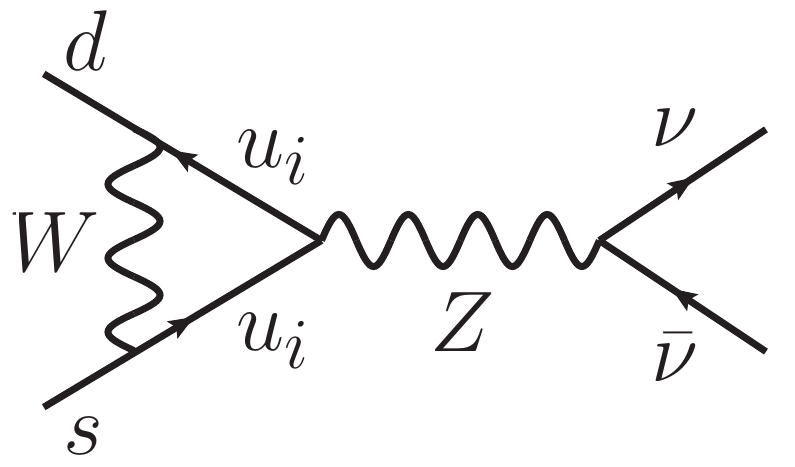}~~~~~~~~~~~~~~
\includegraphics[width=0.23\linewidth]{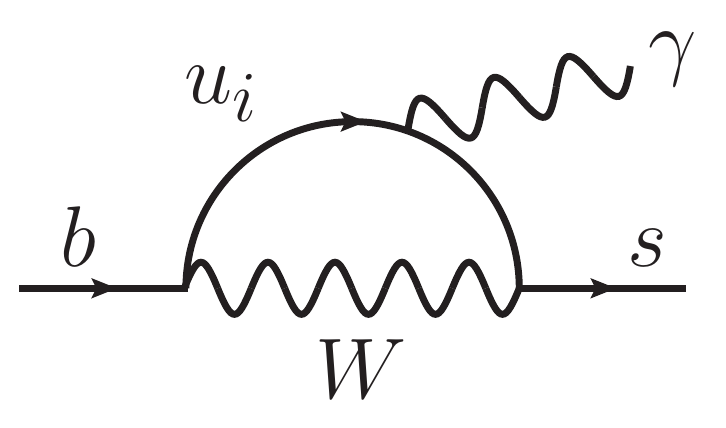}
\caption{The representative SM diagrams for $K\to \pi \nu\bar \nu$ (left) and $b\to s\gamma$ transitions (right). }
\label{fig:Kpinunu}
\end{center}
\end{figure}

The inclusive $b\to s\gamma$ decay is a classic example of a GIM suppressed loop induced SM process. The loop contributions that do not depend on  masses of the quarks running in the loop cancel due to CKM unitarity, $M\propto \sum_i V_{ib}^* V_{is}=0$. The first nonzero contribution is thus proportional to mass differences of the quarks on the internal line in Fig. \ref{fig:Kpinunu} (right). The SM contribution is finite, since it is described by the effective Hamiltonian of dimension 5 \cite{Buchalla:1995vs}
\beq\label{eq:bsgamma:SM}
{\cal H}_{\rm eff}=-\frac{G_F}{\sqrt 2} V_{ts}^* V_{tb} C_{7\gamma}(m_b) \frac{e}{4\pi^2} m_b \big(\bar s_L\sigma^{\mu \nu}b_R\big) F_{\mu\nu}. 
\eeq
In the renormalizable SM Lagrangian there is no such counter-term, thus the contribution needs to be finite.  
The operator in \eqref{eq:bsgamma:SM} is chirality flipping. In the SM the chirality flip occurs on the external leg, and is thus proportional to the $b$ quark mass, $m_b$. NP contributions, on the other hand, can have the chirality flip on the internal line, leading to a relative enhancement of the NP contributions compared to the SM. This happens for instance in the Minimal Supersymmetric Standard Model (MSSM) for the gluino-squark diagram, or for the exchange of a charged Higgs in the loop. The measurements of $b\to s\gamma$ are therefore very sensitive to such NP contributions. A great theoretical effort has thus been devoted to obtain a precise theoretical prediction for the SM $b\to s \gamma$ rate \cite{Misiak:2015xwa}. 

The decays $K^+\to \pi^+\nu\bar \nu$ and $K_L\to \pi^0\nu\bar \nu$ stand out, since these are one of the few rare decays in kaon sector that are the golden modes for NP searches. They are suppressed by a loop factor, CKM factors, and the GIM mechanism. They are also extremely well predicted theoretically \cite{Buras:2015qea}
\beq
\begin{split}
Br(K^+\to \pi^+\nu\bar \nu)=(8.4\pm 1.0)\times 10^{-11},
\\
Br(K_L\to \pi^0\nu\bar \nu)=(3.4\pm0.6)\times 10^{-11},
\end{split}
\eeq
%\jz{cite and numbers}. 
because the hadronic matrix elements, $\langle \pi |(\bar s d)_{V-A}|K\rangle$, are known precisely -- they are extracted from data on $K^+\to \pi^0e^+\nu$ using isospin symmetry. The largest uncertainties are from the CKM inputs, $V_{cb}$ and $\gamma$, which will be improved in the future. 

The experimental challenge is that the two processes are very rare. They are set to be measured by the NA62 experiment at CERN \cite{CortinaGil:2018fkc}, and KOTO at J-PARC \cite{Ahn:2018mvc}, respectively, even if the rates are at the SM values. On the positive note, since these decays are so suppressed, the scales probed are very high, $\sim 10^3$ TeV for $Z'$ models with ${\mathcal O}(1)$ couplings. 
On top of this, $K_L\to \pi^0\nu\bar\nu$ is also CP violating.  
%The two experiments searching \jz{finish}

 \subsection{The future of NP searches with rare decays}
 The NP searches with rare decays, as well as the tests of the CKM unitarity will receive a significant boost with the upcoming Belle II and LHCb upgrades. Belle II expects to collect 50 times the Belle dataset. First collisions were seen in May 2018, and the first $B$ physics run is expected in March 2019. LHCb after upgrade II aims for roughly 100 times the present data set with an upgraded detector. A rule of thumb on the improved NP reach gives, for instance for Belle II, that the reach in $\Lambda_{\rm NP}$ will be improved by $\sim \sqrt[4]{50}=2.7\times$.  Similar if not larger increase applies to LHCb Upgrade II sensitivity improvements. This is a similar jump in energy reach as going from 13TeV LHC to a 35TeV LHC! 
 
 Among other things this also means that, if the two anomalies discussed in Sections \ref{Sec:bsmumu}, \ref{Sec:bctaunu} are not mere statistical fluctuations, we should have available measurements with 5$\sigma$ significance in a relatively near future. 

\section{Higgs and flavour}
In the SM all flavour structure is due to the Higgs Yukawa couplings, $y_f=\sqrt{2}m_f/v$. The very hierarchical values of fermion masses therefore imply similarly very hierarchical Yukawa couplings. How well have we tested this? 
There are a number of tests that are experimentally accessible to different degrees of accuracy\cite{Nir:2016zkd}
\begin{enumerate}
\item 
proportionality: is $y_{ii}\propto m_i$?
\item 
factor of proportionality: is $y_{ii}/m_i=\sqrt{2}/v$?
\item
diagonality (flavour violation): is $y_{ij}=0$ for $i\ne j$?
\item
reality (CP violation): is ${\rm Im}(y_{ij})=0$?
\end{enumerate}
Each of these questions probes a slightly different set of NP models. The proportionality, $y_{ii}\propto m_i$, and factor of proportionality, $y_{ii}/m_i=\sqrt{2}/v$, are relatively well tested for 3rd generation fermions, i.e., the Higgs couplings to top, bottom and tau. Experimentally much more difficult question is how Higgs couples to the first two generations. This is difficult to address since the SM Yukawa couplings are so small. A more modest question is: can we show that the couplings are hierarchical? The answer is already now a positive one, though for quarks this is achieved with some assumptions. Experimentally \cite{Aaboud:2017ojs,Khachatryan:2014aep,Kagan:2014ila,Altmannshofer:2015qra},
\beq
\frac{Y_{e(\mu)}^{\rm exp}}{Y_\tau^{\rm exp}}<0.22(0.10), \qquad \frac{Y_{u(c)}^{\rm exp}}{Y_t^{\rm exp}}\lesssim 0.04 , \qquad \frac{Y_{d(s)}^{\rm exp}}{Y_t^{\rm exp}}<0.7 (6),
\eeq
where the bounds for leptons come from direct measurements, on up quarks from a global fit, and on down quarks from Higgs $p_T$ distributions (global fit).

Pushing these bounds to the SM values would be very challenging, if not impossible. The one bright exception is the muon Yukawa, which will become accessible at the high-luminosity LHC as the only one among the first two generations of fermions \cite{Cepeda:2019klc}. This is quite exciting, since it is easy to imagine that part of the muon mass comes not from the SM Higgs vev, but from new small sources of the electroweak breaking (see, e.g., \cite{Altmannshofer:2015esa}). The muon Yukawa could deviate significantly from the SM, in the extreme case it could even be zero.

\begin{figure}[t]
\begin{center}
\includegraphics[width=0.42\linewidth]{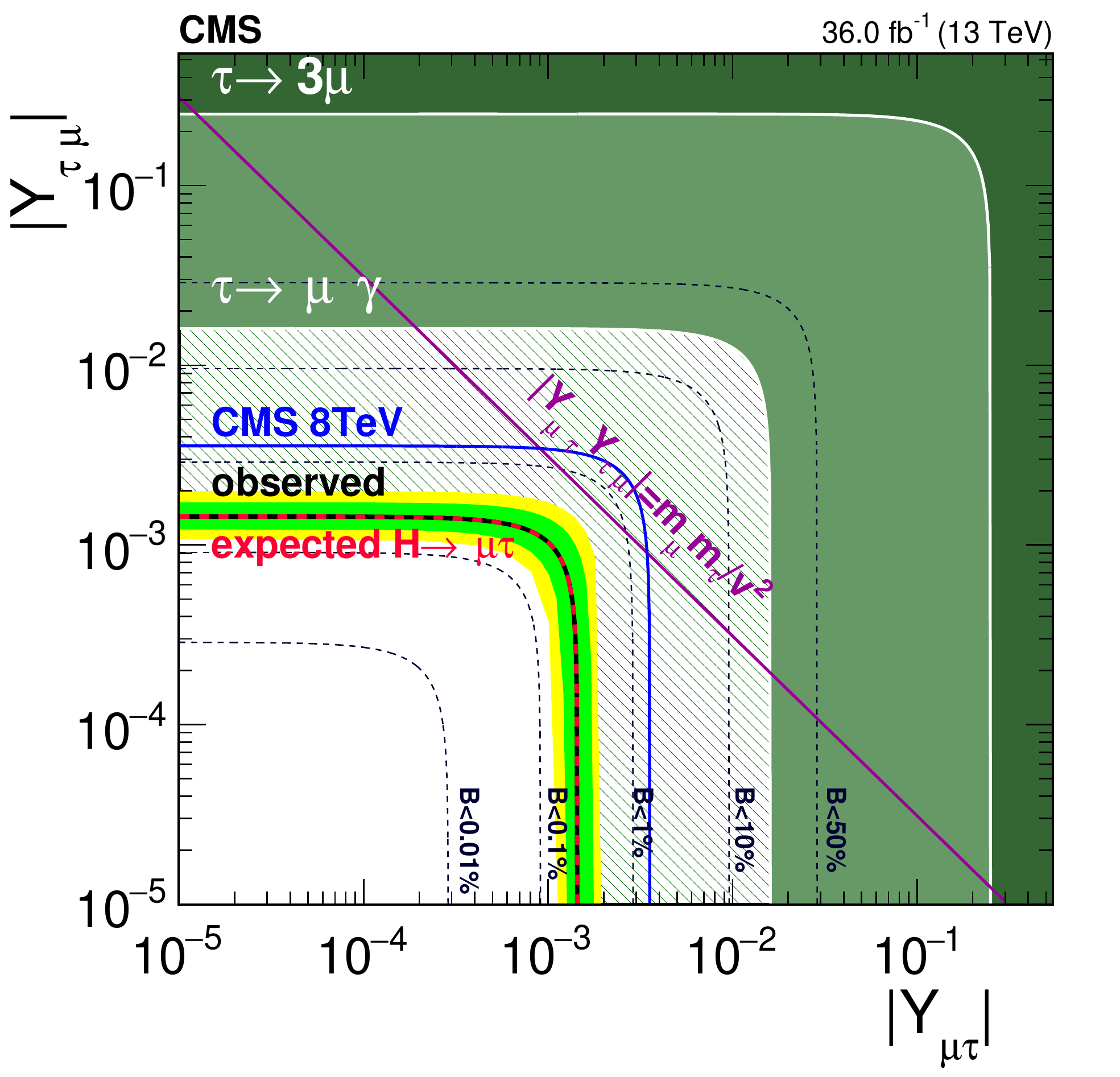}~~~~~~~~~~~~~
\includegraphics[width=0.42\linewidth]{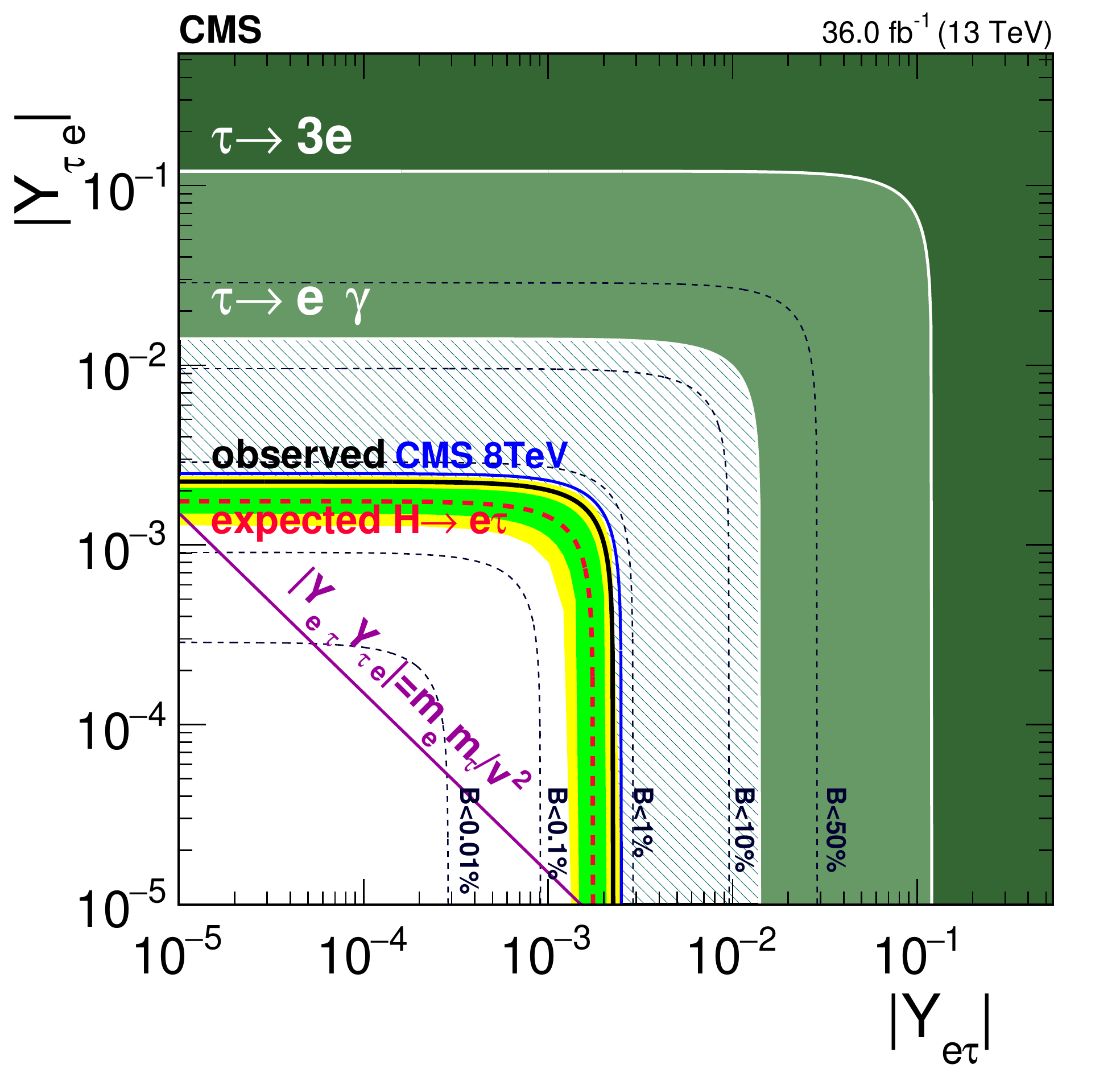}
\caption{The constraints on the Higgs flavour violating couplings to $\tau \mu$ (left) and $\tau e$ (right) (from  \cite{Sirunyan:2017xzt}). }
\label{fig:CMS_HiggsFV}
\end{center}
\end{figure}

\begin{figure}[t]
\begin{center}
\includegraphics[width=0.3\linewidth]{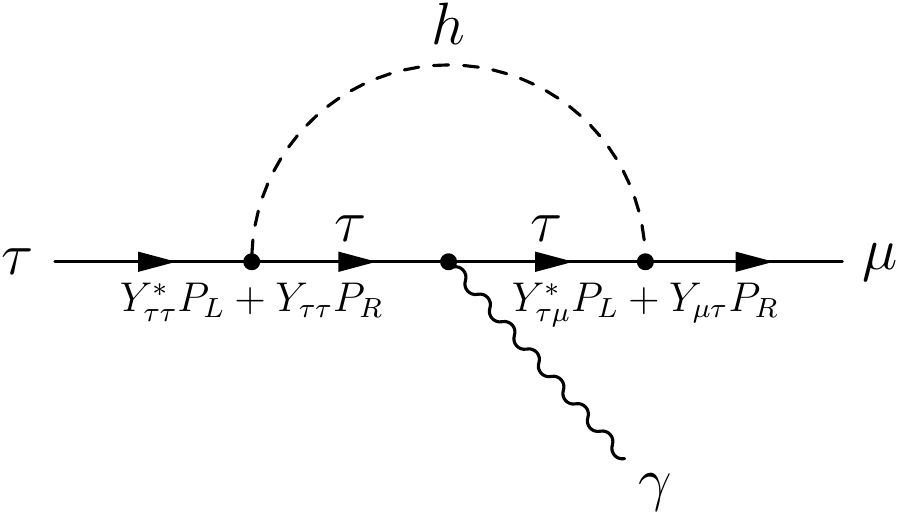}
\includegraphics[width=0.27\linewidth]{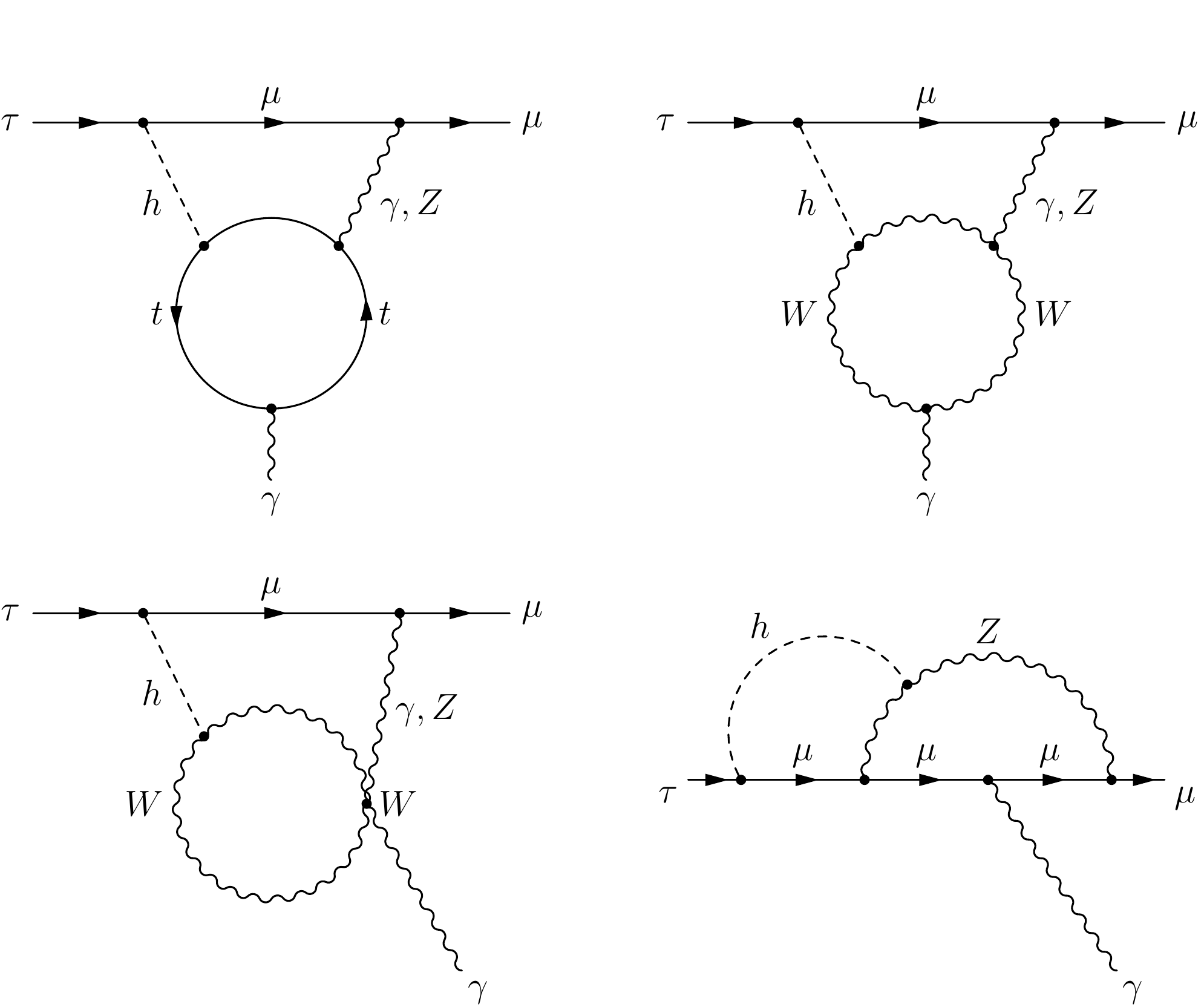}
\includegraphics[width=0.32\linewidth]{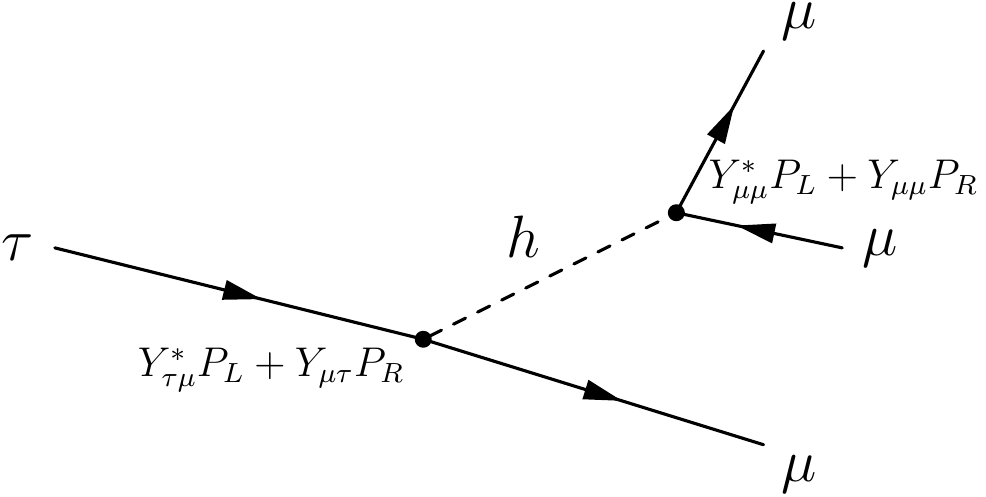}
\caption{Representative diagrams for $\tau\to \mu\gamma$ induced by Higgs flavour violating couplings at 1 loop (left) and 2 loops (middle), as well as the tree level diagram leading to $\tau\to 3\mu$ (from \cite{Harnik:2012pb}). }
\label{fig:taumu}
\end{center}
\end{figure}

Another important NP test are searches for flavour violating Higgs couplings. In the SM Higgs couplings are flavour diagonal (up to very small 1-loop corrections). Discovering flavour violating Higgs couplings would thus immediately mean New Physics. For charged lepton final states these couplings are accessible directly, by searching for $h\to \tau \mu, \tau e$ decays \cite{Harnik:2012pb,Blankenburg:2012ex}. The resulting bounds are shown in Fig. \ref{fig:CMS_HiggsFV}. If the NP corrections come from dimension 6 operators then the Higgs Yukawa couplings are, 
\beq
Y_{ij}=\frac{m_i}{v}\delta_{ij}+\frac{v^2}{\sqrt2 \Lambda^2}\hat \lambda_{ij}.
\eeq
The present bounds give for the NP scale, $\Lambda_{\mu\tau}>5.5$ TeV, $\Lambda_{e\tau}>4.4$ TeV,  taking $\hat \lambda_{ij}=1$. The Higgs decay measurements thus already probe interesting NP scales. There are also indirect bounds on flavour violating Higgs Yukawas that come from charged lepton FCNC transition. The $\tau\to \mu\gamma$ and $\tau \to 3 \mu$
are induced by the diagrams shown in Fig. \ref{fig:taumu}. While these lead to less stringent constraints on flavour violating couplings of the Higgs, see Fig. \ref{fig:CMS_HiggsFV}, this is not the case for $h\to \mu e$ decays, where the bounds on $\mu\to e\gamma$ limits the branching ratio to $Br(h\to \mu e)\lesssim 10^{-8}$, barring cancellations.

\section{Conclusions}
In the SM the flavour violation and CP violation are due to the Higgs couplings to the charged fermions. Experimentally, we know that the CKM is the dominant source of flavour violation in Nature, with the CKM phase responsible for the bulk of the CP violation in quark transitions. New physics contributions at the level of ${\mathcal O}(20\%)$ of the SM amplitude are still allowed, e.g., in the meson mixing. 

Most of the measured flavoured transitions agree with the SM predictions, with the possible exception of two quark level transitions, $b\to s \mu^+\mu^-$ and $b\to c\tau\nu$, which show $\sim 4\sigma$ discrepancies with the SM predictions. If true, this would imply many new signals in both high $p_T$ processes measured by CMS and ATLAS, as well as in precision flavour experiments LHCb, Belle II, NA62, KOTO, the muon $g-2$ experiment, etc.

There are many excellent reviews and books that go beyond the scope of these lectures, some of which were mentioned in the Introduction. A good starting point for exploring the scope of future flavour programmes at LHCb and Belle II can be found in \cite{Kou:2018nap,Bediaga:2018lhg}, and for general flavour physics possibilities at high-luminosity LHC in \cite{Cerri:2018ypt}. A good starting point for a study of new physics models that are bounded by flavour physics measurements is the introductory book \cite{Hou:2019stt}, or the somewhat more detailed, albeit older Ref. \cite{Branco:1999fs}.

{\bf Acknowledgements.} We thank the organizers of the ESHEP 2018, SSI 2018 and the US Belle II summer schools for the excellent organization and the students for the stimulating discussions. We thank Marco Ciuchini for providing the plots in Fig. \ref{fig:mixing projections}, Jorge Martin Camalich for help with diagrams in Fig. \ref{fig:bsmumu:diagrams}, and A. Greljo for careful reading of the manucript. We acknowledge support in part by the DOE grant de-sc0011784.

\appendix
\section{Nonzero neutrino masses}
\label{app:neutrino:masses}
When the neutrino masses are included, the counting of physical parameters in the SM changes from what was given in Section \ref{sec:counting:physical:params}. We show this for (i) the case that the neutrinos are Majorana fermions and (ii) the neutrinos are Dirac. 

For the case of Dirac neutrinos we enlarge the SM field content, Eqs. \eqref{eq:H:rep}, \eqref{eq:fermion:rep} by three right-handed neutrinos that are complete singlets under the SM gauge group, 
\beq
\nu_{R, i}\sim (1,1)_0.
\eeq
The Yukawa interaction Lagrangian is then enlarged by the 
\beq
{\cal L}_{\rm Yukawa}\supset - Y_\nu^{ij} \bar L_L^i H^c \nu_R^j +{\rm h.c.},
\eeq
while we assume that the Majorana mass terms, $m_{ij} \bar \nu_{R,i}^c \nu_{R,j}$, are forbidden by the conservation of total lepton number which this term would violate by two units. The counting of the physical parameters for the leptons is now completely analogous to the counting we did for the quarks in Section \ref{sec:counting:physical:params}. Using unitary transformations 
\beq
L_L\to V_L L_L, \qquad \ell_R\to V_\ell \ell_R, \qquad  \nu_R\to V_\nu \nu_R,
\eeq
one can bring the lepton Yukawa couplings to the form
\beq
Y_\ell=\diag(y_e,y_\mu,y_\tau), \qquad Y_\nu=V_{\rm PMNS}^\dagger \diag(y_{\nu}^1, y_\nu^2,y_\nu^3).
\eeq
The $3\times 3$ unitary Pontecorvo-Maki-Nakagawa-Sakata (PMNS) matrix \cite{Maki:1962mu,Pontecorvo:1957qd} is the analogue of the CKM matrix for the quarks. It has three mixing angles and one physical phase. The remaining six real parameters are the three charged lepton masses and the three neutrino masses.

 This agrees with the counting of physical parameters that follows from the general rule in Eq. \eqref{eq:general:rule}.  The $Y_\ell, Y_\nu$ matrices have $2\times (9{\rm~real}+9{\rm~im.})$ parameters, while the three unitary matrices, $V_{L}, V_\ell, V_\nu$ have in total $3\times (3{\rm~real}+6{\rm~im.})$ parameters. Out of these one corresponds to an unbroken generator, the lepton number, under which all the lepton fields change by the same phase, $L_L\to \exp(i\phi) L_L, \ell_R\to \exp(i\phi) \ell_R,  \nu_R\to \exp(i\phi) \nu_R$. Using  \eqref{eq:general:rule} there are $2\times 9-3\times 3=9$ real parameters, the six leptonic masses and three PMNS mixing angles, and $2\times 9-(3\times 6-1)=1$ imaginary physical parameter, the phase in the PMNS matrix, as anticipated.

If the neutrinos are Majorana, the field content is the same as for the SM with neutrino masses set to zero, Eqs. \eqref{eq:H:rep}, \eqref{eq:fermion:rep}. In this case the 
neutrino masses come from dimension 5 Weinberg operator after the Higgs obtains a vev. In two-component notation this is
\beq
\label{eq:Weinberg}
{\cal L}_{\rm eff}\supset - \frac{c_{ij}}{\Lambda} (H^{c\dagger} L_i) (H^{c\dagger} L_j) +{\rm h.c.}.
\eeq
The coefficient $c_{ij}$ form a $3\times 3$ symmetric complex matrix, which is described by 6 real and 6 imaginary entries. In addition, there are the 9 real and 9 imaginary parameters that describe the charge lepton Yukawa matrix, $Y_\ell$, Eq. \eqref{eq:lepton:Yukawa}. The generators of unitary transformations $L_L\to V_L L_L$, $\ell_R\to V_\ell \ell_R$ are now completely broken by the Weinberg operator in conjuction with the charged lepton Yukawa couplings. This means that we have broken generators described by $2\times (3~\text{real}+6~\text{im.})$ parameters. From the general rule \eqref{eq:general:rule} it then follows that we have $9+6- 2\times 3=9$ real and $9+6-2\times 6=3$ imaginary physical parameters. The nine real parameters are the three charged lepton masses, three neutrino masses, and the three mixing angles of the PMNS matrix. The three physical phases are the phase in the PMNS matrix, and the two Majorana phases in the Majorana mass matrix, $M_\nu=\diag (m_1, m_2 e^{i \phi_2}, m_3 e^{i\phi_3})$.

Other options for neutrino mass matrix are possible. There could be just one, two or more than three sterile neutrinos, $\nu_{R,i}$. For an introduction of phenomenological implications see, e.g., \cite{GonzalezGarcia:2002dz}. In the case where there are only Dirac mass terms, these break 
%mention Weinberg operator, count the parameters for it, the Dirac masses, count parameters then 
the individual lepton flavour numbers $U(1)_e\times U(1)_\mu\times U(1)_\tau$ down to a total lepton number $ U(1)_L$. Majorana mass terms, such as the Weinberg operator in \eqref{eq:Weinberg}, break also the $U(1)_L$. 

\bibliographystyle{h-physrev}
\bibliography{flavour_biblio}

\end{document}